\documentclass[12pt, draftclsnofoot, onecolumn]{IEEEtran}
\ifCLASSINFOpdf

\else

\fi
\usepackage{mathrsfs}%flower

\usepackage{diagbox}
\usepackage{amsmath}
\usepackage{ntheorem}
\allowdisplaybreaks

\usepackage{makeidx}  % allows for indexgeneration
\usepackage{algorithm}
\usepackage{algorithmic}
\usepackage{graphicx}
\usepackage{float}
\usepackage{subfigure}
\usepackage{epstopdf}
\usepackage{extarrows}
\usepackage{bm}
\usepackage{cite}
\usepackage{stfloats}

\usepackage{hyperref}

\usepackage{caption}
\usepackage{amssymb}
\newtheorem{theorem}{Theorem}
\usepackage{url}

\usepackage{tabularx,booktabs}
\newcolumntype{C}{>{\centering\arraybackslash}X} % centered version of "X" type
\usepackage{lipsum}
\UseRawInputEncoding
\usepackage{makecell} % Xhline Xcline

\usepackage{color}

\newtheorem{corollary}{Corollary}

\usepackage{algorithm}
\usepackage{algorithmic}
\usepackage{float}  
\usepackage{lipsum}
\makeatletter
\newenvironment{breakablealgorithm}
{% \begin{breakablealgorithm}
		\begin{center}
			\refstepcounter{algorithm}% New algorithm
			\hrule height.8pt depth0pt \kern2pt% \@fs@pre for \@fs@ruled
			\renewcommand{\caption}[2][\relax]{% Make a new \caption
				{\raggedright\textbf{\ALG@name~\thealgorithm} ##2\par}%
				\ifx\relax##1\relax % #1 is \relax
				\addcontentsline{loa}{algorithm}{\protect\numberline{\thealgorithm}##2}%
				\else % #1 is not \relax
				\addcontentsline{loa}{algorithm}{\protect\numberline{\thealgorithm}##1}%
				\fi
				\kern2pt\hrule\kern2pt
			}
		}{% \end{breakablealgorithm}
		\kern2pt\hrule\relax% \@fs@post for \@fs@ruled
	\end{center}
}
\makeatother
\begin{document}
	
\title{Two-Timescale Transmission Design for RIS-Aided Cell-Free Massive MIMO Systems}
\author{Jianxin Dai, Jin Ge, Kangda Zhi, Cunhua Pan, Zaichen Zhang, Jiangzhou Wang, \textit{Fellow, IEEE}, and Xiaohu You, \textit{Fellow, IEEE} \thanks{
		This work was supported in part by the open research fund of National Mobile Communications Research Laboratory,Southeast University. (No. 2023D03).
		
		Jianxin Dai is with the School of Science, Nanjing University of Posts and Telecommunications, Nanjing 210023, China, and also with the National Mobile Communications Research Laboratory, Southeast University, Nanjing {\color{black}210096}, China. (email: daijx@njupt.edu.cn).
		Jin Ge is with the College of Telecommunications and Information Engineering, Nanjing University of Posts and Telecommunications, Nanjing {\color{black}210003}, China. (email: 1221014211@njupt.edu.cn).
		Kangda Zhi is with the School of Electronic Engineering and Computer Science at Queen Mary University of London, {\color{black}London E1 4NS,} U.K. (email: k.zhi@qmul.ac.uk).
		Cunhua Pan, Zaichen Zhang and Xiaohu You are with the National Mobile Communications Research Laboratory, Southeast University, Nanjing 210096, China. (email: cpan, zczhang, xhyu@seu.edu.cn). Jiangzhou Wang is with the School of Engineering, University of Kent, UK. (email: j.z.wang@kent.ac.uk). (Corresponding author: Cunhua Pan).}}

\maketitle

\begin{abstract}
	This paper investigates the performance of {\color{black}a} two-timescale transmission design for uplink reconfigurable intelligent surface (RIS)-aided cell-free massive multiple-input multiple-output (CF-mMIMO) systems. We consider the Rician channel model and design the passive beamforming of RISs based on the long-time statistical channel state information (CSI), {\color{black} while the central processing unit (CPU) utilizes the maximum ratio combining (MRC) technology to perform fully centralized processing} based on the instantaneous overall channel, which are the superposition of the direct and RIS-reflected channels. Firstly, we derive the closed-form {\color{black}approximate expression of the} uplink achievable rate for arbitrary numbers of {\color{black}access point (AP)} antennas and RIS reflecting elements. 
	Relying on the derived expressions, we theoretically analyze the benefits of {\color{black}deploying RIS into} cell-free mMIMO systems and draw explicit insights. Then, based on {\color{black}the closed-form approximate rate expression} under statistical CSI, {\color{black}we optimize the phase shifts of RISs based on the genetic algorithm (GA) to maximize the sum rate and minimum rate of users, respectively.} Finally, the numerical results demonstrate {\color{black}the correctness of our derived expressions and} the benefits of deploying large-size RISs {\color{black}into cell-free mMIMO systems. Also, we investigate the optimality and convergence behaviors of the GA to verify its effectiveness. To give more beneficial analysis, we give the closed-form expression of the energy efficiency and present numerical results to show the high energy efficiency of the system with the help of RISs. Besides, our results have revealed the benefits of distributed deployment of APs and RISs in the RIS-aided mMIMO system with cell-free networks.}
\end{abstract}
\vspace{0cm}

\IEEEpeerreviewmaketitle

\section{Introduction}
Massive multiple-input multiple-output (mMIMO) technology has been widely envisioned as an essential technique to achieve high spectral efficiency and network throughput in current and future wireless communication systems \cite{bjornson2019massive}. However, in cell-centric multi-cell MIMO systems, all users in one cell are mainly managed by a dedicated {\color{black}access point (AP)}, and thus the users near the cell boundary are susceptible to severe inter-cell interference, which results in the limited throughput of the system. {\color{black}Therefore, a novel user-centric network paradigm called cell-free network has been recently proposed to solve the issue} \cite{7421222}. Unlike the concept of classical cell-centric design, cell-free networks have been proposed as a user-centric implementation that can enable all {\color{black}APs} to coordinate with each other to provide services to all users without cell boundaries \cite{9145164}. As a result, the inter-cell interference can be effectively alleviated, and the network capacity can be improved accordingly. Specifically, in cell-free mMIMO systems, the {\color{black}APs} are connected with the central processing unit (CPU) by optical cables or wireless {\color{black}fronthaul}, which serve all users in the network with a certain level of cooperation. This promising technique is considered as one of the key technologies in the next generation wireless communication systems, which can provide uniform quality of service (QoS) to multiple users through simple signal processing \cite{9586055,8685542,8599043,7827017}. One of the advantages of cell-free mMIMO systems is the lack of cell boundaries, which results in a good QoS for users. Nevertheless, {\color{black}conventional cell-free mMIMO systems require} large-scale deployment of {\color{black}APs}, leading to poor energy efficiency performance due to the high costs of both hardware and power consumption. 
%Nevertheless, the hardware cost and energy consumption of traditional mMIMO systems are too high to satisfy energy efficiency performance due to the large-scale deployment of BSs. 

{\color{black}In this regard, reconfigurable intelligent surface (RIS) (also termed intelligent reflecting surface (IRS)) has emerged as a cost-efficient revolutionary technology to support support high data rate transmission \cite{wu2020intelligent,di2020smart,di2019smart}.} RIS is a passive array composed of a large number of passive reflecting elements, which can intelligently tune the phase shift or amplitude of the incident signal with the help of a controller, thus strengthening the desired signal power or weakening the interference signal. {\color{black}Different from the {\color{black}APs, base station (BSs),} and relays, the RIS does not need active radio frequency (RF) chains and power amplifiers, introduces no additive noise, and operates in a full-duplex (FD) mode without self-interference  \cite{9119122}.}
% Therefore, the RIS is an efficient and cost-effective solution for solving the blockage problem of communication systems. 
%and without with weak enough interference with each other as long as they are deployed sufficiently far apart. 
%Specifically, RISs can constructively strengthen the desired signal from multiple BSs to the target user, which has no direct path to BS due to blockages or deconstructively weaken the interference signal from other users. It can also operate in a full-duplex (FD) mode without self-interference. Therefore, RIS is an efficient and low-cost solution to solve the blockage problem of cell-free massive MIMO systems.
Due to these attractive benefits, {\color{black}RIS is an efficient and cost-effective solution to} increase network capacity, improve transmission reliability \cite{9226616}, reduce {\color{black}transmitted} power \cite{8930608}, and enlarge wireless coverage \cite{pan2020multicell}.
RISs can also bring gains to various emerging systems, such as RIS-aided massive MIMO systems \cite{9838403}, non-orthogonal multiple access (NOMA) networks \cite{9860625}, secure communication systems \cite{9293148}, {\color{black}device-to-device (D2D)} communications \cite{9638679}, and millimeter-wave systems \cite{9314027}. All these studies provide insightful {\color{black}analysis of the} improved performance while exhibiting lower cost and higher efficiency than existing systems.

Motivated by the above background, RISs have also been integrated into cell-free mMIMO systems, {\color{black}and to} maximize the energy efficiency, Zhang \emph{et al.} \cite{9352948} proposed a hybrid beamforming (HBF) scheme that decomposed the original optimization problem into two subproblems, i.e., the digital beamforming subproblem and the RIS-based analog beamforming subproblem. 
For the spatially correlated RIS-aided cell-free mMIMO systems, the authors studied the uplink {\color{black}spectral efficiency (SE) of} the practical system over Rician fading channels \cite{9779130}. Furthermore, the impact of the generalized maximum ratio (GMR) combining was investigated in \cite{9806349}, which showed that the GMR could double the achievable {\color{black}data} rate over the maximum ratio (MR). 
The authors of \cite{9771561} investigated the secure communication in a RIS-aided cell-free mMIMO system in the presence of active eavesdropping, and a RIS-based downlink (DL) transmission scheme was proposed in \cite{9625002} to minimize the information leakage to eavesdropper while maintaining certain QoS requirements for legitimate users. In addition to the downlink, the system performance of RIS-aided cell-free mMIMO uplink was studied in \cite{9322151}. Furthermore, Ge \emph{et al.} \cite{9771957} proposed a generalized superimposed channel estimation scheme for an uplink cell-free mMIMO system, which is aided by several RISs to improve the SE and wireless coverage.

However, most of the above-mentioned contributions considered the design of the passive beamforming at the RIS based on instantaneous channel state information (CSI). In fact, there are two problems associated with the instantaneous CSI-based scheme. The first one is the pilot overhead for the knowledge of the instantaneous CSI, which is proportional to the number of RIS elements {\color{black}in most} of the existing channel estimation schemes \cite{9130088,9087848}. However, the number of reflecting elements should be large enough to serve an excessive number of users, which results in a prohibitively high pilot overhead. Secondly, the beamforming calculation and information feedback need to be performed in each channel coherence interval for the instantaneous CSI-based scheme, which incurs a high computational complexity, power consumption, and feedback overhead. {\color{black}To tackle these two problems, some authors have proposed a novel RIS design based on the two-timescale scheme, which applies the two-timescale transmission design to the cellular MIMO system assisted by only one RIS \cite{9355404,9198125,9408385,9743440}.} 
{\color{black} Specifically, the two-timescale scheme designs the phase shifts of RISs only based on slowly-varying statistical CSI, which means that the phase shift of RISs does not need to be redesigned for a long time.} The phase shifts of the RISs need to be redesigned only when the statistical CSI changes. Therefore, compared with the instantaneous CSI-based scheme that needs to redesign the phase shifts of RIS in each channel coherence interval, the statistical CSI-based strategy can significantly decrease the power consumption and the feedback overhead required by the RIS. {\color{black}This two-timescale design scheme has recently been adopted in RIS-aided cell-free mMIMO systems \cite{9665300}. Due to multiple APs and RISs, the cell-free mMIMO system requires more signal processing for system control and transmission scheduling, resulting in a more challenging design for beamforming at the AP and the phase shift design at RIS. Also, due to the distributed deployment of multiple RISs and APs in a cell-free network, the RIS-aided cascaded channels are complicated and strongly coupled between different APs, RISs, and users. As a result, the derivation of the rate expression is much more challenging than that of the RIS-aided cellular MIMO system with the centralized deployment of RIS and AP. Therefore, it is challenging to derive the closed-form analytical expressions of RIS-aided cell-free mMIMO systems and theoretically characterize the benefits of RISs in the two-timescale transmission design.} Specifically, the authors of \cite{9665300} introduced an aggregated channel estimation approach with low pilot overhead {\color{black}for the RIS-aided cell-free mMIMO systems over spatially-correlated channels.} However, the phase shifts of the RIS are not optimized in \cite{9665300}, and the line-of-sight (LoS) link of the RIS-aided cascaded channel is not considered in the proposed channel model.

Against the above background, we propose {\color{black}a two-timescale design of the uplink} RIS-aided cell-free mMIMO system with the general Rician fading model. {\color{black}Due to the heights of the AP and the RIS and considering the complex environment, both LoS and non-LoS (NLoS) channel components would exist in RIS-aided cell-free mMIMO systems. Meanwhile, by adjusting the ratio between LoS and NLoS components, we can gain some critical insights into the spatial diversity of multi-user communication, which also enables us to provide guidelines for RIS deployment.} {\color{black}Besides, the CPU utilizes a low-complexity maximum-ratio combining (MRC) receiver for fully centralized processing based on instantaneous overall CSI,} while the phase shifts of RISs are designed by utilizing the statistical CSI. In this paper, we first derive the closed-form {\color{black}approximate expression of the achievable rate, which characterizes the impact of various system parameters and provides a comprehensive analysis of the interplay between RISs and cell-free mMIMO systems.} {\color{black}Then,} we optimize the RIS phase shifts relying only on statistical CSI for both the sum rate maximization and minimum user rate maximization problems{\color{black}, which helps us better understand the impact of different optimization objectives. Moreover, we present numerical simulations to validate our analysis and demonstrate the benefits of deploying RISs to cell-free mMIMO systems.} The main contributions of this paper are summarized as follows:
\begin{itemize}
	\item First, we derive the closed-form {\color{black}approximate expression} of the uplink achievable rate by utilizing the general Rician fading model in cell-free {\color{black}mMIMO} systems. {\color{black}Under Rician fading models, the derivation becomes much more complicated due to the coupled random variables and the extensive expanded terms. We successfully propose some decoupling methods and an effective framework to complete the calculation of the expectation. 
		%	Moreover, due to the multiple number of APs and RISs, our work needs to solve the calculation of matrix expectations by dividing different cases. 
		Note that this} analytical rate expression holds for multiple {\color{black}APs} and RISs, and arbitrary numbers of {\color{black}AP} antennas and RIS elements. {\color{black}Based on the analytical rate expression, we draw some useful insights to analyze the impact of various system parameters on the achievable rate and the asymptotic behaviors of the achievable rate. The insights can serve as clear guidelines for the benefits of the proposed RIS-aided cell-free mMIMO systems.}
	\item Then, using the derived rate expression,  {\color{black}we design the optimized phase shifts of the RIS based on statistical CSI.} To guarantee fairness among different users and improve {\color{black}the system capacity, we utilize the GA method to maximize the sum rate and minimum rate of users by optimizing the RIS phase shifts based.}	
	\item Finally, we validate the correctness of the derived closed-form expressions {\color{black}and conduct} extensive simulations based on the closed-form {\color{black}approximate expression} of the achievable rate. {\color{black}To verify the effectiveness of the proposed GA method in optimizing the RIS-aided cell-free mMIMO system, we investigate the optimality and convergence behaviors of the GA. Also,} numerical results reveal the impacts of both the array size of RISs and the large-scale path loss on the performance of the RIS-aided cell-free mMIMO system. {\color{black}As a result,} we validate the promising benefits of integrating large-size RISs into cell-free mMIMO systems. {\color{black}To analyze the energy efficiency and gain more insights, we give the closed-form expression of the energy efficiency and present numerical results to show the high energy efficiency of the RIS-aided system.} {\color{black}Besides, we show the benefits of distributed deployment of APs and RISs in the RIS-aided mMIMO system with cell-free network.}
\end{itemize}

The remainder of this paper is organized as follows. Section \ref{section2} describes the system model of the uplink RIS-aided cell-free mMIMO systems based on perfect overall CSI by exploiting the two-timescale design. Section \ref{section3} derives the closed-form {\color{black}approximate expression of uplink achievable rate and gives some useful insights}. Section \ref{section4} solves the {\color{black}sum rate} maximization and the minimum user rate maximization problems by exploiting the GA-based method. Section \ref{section5} provides extensive numerical results and Section \ref{section6} concludes this paper.

\emph{Notations}: Vectors and matrices are denoted by bold lowercase and uppercase letters, respectively. $\mathbf{A}^H$ and $\mathbf{A}^T$ respectively denote the conjugate transpose and transpose. $\left|\mathbf{a}\right|$  denotes the modulus of the complex number and  $\left\|\mathbf{a}\right\|$ denotes $l_2$-norm of the vector. $\left[\mathbf{a}\right]_{m}$ denotes the $m$-th entry of the vector $\mathbf{a}$. $\left[\mathbf{a}\right]_{m,n}$ denotes the $\big((m-1)N+n\big)$-th entry of the $MN \times 1$ vector $\mathbf{a}$. {\color{black}The real, trace, and expectation operators are denoted by Re$\left\{\cdot\right\}$, Tr$\left\{\cdot\right\}$, and $\mathbb{E}\left\{\cdot\right\}$, respectively.}  $\mathbf{I}_N$ and $\mathbf{0}$ respectively denote an $N \times N$ identity matrix and a zero matrix with appropriate dimension. $\mathbb{C}^{M\times N}$ denotes the space of $M\times N$ complex matrix. Besides, $ x\sim{\cal C}{\cal N}\left( {a,b} \right)$ is a complex Gaussian distributed random variable with mean $a$ and variance $b$. Operation $\left\lfloor n \right\rfloor $ and $\left\lceil n \right\rceil $ respectively denote the nearest integer smaller than and greater than $n$, and operation $\bmod$ means returning the remainder after division.

\section{System Model}\label{section2}
\begin{figure}[h]
	\setlength{\abovecaptionskip}{0pt}
	\setlength{\belowcaptionskip}{-20pt}
	\centering
	\includegraphics[width=4in]{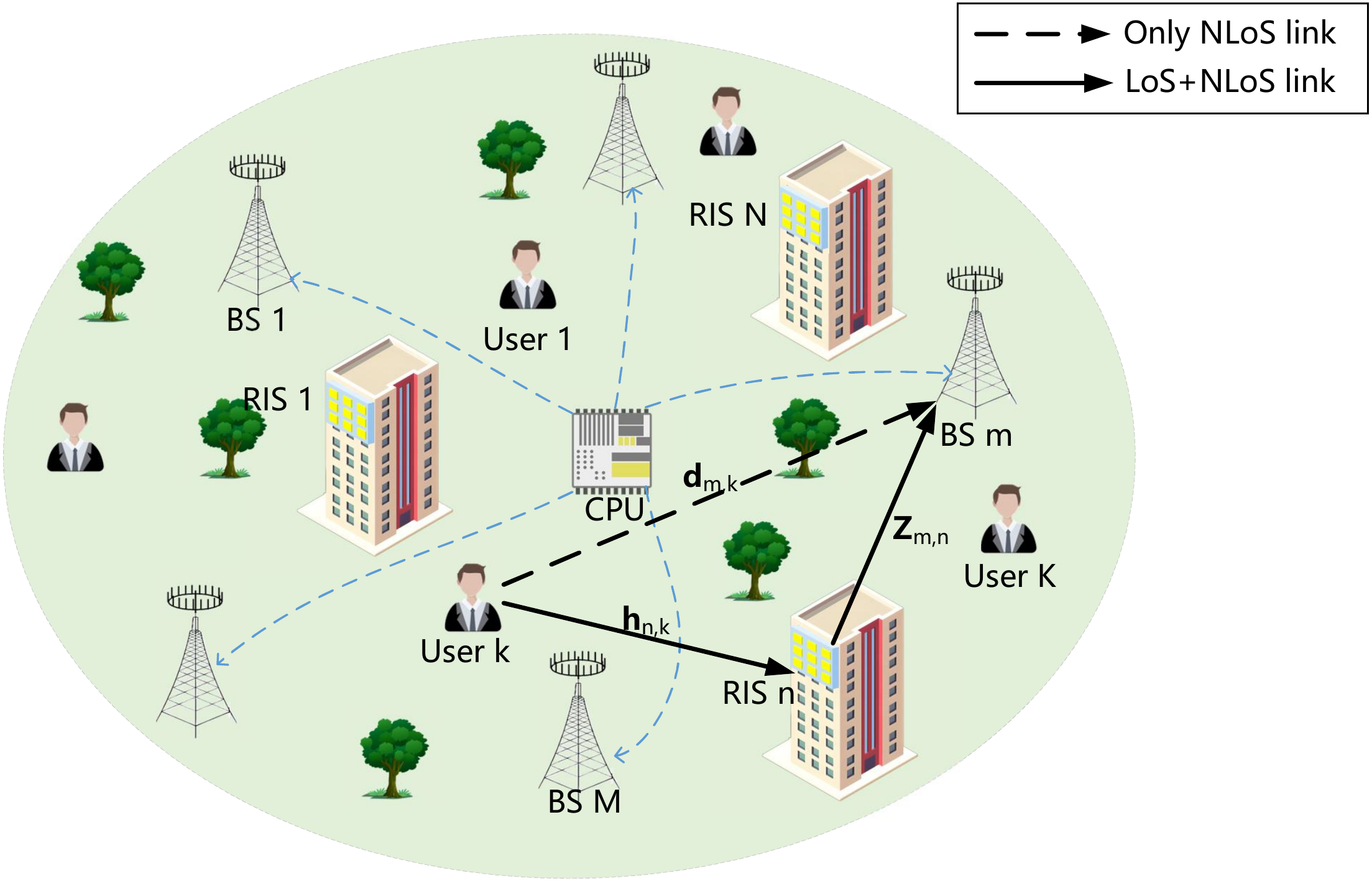}
	\DeclareGraphicsExtensions.
	\caption{The {\color{black}uplink transmission} in the RIS-aided cell-free mMIMO communication system.}
	\label{figure0}
\end{figure}
As shown in Fig. \ref{figure0}, we consider an RIS-aided cell-free mMIMO system, where multiple distributed {\color{black}APs} and RISs are deployed {\color{black}to serve all users cooperatively.} A CPU is deployed for system control and planning, and decides the transmission scheduling based on the locations of {\color{black}APs}, RISs, and users. Specifically, we consider the uplink transmission of the proposed system, where $M$ {\color{black}APs} equipped with $M_{b}$ antennas simultaneously communicate with $K$ single-antenna users with the aid of $N$ RISs. The LoS paths between users and the {\color{black}AP} could be blocked due to a large number of environmental blocking objects in the communication area. As in \cite{pan2020multicell,9355404,han2019large}, we adopt the Rayleigh fading model, and then the direct channel ${{\mathbf{D}}} \in {{\mathbb{C}}^{MM_{b} \times K}}$ can be expressed as $ {\bf {D}} = [{{\bf d}_1},{{\bf d}_2},...,{{\bf d}_K} ] $, $ \mathbf{d}_k^{T} = [{{\bf d}_{1,k}^{T}},{{\bf d}_{2,k}^{T}},...,{{\bf d}_{M,k}^{T}} ]$, ${{\bf d}_{m,k}} = \sqrt{\gamma_{m,k}} \tilde{\mathbf{d}}_{m,k} $, {\color{black}where $\gamma_{m,k}$ is the large-scale path loss, and $\tilde{\mathbf{d}}_{m,k}$ denotes the NLoS of the direct channel between {\color{black}AP} $m$ and user $k$.} The entries of $\tilde{\mathbf{d}}_{m,k}$ are independent and identically distributed (i.i.d.) complex Gaussian random variables, i.e., $ \tilde{\mathbf{d}}_{m,k} \sim \mathcal{CN}\left({\bf 0},\mathbf{I}_{M_{b}}\right)$. 

{\color{black}With the help of RIS, we provide additional paths to assist the communications for users in the rich scattering region. Furthermore, we consider only the signal reflected by the RIS the first time and ignore the signals reflected by the RIS two or more times \cite{9387559}.} Each RIS is composed of $N_{r}$ reflecting elements, and the overall configuration matrix of the RISs can be expressed as $ {\mathbf\Phi } = {\mathrm{diag}}\big\{\mathrm{diag}^{T}\left\{{{\mathbf\Phi}_1}\right\},\mathrm{diag}^{T}\left\{{\mathbf\Phi}_2\right\},...,\mathrm{diag}^{T}\left\{{\mathbf\Phi}_N\right\} \big\}$ $\in {{\mathbb{C}}^{NN_{r} \times NN_{r}}}$, ${\bf \Phi }_n = {\rm{diag}}\left\{ {{e^{j{\theta _{n,1}}}},{e^{j{\theta _{n,2}}}},...,{e^{j{\theta _{n,N_{r}}}}}} \right\}$ $\in {{\mathbb{C}}^{N_{r} \times N_{r}}}$, where ${\theta _{n,r}}\in [0,2\pi)$ represents the phase shift of the $r$-th reflecting element of RIS $n$, $1 \le n \le N$ and $1 \le r \le N_{r}$. Then, the cascaded channels from {\color{black}$K$ users to $M$ APs} can be written as $\mathbf{G} \triangleq \mathbf{Z\bf \Phi{H}} \in \mathbb{C}^{MM_{b} \times K}$, where $\mathbf{Z} \in \mathbb{C}^{MM_{b} \times NN_{r}}$ denotes the channels between the RISs and the {\color{black}APs}, and ${\bf {H}} \in \mathbb{C}^{NN_{r} \times K}$ represents the channels between the users and the RISs.

%To gain a higher line-of-sight (LoS) probability for channels between RISs and users, the RIS is usually installed on tall buildings' facades and placed near users. Besides, the RIS and BS are often deployed high above the ground, meaning LoS paths are likely to exist between the RIS and BS. Therefore, we adopt 
{\color{black}Since the AP and the RIS are often deployed at some height above the ground, and the RIS can be placed on the exterior walls of tall buildings near users, the LoS paths are likely to be present in the user-RIS and RIS-AP channels. As in \cite{pan2020multicell,9355404}, we adopt the general Rician fading model to model the RIS-related cascaded channels as follows$\footnote[1]{The more practical channel model for 6G communication systems will be left for our future work \cite{9727146,9895366,9247315}.}$}:

\begin{align}
	&{\bf {H}} = \left[{{\bf h}_1},{{\bf h}_2},...,{{\bf h}_K} \right],{{\bf h}_k^{T}} = \left[ {{{\bf h}_{1,k}^{T}},{{\bf h}_{2,k}^{T}},...,{{\bf h}_{N,k}^{T}}} \right],\label{rician1}\\
	&\mathbf{Z}=\left[ \mathbf{Z}_1,\mathbf{Z}_2,\ldots,\mathbf{Z}_N\right],  \mathbf{Z}_n^{T} = \left[{{\bf Z}_{1,n}^{T}},{{\bf Z}_{2,n}^{T}},...,{{\bf Z}_{M,n}^{T}} \right],\label{rician4}\\
	&{{\bf h}_{n,k}} = \sqrt {{\alpha _{n,k}}} \left( {\sqrt {\frac{{{\varepsilon _{n,k}}}}{{{\varepsilon _{n,k}} + 1}}} {{\bf  \overline h}_{n,k}} + \sqrt {\frac{1}{{{\varepsilon _{n,k}} + 1}}} {{\bf \tilde h}_{n,k}}} \right),\label{rician2}\\
	&{{\bf Z}_{m,n}} = \sqrt {\beta_{m,n}} \left( {\sqrt {\frac{{{\delta _{m,n}}}}{{{\delta _{m,n}} + 1}}} {{\bf \overline Z}_{m,n}} + \sqrt {\frac{1}{{\delta _{m,n} + 1}}} {{\bf \tilde Z}_{m,n}}} \right),\label{rician3}
\end{align}
where $1 \le n \le N$, $1 \le k \le K$ and $1 \le m \le M$, ${\alpha _{n,k}}$ and $\beta _{m,n}$ represent the distance-dependent large-scale path loss factors. ${\varepsilon _{n,k}}$ and ${\delta _{m,n}}$ are the Rician factors that account for the ratio of the LoS power to the NLoS power. ${{\bf \overline h}_{n,k}} \in {{\mathbb{C}}^{N_{r} \times 1}}$ and ${{\bf \overline Z}_{m,n}} \in {{\mathbb{C}}^{M_{b} \times N_{r}}}$ are deterministic LoS components of user $k$-RIS $n$ link and RIS $n$-{\color{black}AP} $m$ link, respectively. By contrast, ${{\bf \tilde h}_{n,k}} \in {{\mathbb{C}}^{N_{r} \times 1}}$ and ${{\bf \tilde Z}_{m,n}} \in {{\mathbb{C}}^{M_{b} \times N_{r}}}$ are the corresponding non-line-of-sight (NLoS) channel components, whose elements are {\color{black}i.i.d.} complex Gaussian random variables following the distribution of $ {\cal C}{\cal N}\left( {0,1} \right)$ \cite{5281762,6094142}. Furthermore, we characterize the LoS paths of {\color{black}the cascaded channel between the RIS and the {\color{black}AP} under the uniform square planar array (USPA) model \cite{9079457}}. Therefore, the LoS components ${{\bf \overline h}_{n,k}}$ and ${{\bf \overline Z}_{m,n}}$ can be respectively modelled as follows
\begin{align}
	&{{\bf \overline h}_{n,k}} = {{\bf a}_{N_{r}}}\left( {\varphi _{n,k}^a,\varphi _{n,k}^e} \right) \triangleq \mathbf{a}_{N_{r}}(n,k),\\
	&{{\bf \overline Z}_{m,n}} = {{\bf a}_{M_{b}}}\left( {\phi _{m,n}^a,\phi _{m,n}^e} \right){\bf a}_{N_{r}}^H\left( {\varphi _{m,n}^a,\varphi _{m,n}^e} \right)\triangleq \mathbf{a}_{M_{b}}(m,n) \mathbf{a}_{N_{r}}^{H}(m,n),
\end{align}
where ${{\bf a}_X}\left( {\vartheta _{}^a,\vartheta _{}^e} \right) \in {{\mathbb{C}}^{X \times 1}}$ is the array response vector, whose $x$-th entry is 
\begin{align}\label{upa}
	\left[ {{\bf a}_X}\left( {\vartheta _{}^a,\vartheta _{}^e} \right) \right]_{ x} = e^{\left\{j2\pi \frac{d}{\lambda }
		\left( {   \lfloor \left({ x} - 1 \right)/\sqrt{X}\rfloor \sin \vartheta _{}^e\sin \vartheta _{}^a
			+ \left(\left({x}-1\right)\bmod \sqrt{X}\right)  \cos \vartheta _{}^e} \right) \right\}},
\end{align}
where $d$ and $\lambda$ denote the element spacing and carrier wavelength, respectively. $\varphi _{n,k}^a$  and $\varphi _{n,k}^e$  are respectively the azimuth and elevation angles of arrival (AoA) of the incident signal at RIS $n$ from user $k$. $\varphi _{m,n}^a$ and $\varphi _{m,n}^e$ represent the azimuth and elevation angles of departure (AoD) reflected by RIS $n$ towards {\color{black}AP} $m$, respectively. $\phi _{m,n}^a$ and $\phi_{m,n}^e$ respectively denote the AoA at {\color{black}AP} $m$ from RIS $n$. Note that ${\bf \overline h}_{n,k}$ and ${\bf \overline Z}_{m,n}$ only rely on the AoA and AoD, which could be invariant for a long period.

%In the fully centralized operation, the received signals from BSs are fully processed at the CPU. 

Based on the above definitions, {\color{black}the collective signal received by all {\color{black}APs} is given by}
\begin{align}
	{\bf y} = {\bf (G+D) \mathbf{ P} x + n} =  {\bf ({Z}\Phi {H} + D)\mathbf{P} x + n},
\end{align}
where $\mathbf{P}={\rm{diag}}\left\{\sqrt{p_1},\sqrt{p_2},...,\sqrt{p_K}\right\}$ and $p_k$ is transmit power of user $k$. ${\bf x} = {\left[ {{x_1},{x_2},...,{x_K}} \right]^T}\in {{\mathbb{C}}^{K \times 1}}$ represents the transmit symbols of $K$ users, where $\mathbb{E}\left\{\left|x_{k}\right|^{2}\right\}=1$. ${\bf n} \sim {\cal C}{\cal N}\left( {0,{\sigma ^2}{\bf I}_{MM_{b}}} \right)$ denotes the receiver noise vector, which is additional white Gaussian noise (AWGN).

{\color{black}We assume all APs send their received data and pilot signals to the CPU, while the CPU perfectly estimates all channels and utilizes the MRC technology for fully centralized processing \cite{8845768,zhang2014ArRank}. Thus, with perfect CSI, the CPU performs MRC by multiplying the received signal $ \mathbf{y} $ with $(\mathbf {G + D})$ as follows}
\begin{align}
	{{\bf r} = {{\bf (G+D)}^H}{\bf y} }=  { {{\bf (G+D)}^H}{\bf (G+D)\mathbf{P}x} + {{\bf (G+D)}^H}\bf n},
\end{align}
and the received signal corresponding to user $k$ can be expressed as
\begin{align}\label{rek}
	{r_k} = \sqrt {p_k} {({\bf g}_k+{\bf d}_k)^H} {({\bf g}_k+{\bf d}_k)}{x_k} + \sum\limits_{i = 1,i \ne k}^K {\sqrt {p_i} {{({\bf g}_k+{\bf d}_k)^H}} {({\bf g}_i+{\bf d}_i)}{x_i}} + {{({\bf g}_k+{\bf d}_k)^H}}{\bf n},
\end{align}
where ${{\bf g}_k}=[{{\bf g}_{1,k}^{T}},{{\bf g}_{2,k}^{T}},...,{{\bf g}_{M,k}^{T}} ]^{T} \buildrel \Delta \over = {{\bf Z}}{\bf \Phi }{{\bf h}_k} \in {{\mathbb{C}}^{MM_{b} \times 1}}$ is the $k$-th column of $\bf G$ denoting the cascaded channel between all {\color{black}APs} and user $k$, and ${\bf g}_{m,k}\in {{\mathbb{C}}^{M_{b} \times 1}}$ represents the cascaded channel between {\color{black}AP} $m$ and user $k$. 
Based on (\ref{rek}), the uplink achievable rate expression of user $k$ can be written as {\color{black} $R_{k}=\mathbb{E}\left\{\mathrm{log_{2}}\left(1+\mathrm{SINR}_{k}\right)\right\}$, where} the signal-to-interference-plus-noise ratio (SINR) of user $k$ is given by
\begin{align}
	\mathrm{SINR}_{k} = \frac{{p_k}\left\|{\bf g}_k+{\bf d}_k\right\|^{4}}{ \sum\limits_{i=1, i\neq k}^{K}{p_i}\left|({\bf g}_k+{\bf d}_k)^{H} ({\bf g}_i+{\bf d}_i)\right|^{2}+\sigma^{2}\left\|{\bf g}_k+{\bf d}_k\right\|^{2}}.
\end{align}
\section{Uplink Achievable Rate Analysis}\label{section3}
In this section, we first derive the {\color{black}closed-form approximate} expression of the achievable rate and then draw some useful insights {\color{black}to serve as clear guidelines for the benefits of the uplink RIS-aided cell-free mMIMO systems.}
%Theoretically, we can analyze the closed-form expression to capture the impacts of various variables, including the number of the RIS elements, the number of BS antennas, the transmit power at the users. 
\begin{theorem}\label{lemma1}
In the RIS-aided cell-free mMIMO system, the {\color{black}closed-form} expression of the uplink achievable rate of user $k$ {\color{black}can be approximated as}
\begin{align}\label{rate}
   {\color{black}R_{k} \approx \log _{2}\left(1+\frac{p_k {E}_{k}^{(signal)}({\bf\Phi})}{ \sum\limits_{i=1, i \neq k}^{K} p_i I_{k i}({\bf\Phi})+\sigma^{2} E_{k}^{(noise)}({\bf\Phi})}\right),}
\end{align}
where
$E_{k}^{(noise)}({\bf\Phi})$, ${E}_{k}^{(signal)}({\bf\Phi})$, and $ I_{k i}({\bf\Phi}) $ are given by
\begin{align}\label{noise}
	\begin{array}{l}
		E_{k}^{(noise)}({\bf\Phi})\\=\mathbb{E}\left\{\left\|{\bf g}_k+{\bf d}_k\right\|^{2}\right\} \\=  {\sum\limits_{m=1}^M \sum\limits_{n_{1}=1 }^N \sum\limits_{n_{2}=1 }^N} {\sqrt{c_{m,n_{1},k}c_{m,n_{2},k}\delta _{m,n_{1}}\delta _{m,n_{2}}\varepsilon _{n_{1},k}\varepsilon _{n_{2},k}} {f_{m,n_{1},k}^{H}({\bf\Phi})}{f_{m,n_{2},k}({\bf\Phi})}} \\{\mathbf{a}_{M_{b}}^{H}(m,n_{1})} {\mathbf{a}_{M_{b}}(m,n_{2})} + 
		{\sum\limits_{m=1}^M \sum\limits_{n=1}^N c_{m,n,k} M_{b} N_{r} (\delta _{m,n}+\varepsilon _{n,k}+1)} +    {\sum\limits_{m=1}^M \gamma_{m,k} M_{b}},
	\end{array}
\end{align}
\begin{align}\label{signal power}
	\begin{array}{l}
		E_{k}^{(signal)}({\bf\Phi})\\= \mathbb{E}\left\{\left\|{\bf g}_k+{\bf d}_k\right\|^{4}\right\} \\ %等价式1
		{= \sum\limits_{m_{1}=1}^{M} \sum\limits_{m_{2}=1}^{M} \sum\limits_{n_{1}=1}^{N} \sum\limits_{n_{2}=1}^{N} \sum\limits_{n_{3}=1}^{N} \sum\limits_{n_{4}=1}^{N} \sqrt{c_{m_{1},n_{1},k}c_{m_{1},n_{2},k}c_{m_{2},n_{3},k}c_{m_{2},n_{4},k} \delta_{m_{1},n_{1}}\delta_{m_{1},n_{2}}
				\delta_{m_{2},n_{3}}\delta_{m_{2},n_{4}}}}  \\\sqrt{\varepsilon_{n_{1},k}\varepsilon_{n_{2},k}\varepsilon_{n_{3},k}\varepsilon_{n_{4},k}} {f_{m_{1},n_{1},k}^{H}({\bf\Phi})f_{m_{1},n_{2},k}({\bf\Phi})f_{m_{2},n_{3},k}^{H}({\bf\Phi})f_{m_{2},n_{4},k}}({\bf\Phi}) {\mathbf{a}_{M_{b}}^{H}(m_{1},n_{1})}{\mathbf{a}_{M_{b}}(m_{1},n_{2})}\\{\mathbf{a}_{M_{b}}^{H}(m_{2},n_{3})}{\mathbf{a}_{M_{b}}(m_{2},n_{4})} \\ +  %式子2
		{\sum\limits_{m_{1}=1}^{M} \sum\limits_{m_{2}=1}^{M} \sum\limits_{n_{1}=1}^{N} \sum\limits_{n_{2}=1}^{N} \sum\limits_{n_{3}=1}^{N}} \Big(2{\sqrt{c_{m_{1},n_{1},k}c_{m_{1},n_{2},k} \delta_{m_{1},n_{1}}\delta_{m_{1},n_{2}} \varepsilon_{n_{1},k}\varepsilon_{n_{2},k}}} c_{m_{2},n_{3},k}M_{b}N_{r}\\ (\delta_{m_{2},n_{3}}+\varepsilon_{n_{3},k}+1) {\mathrm{Re}\left\{f_{m_{1},n_{1},k}^{H}({\bf\Phi}) f_{m_{1},n_{2},k}({\bf\Phi}) {\mathbf{a}_{M_{b}}^{H}(m_{1},n_{1})} {\mathbf{a}_{M_{b}}(m_{1},n_{2})} \right\}} \\+  %式子3
		2\sqrt{c_{m_{1},n_{1},k}c_{m_{1},n_{2},k}c_{m_{2},n_{3},k}c_{m_{2},n_{1},k} \delta_{m_{1},n_{1}}\delta_{m_{1},n_{2}}
			\delta_{m_{2},n_{3}}\delta_{m_{2},n_{1}} \varepsilon_{n_{2},k}\varepsilon_{n_{3},k}}  \mathrm{Re}\left\{f_{m_{2},n_{3},k}^{H}({\bf\Phi})\right. \\ \left.f_{m_{1},n_{2},k}({\bf\Phi}) {\mathbf{a}_{M_{b}}^{H}(m_{2},n_{3})} {\mathbf{a}_{M_{b}}(m_{2},n_{1})}
		{\mathbf{a}_{N_{r}}^{H}(m_{2},n_{1})} {\mathbf{a}_{N_{r}}(m_{1},n_{1})} {\mathbf{a}_{M_{b}}^{H}(m_{1},n_{1})} {\mathbf{a}_{M_{b}}(m_{1},n_{2})}
		\right\}\Big) \\+ %式子4
		{\sum\limits_{m_{1}=1}^{M} \sum\limits_{m_{2}=1}^{M} \sum\limits_{n_{1}=1}^{N} \sum\limits_{n_{2}=1}^{N} \Big(2\sqrt{c_{m_{1},n_{1},k}c_{m_{1},n_{2},k}\delta_{m_{1},n_{1}}\delta_{m_{1},n_{2}} \varepsilon_{n_{1},k}\varepsilon_{n_{2},k}}M_{b}  \big(2c_{m_{2},n_{2},k} \mathrm{Re}\big\{f_{m_{1},n_{1},k}^{H}({\bf\Phi})}\\ {f_{m_{1},n_{2},k}({\bf\Phi}) {\mathbf{a}_{M_{b}}^{H}(m_{1},n_{1})} {\mathbf{a}_{M_{b}}(m_{1},n_{2})}}\big\}+ \gamma_{m_{2},k} f_{m_{1},n_{1},k}^{H}({\bf\Phi}) {f_{m_{1},n_{2},k}({\bf\Phi}) {\mathbf{a}_{M_{b}}^{H}(m_{1},n_{1})} {\mathbf{a}_{M_{b}}(m_{1},n_{2})}}\big)
		\\+ %式子5
		{\sqrt{c_{m_{1},n_{1},k}c_{m_{1},n_{2},k} c_{m_{2},n_{2},k}c_{m_{2},n_{1},k}\delta_{m_{1},n_{1}} \delta_{m_{1},n_{2}}\delta_{m_{2},n_{1}} \delta_{m_{2},n_{2}}} {\mathbf{a}_{M_{b}}^{H}(m_{1},n_{1})} {\mathbf{a}_{M_{b}}(m_{1},n_{2})}} \\ {{\mathbf{a}_{N_{r}}^{H}(m_{1},n_{2})}
			{\mathbf{a}_{N_{r}}(m_{2},n_{2})} {\mathbf{a}_{M_{b}}^{H}(m_{2},n_{2})} {\mathbf{a}_{M_{b}}(m_{2},n_{1})}  {\mathbf{a}_{N_{r}}^{H}(m_{2},n_{1})} {\mathbf{a}_{N_{r}}(m_{1},n_{1})}} \\ + %式子6 
		{{c_{m_{1},n_{1},k}c_{m_{2},n_{2},k}}M_{b}^{2}N_{r}^{2}\big(2(\delta_{m_{1},n_{1}}\varepsilon_{n_{2},k}+ \delta_{m_{1},n_{1}}+\varepsilon_{n_{1},k})+ \delta_{m_{1},n_{1}}\delta_{m_{2},n_{2}}+
			\varepsilon_{n_{1},k}\varepsilon_{n_{2},k}+1\big)}\Big) 
		\\ + %式子7
		{\sum\limits_{m=1}^{M} \sum\limits_{n_{1}=1}^{N} \sum\limits_{n_{2}=1}^{N} \sum\limits_{n_{3}=1}^{N}}
		2\sqrt{c_{m,n_{1},k}c_{m,n_{3},k} \delta_{m,n_{1}}\delta_{m,n_{3}} \varepsilon_{n_{1},k}\varepsilon_{n_{3},k}}  c_{m,n_{2},k}N_{r}(\varepsilon_{n_{2},k}+1)\\ f_{m,n_{1},k}^{H}({\bf\Phi}) f_{m,n_{3},k}({\bf\Phi}) {\mathbf{a}_{M_{b}}^{H}(m,n_{1})} {\mathbf{a}_{M_{b}}(m,n_{3})}
		\\ + %式子8
		{\sum\limits_{m_{1}=1}^{M} \sum\limits_{m_{2}=1}^{M} \sum\limits_{n=1}^{N}}   c_{m_{1},n,k}M_{b}^{2}N_{r}\big(2(c_{m_{2},n,k}+\gamma_{m_{2},k})(\delta_{m_{1},n}+\varepsilon_{n,k})+2\gamma_{m_{2},k}+c_{m_{2},n,k}\big)
		\\ +  %式子9
		{\sum\limits_{m=1}^{M} \sum\limits_{n_{1}=1}^{N} \sum\limits_{n_{2}=1}^{N}}\Big( c_{m,n_{1},k} c_{m,n_{2},k}M_{b} N_{r}^{2}\big( 2(\delta_{m,n_{1}}\varepsilon_{n_{2},k}+ \delta_{m,n_{1}}+\varepsilon_{n_{1},k})+ \varepsilon_{n_{1},k}\varepsilon_{n_{2},k}+1\big)
		\\ + %式子10
		2\sqrt{c_{m,n_{1},k}c_{m,n_{2},k}\delta_{m,n_{1}}\delta_{m,n_{2}}\varepsilon_{n_{1},k}\varepsilon _{n_{2},k}} \gamma_{m,k} {f_{m,n_{1},k}^{H}({\bf\Phi})}{f_{m,n_{2},k}({\bf\Phi})} {\mathbf{a}_{M_{b}}^{H}(m,n_{1})} {\mathbf{a}_{M_{b}}(m,n_{2})}
		\\ + %式子11
		4\sqrt{c_{m,n_{1},k}c_{m,n_{2},k} \delta_{m,n_{1}}\delta_{m,n_{2}} \varepsilon_{n_{1},k}\varepsilon_{n_{2},k}} c_{m,n_{2},k} \mathrm{Re}\big\{
		f_{m,n_{1},k}^{H}({\bf\Phi}) f_{m,n_{2},k}({\bf\Phi}) {\mathbf{a}_{M_{b}}^{H}(m,n_{1})} {\mathbf{a}_{M_{b}}(m,n_{2})}\big\}\Big) 
		\\ + %式子12
		{\sum\limits_{m=1}^{M} \sum\limits_{n=1}^{N}}  c_{m,n,k}M_{b}N_{r}\big( 2(\delta_{m,n}+\varepsilon_{n,k})(c_{m,n,k}+\gamma_{m,k})+c_{m,n,k}+2\gamma_{m,k}\big)
		\\+ %式子13
		\left(\sum\limits_{m=1}^{M}\gamma_{m,k}M_{b}\right)^{2}+ \sum\limits_{m=1}^{M} \gamma_{m,k}^{2}M_{b},
	\end{array}
\end{align}
\begin{align}\label{interference}
	\begin{array}{l}
		I_{ki}({\bf\Phi}) \\=
		\mathbb{E}\left\{\left|({\bf g}_k+{\bf d}_k)^{H} ({\bf g}_i+{\bf d}_i)\right|^{2}\right\}\\%等价式1
		{= \sum\limits_{m_{1}=1}^{M} \sum\limits_{m_{2}=1}^{M} \sum\limits_{n_{1}=1}^{N} \sum\limits_{n_{2}=1}^{N} \sum\limits_{n_{3}=1}^{N} \sum\limits_{n_{4}=1}^{N} \sqrt{c_{m_{1},n_{1},k}c_{m_{1},n_{2},i}c_{m_{2},n_{3},i}c_{m_{2},n_{4},k} \delta_{m_{1},n_{1}}\delta_{m_{1},n_{2}}
				\delta_{m_{2},n_{3}}\delta_{m_{2},n_{4}}}}\\\sqrt{\varepsilon_{n_{1},k}\varepsilon_{n_{2},i}\varepsilon_{n_{3},i}\varepsilon_{n_{4},k}}  {f_{m_{1},n_{1},k}^{H}({\bf\Phi})f_{m_{1},n_{2},i}({\bf\Phi})f_{m_{2},n_{3},i}^{H}({\bf\Phi})f_{m_{2},n_{4},k}}({\bf\Phi}) {\mathbf{a}_{M_{b}}^{H}(m_{1},n_{1})}{\mathbf{a}_{M_{b}}(m_{1},n_{2})}\\{\mathbf{a}_{M_{b}}^{H}(m_{2},n_{3})}{\mathbf{a}_{M_{b}}(m_{2},n_{4})} \\ +  %式子2
		{\sum\limits_{m_{1}=1}^{M} \sum\limits_{m_{2}=1}^{M} \sum\limits_{n_{1}=1}^{N} \sum\limits_{n_{2}=1}^{N} \sum\limits_{n_{3}=1}^{N}} \Big(2{\sqrt{c_{m_{1},n_{1},k}c_{m_{1},n_{2},i}c_{m_{2},n_{3},i}c_{m_{2},n_{3},k} \delta_{m_{1},n_{1}}\delta_{m_{1},n_{2}} \varepsilon_{n_{1},k}\varepsilon_{n_{2},i} \varepsilon_{n_{3},i}\varepsilon_{n_{3},k}}}M_{b}\\ {\mathrm{Re}\left\{f_{m_{1},n_{1},k}^{H}({\bf\Phi}) f_{m_{1},n_{2},i}({\bf\Phi}) {\mathbf{a}_{M_{b}}^{H}(m_{1},n_{1})} {\mathbf{a}_{M_{b}}(m_{1},n_{2})} \overline{\mathbf{h}}_{n_{3},i}^{H} \overline{\mathbf{h}}_{n_{3},k} \right\}} \\+  %式子3
		\sqrt{\delta_{m_{1},n_{1}}\delta_{m_{1},n_{2}}
			\delta_{m_{2},n_{2}}\delta_{m_{2},n_{3}}} \big(\sqrt{c_{m_{1},n_{1},k}c_{m_{1},n_{2},i}c_{m_{2},n_{2},i}c_{m_{2},n_{3},k} \varepsilon_{n_{1},k}\varepsilon_{n_{3},k}} f_{m_{1},n_{1},k}^{H}({\bf\Phi}) f_{m_{2},n_{3},k}({\bf\Phi})+ \\ %换行
		\sqrt{c_{m_{1},n_{1},i}c_{m_{1},n_{2},k}c_{m_{2},n_{2},k}c_{m_{2},n_{3},i} \varepsilon_{n_{1},i}\varepsilon_{n_{3},i}} f_{m_{1},n_{1},i}^{H}({\bf\Phi}) f_{m_{2},n_{3},i}({\bf\Phi})\big) {\mathbf{a}_{M_{b}}^{H}(m_{1},n_{1})} {\mathbf{a}_{M_{b}}(m_{1},n_{2})}\\ %换行
		{\mathbf{a}_{N_{r}}^{H}(m_{1},n_{2})} {\mathbf{a}_{N_{r}}(m_{2},n_{2})} {\mathbf{a}_{M_{b}}^{H}(m_{2},n_{2})} {\mathbf{a}_{M_{b}}(m_{2},n_{3})}
		\Big) \\+ %式子4
		{\sum\limits_{m_{1}=1}^{M} \sum\limits_{m_{2}=1}^{M} \sum\limits_{n_{1}=1}^{N} \sum\limits_{n_{2}=1}^{N} \Big(2\sqrt{c_{m_{1},n_{1},k}c_{m_{1},n_{2},i}\delta_{m_{1},n_{1}}\delta_{m_{1},n_{2}}}M_{b}\mathrm{Re}\left\{\left(\sqrt{c_{m_{2},n_{2},i}c_{m_{2},n_{2},k}\varepsilon_{n_{1},k}\varepsilon_{n_{2},k}}\right.\right.}   \\
		{\left.f_{m_{1},n_{1},k}^{H}({\bf\Phi}) f_{m_{1},n_{2},k}({\bf\Phi}) + \sqrt{c_{m_{2},n_{1},i}c_{m_{2},n_{1},k} \varepsilon_{n_{1},i}\varepsilon_{n_{2},i}} f_{m_{1},n_{1},i}^{H}({\bf\Phi}) f_{m_{1},n_{2},i}({\bf\Phi})\right)} {\left.{\mathbf{a}_{M_{b}}^{H}(m_{1},n_{1})} {\mathbf{a}_{M_{b}}(m_{1},n_{2})}\right\} } \\+ %式子5
		{\sqrt{c_{m_{1},n_{1},k}c_{m_{1},n_{2},i} c_{m_{2},n_{2},i}c_{m_{2},n_{1},k}\delta_{m_{1},n_{1}} \delta_{m_{1},n_{2}}\delta_{m_{2},n_{1}} \delta_{m_{2},n_{2}}} {\mathbf{a}_{M_{b}}^{H}(m_{1},n_{1})} {\mathbf{a}_{M_{b}}(m_{1},n_{2})}} \\ {{\mathbf{a}_{N_{r}}^{H}(m_{1},n_{2})}
			{\mathbf{a}_{N_{r}}(m_{2},n_{2})} {\mathbf{a}_{M_{b}}^{H}(m_{2},n_{2})} {\mathbf{a}_{M_{b}}(m_{2},n_{1})}  {\mathbf{a}_{N_{r}}^{H}(m_{2},n_{1})} {\mathbf{a}_{N_{r}}(m_{1},n_{1})}} \\ + %式子6 
		{\sqrt{c_{m_{1},n_{1},k}c_{m_{1},n_{2},i} c_{m_{2},n_{2},i}c_{m_{2},n_{1},k} \varepsilon_{n_{1},k}\varepsilon_{n_{1},i}
				\varepsilon_{n_{2},i}\varepsilon_{n_{2},k}}M_{b}^{2}\overline{\mathbf{h}}_{n_{1},k}^{H} \overline{\mathbf{h}}_{n_{1},i} \overline{\mathbf{h}}_{n_{2},i}^{H} \overline{\mathbf{h}}_{n_{2},k}}\Big) \\ + %式子7
		{\sum\limits_{m=1}^{M} \sum\limits_{n_{1}=1}^{N} \sum\limits_{n_{2}=1}^{N} \sum\limits_{n_{3}=1}^{N}}
		\Big(\sqrt{c_{m,n_{1},k}c_{m,n_{3},k} \delta_{m,n_{1}}\delta_{m,n_{3}} \varepsilon_{n_{1},k}\varepsilon_{n_{3},k}}  c_{m,n_{2},i}N_{r}(\varepsilon_{n_{2},i}+1) f_{m,n_{1},k}^{H}({\bf\Phi})\\ f_{m,n_{3},k}({\bf\Phi}) {\mathbf{a}_{M_{b}}^{H}(m,n_{1})} {\mathbf{a}_{M_{b}}(m,n_{3})} + %式子8
		\sqrt{c_{m,n_{2},i}c_{m,n_{3},i} \delta_{m,n_{2}}\delta_{m,n_{3}} \varepsilon_{n_{2},i}\varepsilon_{n_{3},i}}  c_{m,n_{1},k}N_{r}(\varepsilon_{n_{1},k}+1)\\ f_{m,n_{3},i}^{H}({\bf\Phi}) f_{m,n_{2},i}({\bf\Phi}) {\mathbf{a}_{M_{b}}^{H}(m,n_{3})} {\mathbf{a}_{M_{b}}(m,n_{2})}\Big)
		\\+
		{\sum\limits_{m_{1}=1}^{M} \sum\limits_{m_{2}=1}^{M} \sum\limits_{n=1}^{N}} \sqrt{c_{m_{1},n,k}c_{m_{1},n,i} c_{m_{2},n,i}c_{m_{2},n,k}} M^{2}_{b} N_{r} (2\delta_{m_{1},n}+\varepsilon_{n,k}+\varepsilon_{n,i}+1) \\ +  %式子10
		{\sum\limits_{m=1}^{M} \sum\limits_{n_{1}=1}^{N} \sum\limits_{n_{2}=1}^{N}} \Big( 
		\sqrt{\delta_{m,n_{1}}\delta_{m,n_{2}}} \big(\sqrt{c_{m,n_{1},k}c_{m,n_{2},k}\varepsilon _{n_{1},k}\varepsilon _{n_{2},k}} \gamma_{m,i} {f_{m,n_{1},k}^{H}({\bf\Phi})}{f_{m,n_{2},k}({\bf\Phi})}
		\\+ \sqrt{c_{m,n_{1},i}c_{m,n_{2},i}\varepsilon _{n_{1},i}\varepsilon _{n_{2},i}} \gamma_{m,k} {f_{m,n_{1},i}^{H}({\bf\Phi})}{f_{m,n_{2},i}({\bf\Phi})}\big) {\mathbf{a}_{M_{b}}^{H}(m,n_{1})} {\mathbf{a}_{M_{b}}(m,n_{2})}
		\\+
		c_{m,n_{1},k} c_{m,n_{2},i} M_{b} N_{r}^{2}( \delta_{m,n_{1}}\varepsilon_{n_{2},i}+ \delta_{m,n_{2}}\varepsilon_{n_{1},k}+ \varepsilon_{n_{1},k}\varepsilon_{n_{2},i}+ \delta_{m,n_{1}}+\delta_{m,n_{2}}+ \varepsilon_{n_{1},k}+\varepsilon_{n_{2},i}+1)\Big)
		\\+ %式子12
		{\sum\limits_{m=1}^M \sum\limits_{n=1}^N} 
		{M_{b} N_{r} \big(c_{m,n,k} (\delta _{m,n}+\varepsilon _{n,k}+1)\gamma_{m,i}+c_{m,n,i} (\delta _{m,n}+\varepsilon _{n,i}+1)\gamma_{m,k}}\big)
		\\+ %式子13
		{\sum\limits_{m=1}^M \gamma_{m,k}\gamma_{m,i}M_{b}},
	\end{array}
\end{align}
with
\begin{align}\label{C_mnk}
	c_{m,n,k} \triangleq \frac{\beta _{m,n}\alpha _{n,k}}{\left({\delta _{m,n}+1} \right) \left({\varepsilon _{n,k}+1}\right) },
\end{align}
\begin{align}\label{f_k_Phi_1}
	f_{m,n,k}({\bf\Phi}) \triangleq &\mathbf{a}_{N_{r}}^{H}(m,n) {\bf\Phi}_{n} \overline{\mathbf{h}}_{n,k}=\sum\nolimits_{r=1}^{N_{r}} e^{j\left(\zeta_{r}^{m,n,k}+\theta_{n,r}\right)},
\end{align}
and
\begin{align}\label{f_k_Phi_2}
	\zeta_{r}^{m,n,k}=&2 \pi \frac{d}{\lambda}  \left(   \lfloor(r-1) / \sqrt{N_{r}}\rfloor  \left(\sin \varphi_{n,k}^{e} \sin \varphi_{n,k}^{a}-\sin \varphi_{m,n}^{e} \sin \varphi_{m,n}^{a}\right)\right.\nonumber\\
	&\left.+\big((r-1) \bmod \sqrt{N_{r}}\big)\left( \cos \varphi_{n,k}^{e}- \cos \varphi_{m,n}^{e}\right)\right).
\end{align}
\end{theorem}

\itshape \textbf{Proof:}  \upshape Please refer to Appendix \ref{appB}.  \hfill $\blacksquare$

%The closed-form expression in Lemma \ref{lemma1} does not involve the calculation of inverse matrices and the numerical computation of integrals. In contrast to time-consuming Monte Carlo simulations, the evaluation of the rate based on Theorem 2 has a low computational complexity even if M and N are large numbers, as usually is in RIS-aided massive MIMO systems.

Theorem \ref{lemma1} shows that the closed-form {\color{black}approximate} expression (\ref{rate}) does not rely on the instantaneous CSI $ {{\bf \tilde h}_k} $, $ \bf \tilde Z$, and $ {{\bf \tilde d}_k} $, but it only depends on the slowly-varying statistical CSI, i.e., the AoA and AoD in $ {{\bf \overline h}_k} $ and $ \bf \overline Z$, the large-scale path-loss factors ${\alpha _{n,k}}$, $\beta _{m,n}$, $\gamma_{m,k}$, Rician factors ${\varepsilon _{n,k}}$ and ${\delta _{m,n}}$. {\color{black}Therefore, based on the analytical rate  expression in (\ref{rate}) as an objective function for the RIS-aided system design, we can exploit long-term statistical CSI to design and optimize the phase shifts of the RISs. Unlike the time-consuming Monte Carlo (MC) simulation that requires repeating the calculation $10^{5}$ times to calculate the expectation, our expression can help evaluate the system performance more quickly, even if $M$ and $N$ are large numbers. Even though the derived analytical expressions may seem long and cumbersome at the first sight, they provide clear insights in terms of the key system parameters, i.e., the number of APs $M$, the number of RISs $N$, the number of AP antennas $M_{b}$, the number of reflecting elements $N_{r}$, and the phase shifts of RISs ${\bf\Phi}$. For example, due to the multi-user interference term $I_{k i}({\bf\Phi})$ scaled as $\mathcal{O}(M_{b}^2)$, we conclude that the RIS-aided cell-free mMIMO system suffers from stronger multi-user interference than the conventional cell-free mMIMO system. Also, we find that parameters $M$ and $N$ affect the rate only as a role of the number of cumulative terms, which clarifies the impacts of these parameters. In the following, we draw some useful analysis of RIS-aided cell-free mMIMO systems based on the derived closed-form rate expression, including the asymptotic behavior of the achievable rate for large $M_{b}$ and $N_{r}$, the impact of the Rician factors, and the significance of optimizing the RIS phase shift.} 

\begin{corollary}\label{corollary1}
	The achievable {\color{black}rate} for {\color{black}RIS-free cell-free mMIMO systems} can be obtained by setting $c_{m,n,k}=0$, $ \forall m,n,k$, which is given by {\color{black}$R_{k}^{(\mathrm{w})} \triangleq \log _{2}\Big(1+\mathrm{SINR}_{k}^{(\mathrm{w})}\Big)$} with
	\begin{align}\label{SINK_w}
		\mathrm{SINR}_{k}^{(\mathrm{w})}{\color{black}\approx}  
		\frac{p_k \bigg(\bigg(\sum\limits_{m=1}^{M}\gamma_{m,k}M_{b}\bigg)^{2}+\sum\limits_{m=1}^{M}\gamma_{m,k}^{2}M_{b}\bigg)}       { \sum\limits_{i=1, i \neq k}^{K}  {\sum\limits_{m=1}^M p_i \gamma_{m,k}\gamma_{m,i}M_{b}}       +\sigma^{2} {\sum\limits_{m=1}^M \gamma_{m,k} M_{b} }}.
	\end{align}
\end{corollary}

{\color{black}\itshape \textbf{Proof:}  \upshape  We can ignore the zero terms by substituting $c_{m,n,k}=0, \forall m,n,k$ into the rate expression in (\ref{rate}) and then complete the proof after some simplifications. \hfill $\blacksquare$}

Corollary \ref{corollary1} shows that in the conventional cell-free mMIMO system, {\color{black}i.e., the RIS-free cell-free mMIMO system,} when the number of {\color{black}AP} antennas $M_{b}$ is large, the interference power could be ignored compared with the desired signal. However, this feature no longer holds for the RIS-aided cell-free mMIMO system. We can find that both the signal term in (\ref{signal power}) and the interference term in (\ref{interference}) scale as $\mathcal{O}(M_{b}^2)$, which means that a large $M_{b}$ will make {\color{black}the SINR} converge to a constant. Therefore, a question arises as to whether {\color{black}this additional interference will limit the gains of RISs.} To answer this question and get more insights, we consider some special cases to explore the gains of deploying RISs into the cell-free mMIMO system.

\begin{corollary}\label{corollary2}
	When the {\color{black}RIS-aided} cascaded channels are pure NLoS (i.e., $\varepsilon_{n,k}=0$ and $\delta_{m,n}=0$, $ \forall m,n,k$), the achievable rate is {\color{black}$R_{k}^{(\mathrm{NL})} \triangleq \log _{2}\Big(1+\mathrm{SINR}_{k}^{(\mathrm{NL})}\Big)$} and $\mathrm{SINR}_{k}^{(\mathrm{NL})}$ is given by
	\begin{align}\label{SINK_NL}
		\mathrm{SINR}_{k}^{(\mathrm{NL})}{\color{black}\approx} 
		\frac{p_k {E}_{k}^{(signal,\mathrm{NL})}}{ \sum\limits_{i=1, i \neq k}^{K} p_i I_{k i}^{(\mathrm{NL})}+\sigma^{2} E_{k}^{(noise,\mathrm{NL})}},
	\end{align}
	where
	\begin{align}\label{noise_NL}
		\begin{array}{l}
			E_{k}^{(noise,\mathrm{NL})} =  
			{\sum\limits_{m=1}^M \sum\limits_{n=1}^N   c_{m,n,k}^{(\mathrm{NL})} M_{b}N_{r}} +    {\sum\limits_{m=1}^M \gamma_{m,k}M_{b} },
		\end{array}
	\end{align}
	\begin{align}\label{signal power_NL}
		\begin{array}{l}
			E_{k}^{(signal,\mathrm{NL})} \\
			=
			{\sum\limits_{m_{1}=1}^{M} \sum\limits_{m_{2}=1}^{M} \sum\limits_{n_{1}=1}^{N} \sum\limits_{n_{2}=1}^{N} }
			{c_{m_{1},n_{1},k}^{(\mathrm{NL})}c_{m_{2},n_{2},k}^{(\mathrm{NL})}}M_{b}^{2}N_{r}^{2} 
			+  %式子9
			{\sum\limits_{m=1}^{M} \sum\limits_{n_{1}=1}^{N} \sum\limits_{n_{2}=1}^{N}} c_{m,n_{1},k}^{(\mathrm{NL})} c_{m,n_{2},k}^{(\mathrm{NL})} M_{b} N_{r}^{2}
			\\+ 
			{\sum\limits_{m_{1}=1}^{M} \sum\limits_{m_{2}=1}^{M} \sum\limits_{n=1}^{N}}   c_{m_{1},n,k}^{(\mathrm{NL})} M_{b}^{2}N_{r}\left(2\gamma_{m_{2},k}+c_{m_{2},n,k}^{(\mathrm{NL})}\right)
			+
			{\sum\limits_{m=1}^{M} \sum\limits_{n=1}^{N}}  c_{m,n,k}^{(\mathrm{NL})}M_{b}N_{r} \left(2\gamma_{m,k}+c_{m,n,k}^{(\mathrm{NL})}\right)
			\\ + %式子13
			\left(\sum\limits_{m=1}^{M}\gamma_{m,k}M_{b}\right)^{2}+ \sum\limits_{m=1}^{M} \gamma_{m,k}^{2}M_{b},
		\end{array}
	\end{align}
	\begin{align}\label{interference_NL}
		\begin{array}{l}
			I_{ki}^{(\mathrm{NL})} \\
			= 
			{\sum\limits_{m_{1}=1}^{M} \sum\limits_{m_{2}=1}^{M} \sum\limits_{n=1}^{N}} \sqrt{c_{m_{1},n,k}^{(\mathrm{NL})}c_{m_{1},n,i}^{(\mathrm{NL})} c_{m_{2},n,i}^{(\mathrm{NL})}c_{m_{2},n,k}^{(\mathrm{NL})}} M^{2}_{b} N_{r}  +  %式子10
			{\sum\limits_{m=1}^{M} \sum\limits_{n_{1}=1}^{N} \sum\limits_{n_{2}=1}^{N}}  c_{m,n_{1},k}^{(\mathrm{NL})} c_{m,n_{2},i}^{(\mathrm{NL})} M_{b} N_{r}^{2}
			\\+ %式子12
			{\sum\limits_{m=1}^M \sum\limits_{n=1}^N} 
			{M_{b} N_{r} \left(c_{m,n,k}^{(\mathrm{NL})} \gamma_{m,i} +c_{m,n,i}^{(\mathrm{NL})}\gamma_{m,k} \right)}
			+ %式子13
			{\sum\limits_{m=1}^M \gamma_{m,k}\gamma_{m,i}M_{b}},
		\end{array}
	\end{align}
	with
	\begin{align}\label{C_mnk_NL}
		c_{m,n,k}^{(\mathrm{NL})} \triangleq {\beta _{m,n}\alpha _{n,k}}.
	\end{align}
\end{corollary}

{\color{black}\itshape \textbf{Proof:}  \upshape  We can ignore the zero terms by substituting $\delta_{m,n}=0, \forall m,n$ into the rate expression in (\ref{rate}). Then, we can complete the proof by simplifying the expressions and noting that $c_{m,n,k}^{(\mathrm{NL})}\left(\varepsilon_{n,k}+1\right)=\beta_{m,n}\alpha_{n,k}, \forall m,n,k$.  \hfill $\blacksquare$}

Corollary \ref{corollary2} represents the scenario where the cascaded channels are in a rich scattering environment, and the Rician channel reduces to the Rayleigh channel. It can be seen that {\color{black}the rate expression $R_{k}^{(\mathrm{NL})}$ does not contain the parameter ${\bf\Phi}$, which means the rate performance is not affected by the phase shifts of RISs. Therefore, there is no need to design the phase shifts of RISs when the communication area around RIS is a rich scattering environment.}

{\color{black}Besides, when $M_{b}\rightarrow\infty$, we can get the result by ignoring the insignificant terms which do not scale with $M_{b}$. Note that ${\mathbf{a}_{M_{b}}^{H}(m_{1},n_{1})} {\mathbf{a}_{M_{b}}(m_{2},n_{2})}=M_{b}$ when $m_{1}=m_{2}$ and $n_{1}=n_{2}$. Then, as $M_{b}\rightarrow\infty$, we have}
%or a large number of reflecting elements
\begin{align}
	{R}_{k}^{\mathrm{(NL)}}\rightarrow {R}_{k}^{\mathrm{(NL)}(1)}{\color{black}\approx  \log _{2}\left(1+\frac{p_k {E}_{k}^{(signal,\mathrm{NL})(1)}}  {\sum\limits_{i=1, i \neq k}^{K} p_i I_{k i}^{(\mathrm{NL})(1)}}\right), }
\end{align}
where 
\begin{align}
	\begin{array}{l}
		{E}_{k}^{(signal,\mathrm{NL})(1)}\\
		=
		{\sum\limits_{m_{1}=1}^{M} \sum\limits_{m_{2}=1}^{M} \sum\limits_{n_{1}=1}^{N} \sum\limits_{n_{2}=1}^{N} }
		{c_{m_{1},n_{1},k}^{(\mathrm{NL})}c_{m_{2},n_{2},k}^{(\mathrm{NL})}}N_{r}^{2}
		+
		{\sum\limits_{m_{1}=1}^{M} \sum\limits_{m_{2}=1}^{M} \sum\limits_{n=1}^{N}}   c_{m_{1},n,k}^{(\mathrm{NL})} N_{r}\left(2\gamma_{m_{2},k}+c_{m_{2},n,k}^{(\mathrm{NL})}\right)
		+
		\left(\sum\limits_{m=1}^{M}\gamma_{m,k}\right)^{2},
	\end{array}
\end{align}
\begin{align}
	\begin{array}{l}
		I_{k i}^{(\mathrm{NL})(1)}
		=
		{\sum\limits_{m_{1}=1}^{M} \sum\limits_{m_{2}=1}^{M} \sum\limits_{n=1}^{N}} \sqrt{c_{m_{1},n,k}^{(\mathrm{NL})}c_{m_{1},n,i}^{(\mathrm{NL})} c_{m_{2},n,i}^{(\mathrm{NL})}c_{m_{2},n,k}^{(\mathrm{NL})}} N_{r}.
	\end{array}
\end{align}
%\begin{align}
%	{R}_{k}^{\mathrm{(NL)}}\rightarrow {R}_{k}^{\mathrm{(NL)}(2)}= \frac{\tau_c-\tau}{\tau_c}\log _{2}\left(1+\frac{p_k {E}_{k}^{(signal,\mathrm{NL})(2)}({\bf\Phi})}  {\sum\limits_{i=1, i \neq k}^{K} p_i I_{k i}^{(\mathrm{NL})(2)}({\bf\Phi})}\right), \mathrm{as} \; N_{r}\rightarrow\infty,
%\end{align}
%where
%\begin{align}
%	\begin{array}{l}
%		{E}_{k}^{(signal,\mathrm{NL})(2)}({\bf\Phi})
%		\\=
%		{\sum\limits_{m_{1}=1}^{M} \sum\limits_{m_{2}=1}^{M} \sum\limits_{n_{1}=1}^{N} \sum\limits_{n_{2}=1}^{N} }
%		{c_{m_{1},n_{1},k}^{(\mathrm{NL})}c_{m_{2},n_{2},k}^{(\mathrm{NL})}}M_{b}^{2}
%		+
%		{\sum\limits_{m=1}^{M} \sum\limits_{n_{1}=1}^{N} \sum\limits_{n_{2}=1}^{N}} c_{m,n_{1},k}^{(\mathrm{NL})} c_{m,n_{2},k}^{(\mathrm{NL})} M_{b},
%	\end{array}
%\end{align}
%\begin{align}
%	\begin{array}{l}
%		I_{k i}^{(\mathrm{NL})(2)}({\bf\Phi})
%		=
%		{\sum\limits_{m=1}^{M} \sum\limits_{n_{1}=1}^{N} \sum\limits_{n_{2}=1}^{N}}  c_{m,n_{1},k}^{(\mathrm{NL})} c_{m,n_{2},i}^{(\mathrm{NL})} M_{b}.
%	\end{array}
%\end{align}

We can find that $ {R}_{k}^{\mathrm{(NL)}(1)}\rightarrow\infty$ when  $N_{r}\rightarrow\infty$, which means the rate  ${R}_{k}^{\mathrm{(NL)}}$ will increase without {\color{black}limit} when both $N_{r}\rightarrow\infty$ and $M_{b}\rightarrow\infty$.
Therefore, even though there are no LoS paths in the cascaded channels, significant performance benefits can be obtained by deploying RISs with large numbers of reflecting elements in cell-free mMIMO systems.

\begin{corollary}\label{corollary3}
	When the cascaded channels are only LoS (i.e., $\varepsilon_{n,k}\to\infty$ and $\delta_{m,n}\to\infty$, $ \forall m,n,k$), and $M_{b}\to\infty$, the achievable rate is {\color{black}$R_{k}^{(\mathrm{OL})} \triangleq \log _{2}\Big(1+\mathrm{SINR}_{k}^{(\mathrm{OL})}\Big)$} and $\mathrm{SINR}_{k}^{(\mathrm{OL})}$ is given by
	{\color{black}\begin{align}\label{SINK_OL_1}
		\mathrm{SINR}_{k}^{(\mathrm{OL})}\approx \frac{p_k {E}_{k}^{(signal,\mathrm{OL})}({\bf\Phi})}  {\sum\limits_{i=1, i \neq k}^{K} p_i I_{k i}^{(\mathrm{OL})}({\bf\Phi})}, 
	\end{align}}
	where 
	\begin{align}
		\begin{array}{l}
			E_{k}^{(signal,\mathrm{OL})}({\bf\Phi}) \\
			{= \sum\limits_{m_{1}=1}^{M} \sum\limits_{m_{2}=1}^{M} \sum\limits_{n_{1}=1}^{N} \sum\limits_{n_{2}=1}^{N} } \beta_{m_{1},n_{1}}\beta_{m_{2},n_{2}}\alpha_{n_{1},k}\alpha_{n_{2},k}  {\left|f_{m_{1},n_{1},k}({\bf\Phi})\right|^{2}}{\left|f_{m_{2},n_{2},k}({\bf\Phi})\right|^{2}}\\ + %式子2
			{\sum\limits_{m_{1}=1}^{M} \sum\limits_{m_{2}=1}^{M} \sum\limits_{n=1}^{N} 2{\beta_{m_{1},n}\alpha_{n,k}}  \gamma_{m_{2},k} {\left|f_{m_{1},n,k}({\bf\Phi})\right|^{2}}}
			+ %式子13
			\left(\sum\limits_{m=1}^{M}\gamma_{m,k}\right)^{2},
		\end{array}
	\end{align}
	\begin{align}
		\begin{array}{l}
			I_{ki}^{(\mathrm{OL})}({\bf\Phi}) \\
			{= \sum\limits_{m_{1}=1}^{M} \sum\limits_{m_{2}=1}^{M} \sum\limits_{n_{1}=1}^{N} \sum\limits_{n_{2}=1}^{N} \sqrt{\alpha_{n_{1},k}\alpha_{n_{1},i}\alpha_{n_{2},i}\alpha_{n_{2},k}}}{\beta_{m_{1},n_{1}}\beta_{m_{2},n_{2}}}  {f_{m_{1},n_{1},k}^{H}({\bf\Phi})f_{m_{1},n_{1},i}({\bf\Phi})f_{m_{2},n_{2},i}^{H}({\bf\Phi})f_{m_{2},n_{2},k}}({\bf\Phi}).
		\end{array}
	\end{align}
\end{corollary}

{\color{black}\itshape \textbf{Proof:}  \upshape  As all the Rician factors go to infinity, we have $c_{m,n,k}=0$ but $c_{m,n,k}\delta_{m,n}\varepsilon_{n,k}=\alpha_{n,k}\beta_{m,n}$, $\forall m,n,k$, and the zero terms can be ignored. When $M_{b}\rightarrow\infty$, we can ignore the insignificant terms which do not scale with $M_{b}$. Note that ${\mathbf{a}_{M_{b}}^{H}(m_{1},n_{1})} {\mathbf{a}_{M_{b}}(m_{2},n_{2})}=M_{b}$ when $m_{1}=m_{2}$ and $n_{1}=n_{2}$. Then, we can complete the proof by some simplifications. \hfill $\blacksquare$}

Corollary \ref{corollary3} represents the scenario where {\color{black}where no obstacles exist} and only LoS paths exist. We can find that in the RIS-aided cell-free mMIMO systems with a low-complexity MRC scheme, when the number of {\color{black}AP} antennas $M_{b}$ is large, the achievable rate in  (\ref{SINK_OL_1}) will converge to a constant due to the existence of interference. However, this rate degradation can be compensated by properly designing phase shifts $\mathbf{ \Phi}$. For example,  when the phase shifts are aligned to user $k$, the interference imposed on user $k$ becomes negligible compared with the desired signal received by user $k$. This observation emphasizes the importance of optimizing the {\color{black}phase shifts $\bf\Phi$.}

\begin{corollary}\label{corollary4}
	Assume that the phase shifts of RISs are set randomly in each coherence interval. When $M_{b}\to\infty$ and $N_{r}\to\infty$, the achievable rate is given by {\color{black}$R_{k}^{(\mathrm{rm})} \triangleq \log _{2}\Big(1+\mathrm{SINR}_{k}^{(\mathrm{rm})}\Big)$} with
	\begin{align}\label{SINK_rm_1}
		\mathrm{SINR}_{k}^{(\mathrm{rm})} {\color{black}\approx} 
		\frac{p_k {E}_{k}^{(signal,\mathrm{rm})}}{ \sum\limits_{i=1, i \neq k}^{K} p_i I_{k i}^{(\mathrm{rm})}},
	\end{align}
	where 
	\begin{align}\label{signal power_rm_1}
		\begin{array}{l}
			E_{k}^{(signal,\mathrm{rm})} \\
			{=\sum\limits_{m_{1}=1}^{M} \sum\limits_{m_{2}=1}^{M} \sum\limits_{n_{1}=1}^{N} \sum\limits_{n_{2}=1}^{N}} 
			{{c_{m_{1},n_{1},k}c_{m_{2},n_{2},k}}\Big(2\big({\delta_{m_{1},n_{1}} \varepsilon_{n_{1},k}}  (\delta_{m_{2},n_{2}}+\varepsilon_{n_{2},k}+1)+\delta_{m_{1},n_{1}}\varepsilon_{n_{2},k}}\\+ {\delta_{m_{1},n_{1}}+\varepsilon_{n_{1},k}\big)+\delta_{m_{1},n_{1}}\delta_{m_{2},n_{2}}(\varepsilon_{n_{1},k}\varepsilon_{n_{2},k}+1)+
				\varepsilon_{n_{1},k}\varepsilon_{n_{2},k}+1\Big)}
			\\ + %式子7
			{\sum\limits_{m=1}^{M}  \sum\limits_{n=1}^{N} } 
			{{c_{m,n,k}c_{m,n,k} \delta_{m,n} \delta_{m,n}}(\varepsilon_{n,k}\varepsilon_{n,k}+2\varepsilon_{n,k}+1)} ,
		\end{array}
	\end{align}
	and
	\begin{align}\label{interference_rm_1}
		\begin{array}{l}
			I_{ki}^{(\mathrm{rm})} 
			{=\sum\limits_{m=1}^{M}  \sum\limits_{n=1}^{N} }
			{c_{m,n,k}c_{m,n,i}{\delta_{m,n}\delta_{m,n}} (\varepsilon_{n,k}\varepsilon_{n,i}+\varepsilon_{n,k}+\varepsilon_{n,i}+1)}.
		\end{array}
	\end{align}
\end{corollary}
\itshape \textbf{Proof:}  \upshape Please refer to Appendix \ref{appA}.  \hfill $\blacksquare$

Corollary \ref{corollary4} shows that when the {\color{black}AP} antennas $M_{b}$ and RIS reflecting elements $N_{r}$ are large, the achievable rate is still {\color{black}limited} in random phase shifts-based RIS-aided systems. This conclusion emphasizes the necessity of optimizing the phase shifts ${\bf\Phi}$ in the RIS-aided cell-free mMIMO systems. Nevertheless, it can be found that as the {\color{black}numbers of APs and RISs increase} (i.e., as $M$ and $N$ increase), the rate in (\ref{SINK_rm_1}) could be improved effectively, {\color{black}demonstrating} the gains of applying the cell-free and multi-RIS structures. Besides, unlike the case with $\delta_{m,n}=0$ in Corollary \ref{corollary2}, we can find that the rate in (\ref{SINK_rm_1}) decreases as $\delta_{m,n}$ increases. This is because when the phase shifts are designed randomly in each coherence interval, it tends to distribute the passive beamforming gain equally among all users. 
%However, when $\delta\rightarrow\infty$, the channel with unit rank will be unable to support the multi-user communications.

\section{Phase Shifts Design}\label{section4}
In this section, we design the phase shifts of RISs to maximize the achievable {\color{black}rate} derived in Theorem \ref{lemma1}. Since the derived {\color{black}rate expression} only depends on long-term statistical CSI, we only need to redesign the phase shifts of RISs when the long-term CSI changes, which could effectively decrease the channel estimation overhead and the update frequency at RISs. To make a more general investigation into the RIS-aided cell-free mMIMO system, we consider two optimization problems with different objective functions, i.e., the {\color{black}sum rate} maximization and the minimum user rate maximization problems. The {\color{black}sum rate} maximization problem, which {\color{black}improves} the system capacity, is formulated as
\begin{subequations}\label{p1}
	\begin{equation}\label{objective1}
		\;\;\;\;\max\limits_{\mathbf{\Phi}}  \;\; \sum\limits_{k=1}^{K} {\color{black}R_{k}(\mathbf{\Phi})},\qquad\qquad\quad
	\end{equation}
	\begin{equation}\text {s.t. } \;\;{\theta_{n,r}} \in[0,2 \pi), \forall n,r, \label{constraint1}
	\end{equation}
\end{subequations}
where ${\color{black}R_{k}(\mathbf{\Phi})}$ is given by (\ref{rate}). Then, the minimum user rate maximization problem, which could guarantee user fairness and represent systems spatial multiplexing, is formulated as
\begin{subequations}\label{p2}
	\begin{equation}\label{objective2}
		\max\limits_\mathbf{\Phi}  \;\; \min\limits_{k} \;{\color{black}R_{k}(\mathbf{\Phi})},\qquad\;\;\;\end{equation}
	\begin{equation}\text {s.t. } \;\;(\text{\ref{constraint1}}).\nonumber \qquad\qquad\quad\end{equation}
\end{subequations}

%%\[\mathop {\max }\limits_\Phi  \]
%\subsection{Special Cases}
%To begin with, we will discuss phase shifts design in some special cases.
%\begin{proposition}
%
%\end{proposition}
%
%\itshape \textbf{Proof:}  \upshape \hfill $\blacksquare$
%
%\subsection{General Case}
Due to the complex expression of the rate and the tightly coupled phase shifts ${\bf\Phi}_{n}$ from different RISs, it is challenging to obtain a globally optimal solution in (\ref{p1}) and (\ref{p2}) by exploiting conventional optimization methods, such as semi-definite programming (SDP), {\color{black}the majorization-minimization (MM) algorithms, and the gradient ascent-based algorithms.}
Therefore, we solve the two optimization problems by exploiting the GA method, whose effectiveness in optimizing RIS-aided systems has been validated in \cite{peng2020analysis}. {\color{black}To further verify the effectiveness of the GA method in optimizing RIS-aided cell-free mMIMO systems, we provide the details for the optimality and convergence behaviors of the proposed method in Section \ref{section5}.}

The GA-based method simulates the evolution of a population \cite{matlab2003ga,mitchell1998introduction}, and its main idea is to regard the RIS phase shifts as the gene of a population. GA is initialized by generating a population with $S$ individuals, {\color{black}where} $S_{e}$ individuals will be selected as elites, $2S_{c}$ individuals will be selected as crossover parents to generate offspring, and the remaining $S_{m}=S-S_{e}-S_{c}$ individuals will be used for mutation operation. The main processes of the proposed GA method are summarized in Algorithm \ref{GA}, where $N_{r}$ is the number of RIS reflecting elements, and $N$ is the number of RISs.
\begin{breakablealgorithm}
	\caption{\color{black}GA-Based Method}
	\begin{algorithmic}[1]\label{GA}    %从1：开始标序号
		{\color{black}\STATE Initialization: {\color{black}Generate} a population with $S=S_e+S_c+S_m$ individuals, where the $t$-th individual randomly generates a chromosome $\mathbf{\Phi}_t$; {\color{black}Define} the iteration number as $i$ = 1;
			\WHILE{$i \le NN_{r}*100$}
			\STATE Fitness evaluation: {Define the objective function in (\ref{objective1}) or (\ref{objective2}) as the fitness of each individual in the current population, then calculate each individual's fitness and sort them in descending order;}
			\STATE  {Selection: Retain the top $S_e$ individuals with higher fitness as elites to the next generation;}\\
			\STATE  {Mutation:  Select the last $S_m$ individuals with lower fitness to perform the mutation operation with probability $p_{m}$ and create $S_m$ offspring};
			\STATE Crossover: Generate $2S_c$ crossover parents from the remaining $S_c$ individuals based on stochastic universal sampling. Then adopt the two-point crossover method to generate $S_c$ offspring from $2S_c$ crossover parents;
			\STATE  Combine $S_e$ elites, $S_m+S_c$ offspring to evolve the next generation population;
			\STATE  If the change of the best fitness value is less than $10^{5}$, stop the iteration of the GA. Otherwise, $i=i+1$
			\ENDWHILE
			\STATE  Output the chromosome of the individual with the highest fitness in the current population.}
	\end{algorithmic}
\end{breakablealgorithm} 

Specifically, we describe the implementation details of the GA for both problems (\ref{p1}) and (\ref{p2}) in the following five parts.
\subsubsection{Initial Population}
GA generates an initial population with $S$ individuals, where each individual contains $N  N_{r}$ chromosomes, and the $\big((n-1)N_{r}+r\big)$-th chromosome corresponds to the phase shift $\theta_{n,r}$ of the $r$-th reflecting element of RIS $n$. In general, the chromosomes of individuals are randomly generated in the population in $\left[0,2\pi\right)$. 
\subsubsection{Fitness Evaluation and Scaling}
We first define the initial fitness of each individual in the current population as the objective function in (\ref{objective1}) or (\ref{objective2}). Next, we will scale the initial fitness value of individuals in terms of their rank in the population. 
In the population, the individuals with extensive fitness may reproduce their chromosomes too frequently and have premature convergence. Therefore, we convert initial fitness values to a more suitable range by utilizing a rank scaling method to avoid prematurity.
We sort the initial fitness of individuals and compute their scaled fitness as follows
\begin{align}
	f_{i} = 2 S_{c} \frac{\operatorname{rank}_{i}^{-0.5}}{\sum_{i=1}^{S} \operatorname{rank}_{i}^{-0.5} },1\leq i \leq S,
\end{align}
where $f_{i}$ is the scaled fitness of individual $i$, $\operatorname{rank}_{i}$ is the index of initial fitness of individual $i$ after the descending sort, $S_c$ is the number of crossover parents to be used in the next operation.
\subsubsection{Selection}\label{select}
Then, some individuals are selected as elites from the current population, and some are chosen as parents to generate offspring. First, we will select $ S_{e}$ individuals with larger $f_{i}$ as elite individuals, which will be directly passed on to the next generation. Then, $ 2S_{c} $ individuals will be selected as crossover parents based on stochastic universal sampling. To perform this method, we form a roulette wheel with $ 2S_{c} $ slots corresponding to the crossover parent, and the size of slot $ i $ is proportional to $f_{i}$ as follows 
\begin{align}
	\operatorname{slot}_{i}=\frac{f_{i}}{2 S_{c}},
\end{align}
with $ \sum_{i=1}^{S} \operatorname{slot}_{i}=1 $. Next, we rotate the roulette wheel $ 2S_c $ times in an equal step $\frac{1}{2S_c}$ and select the individual that the pointer points to as a parent. Finally, we will use the remaining $S_m = S - S_e -S_c$ individuals for mutation operation.
\subsubsection{Crossover}
Based on previously selected $ 2S_c $ parents, we perform the crossover operation that can extract and recombine the best chromosome from parents to generate superior $ S_c $ offspring. Specifically, we adopted the two points crossover method and its processes are sketched in Algorithm \ref{algorithm1}.
\begin{breakablealgorithm}
	\caption{Crossover Algorithm}
	\begin{algorithmic}[1]\label{algorithm1}
		\STATE Set $a_1=1$, $a_2=2$;
		\FOR {$i=1:S_c$}
		\STATE Select the $a_1$-th and the $a_2$-th parents from the $2S_c$ combination;
		\STATE  Generate two different integer crossover points $i_1$ and $i_2$ randomly from $\left[1,NN_{r}-1\right]$;
		\IF{$i_1>i_2$}
		\STATE Swap $i_1$ and $i_2$, swap parents $a_1$ and $a_2$;
		\ENDIF		
		\STATE Generate the $i$-th offspring by  $\left[ a_1 (1:i_1) , a_2 (i_1+1:i_2) , a_1 (i_2+1:NN_{r})  \right]$;
		\STATE $a_1=2+a_1$, $a_2=2+a_2$;
		\ENDFOR
	\end{algorithmic}
\end{breakablealgorithm}
\subsubsection{Mutation}
To increase the diversity of the population and improve the possibility of generating offspring with better fitness, we use the remaining $S_m$ individuals to perform the mutation operation with probability $p_m$ and produce $S_m$ offspring. In this paper, we use the uniform mutation method and sketch the process in Algorithm \ref{algorithm2}.
\begin{breakablealgorithm}
	\caption{Mutation Algorithm}
	\begin{algorithmic}[1]\label{algorithm2}
		\FOR{$i=1:S_m$}
		\FOR{$ n=1:NN_{r} $}
		\IF{$p_m>\text{rand}\left(1\right)$}
		\STATE The $n$-th chromosome $\theta_{n}$ of parent $i$ mutates to $2\pi\times\text{rand}(1)$;
		\ENDIF		
		\ENDFOR
		\ENDFOR
	\end{algorithmic}
\end{breakablealgorithm}

\section{Numerical Results}\label{section5}
{
	\begin{figure}
		\setlength{\abovecaptionskip}{0pt}
		\setlength{\belowcaptionskip}{-20pt}
		\centering
		\includegraphics[width=4in]{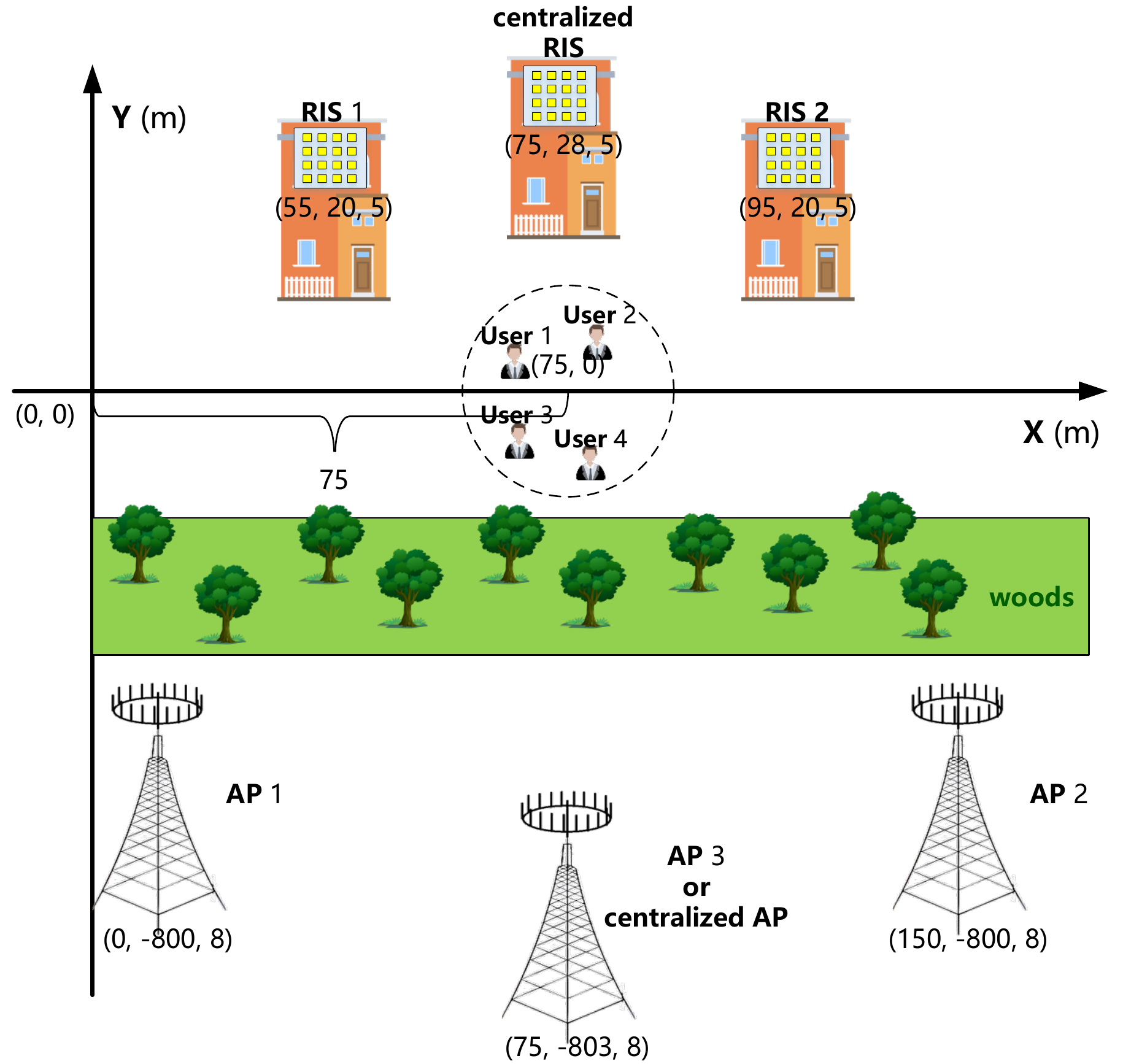}
		\DeclareGraphicsExtensions.
		\caption{\color{black}The simulation scenario of RIS-aided mMIMO systems in cell-free and single-cell networks.}
		\label{sim_fig}
\end{figure}}
In this section, several numerical simulations will be presented to study the performance of the RIS-aided cell-free mMIMO system. By exploiting Theorem \ref{lemma1} and the GA, we can solve these two optimization problems in Section \ref{section4} and obtain the analytical results. 
Unless otherwise specified, we consider a cell-free network with the topology shown in Fig. \ref{sim_fig}. 
In this setup, a cell-free {\color{black}mMIMO} system with {\color{black}$M = 3$ APs serves $K = 4$ users,} while the system throughput is limited due to the obstruction of the woods. To address the issue, {\color{black}we deploy $N = 2$ RISs separately on two building surfaces}, which are high enough to construct extra reflection links. Specifically, we {\color{black}assume the APs} are located at {\color{black}$(0\;\text{m},-800\;\text{m},8\;\text{m})$, $(75\;\text{m},-803\;\text{m},8\;\text{m})$, and $(150\;\text{m},-800\;\text{m},8\;\text{m})$,} respectively. The RISs are located at {\color{black}$(55\;\text{m},20\;\text{m},5\;\text{m})$ and $(95\;\text{m},20\;\text{m},5\;\text{m})$, respectively. Similar to \cite{9154244}, we assume that four users are randomly distributed in a circle centered at $(75\;\text{m},0\;\text{m})$ {\color{black}with a radius} of $3$ m and height of $0$ m.}  The AoA and AoD of {\color{black}APs}, RISs, and users are generated randomly from $[0,2\pi]$\cite{pan2020intelligent,pan2020multicell}, and these angles will be fixed after the initial generation. Large-scale path-loss is calculated as $ \alpha_{n,k}=10^{-3} \left(d^{\mathrm{UR}}_{n,k}\right)^{-\alpha_{UR}} $, {\color{black}$ \beta_{m,n}=10^{-3} \left(d^{\mathrm{RA}}_{m,n}\right)^{-\beta_{RA}}  $, and $\gamma_{m,k}=10^{-3}\left(d_{m,k}^{\mathrm{UA}}\right)^{-\gamma_{UA}}$} \cite{pan2020multicell}, where $d^{\mathrm{UR}}_{n,k}$, {\color{black}$d^{\mathrm{RA}}_{m,n}$, and $d_{m,k}^{\mathrm{UA}}$} are respectively the distances of user $k$-RIS $n$, RIS $n$-{\color{black}AP} $m$, and user $k$-{\color{black}AP} $m$, and the path-loss exponents are $\alpha_{UR}=2$, {\color{black}$\beta_{RA}=2.5$, and $\gamma_{UA}=4$,} $\forall m,n,k$ {\color{black}\cite{wu2019intelligent,9847080}. Moreover,} other simulation parameters (unless otherwise stated) are defined in Table \ref{tab1}. The MC simulations are obtained by averaging $ 10^{5} $ random channel realizations. {\color{black}To verify that our derived rate expression can evaluate system performance faster than MC simulation, we present the needed computing time taken by them with changing $ M $ or $ N $ in Table \ref{tab2}. The results in Table \ref{tab2} show that the computing time taken by the MC simulation is larger than that taken by using the derived expressions, which demonstrates that the rate expression is quicker than the MC simulation in evaluating system performance.}
\begin{table}[t]
	\centering 
	\captionsetup{font={small}}
	\caption{Simulation parameters.}	
	\vspace{-5pt}
	\begin{tabular}{|c|c|c|c|c|c|c}
		\hline
		{\color{black}AP} antennas & {\color{black}$M_{b}=9$} & RIS elements &	$N_{r}=49$\\	
		\hline
		Transmit power & {\color{black}$p_{k}=P=30$ dBm, $\forall k$}  & Noise power& $\sigma^2=-104$ dBm\\
		\hline
		\color{black}Antenna spacing & \color{black}$d=\lambda/2$ & Rician factors&$\delta_{m,n}=1$, $\varepsilon_{n,k}=10,\forall m,n,k$\\
		\hline
		GA parameters &\multicolumn{3}{c| } {$S=200$, $S_e=10$, $S_c=160$, $S_m=30$, $p_m=0.2$}\\
		\hline
	\end{tabular}\label{tab1}
\end{table}
	\begin{table}[t]
	\centering 
	\setcounter{table}{1}
	\color{black}\captionsetup{font={small}}
	\caption{\color{black}The needed computing time (s).}	
	\vspace{-5pt}
	\begin{tabular}{|c|c|c|c|c|c|c|c|c|c|c|c|}
\hline
\multicolumn{2}{|c|}{{\diagbox{Scheme}{$M$}}}& 3 & 6 & 9 & 12 & 15 & 18 & 21 & 24& 27& 30\\
\hline
\multicolumn{2}{|c|}{MC simulation}& 41.859 & 51.196 & 61.802 & 68.181 & 86.364 & 96.858 & 105.293 & 127.731& 147.679 & 176.980\\
\hline
\multicolumn{2}{|c|}{Rate expression}& 0.083 & 0.286 & 0.613 & 1.088 & 1.526 & 2.113 & 2.693 & 4.135& 5.197& 6.070\\
\hline
\multicolumn{2}{|c|}{{\diagbox{Scheme}{$N$}}}& 1 & 2 & 3 & 4 & 5 & 6 & 7 & 8& 9& 10\\
\hline
\multicolumn{2}{|c|}{MC simulation}& 29.269 & 41.959 & 54.004 & 73.216 & 97.882 & 126.473 & 167.065 & 225.210& 289.685 & 343.720\\
\hline
\multicolumn{2}{|c|}{Rate expression}& 0.013 & 0.083 & 0.241 & 0.656 & 1.302 & 2.403 & 3.653 & 7.923& 11.003 & 16.146\\
\hline
	\end{tabular}\label{tab2}
\end{table}

\begin{figure}
	\setlength{\abovecaptionskip}{0pt}
	\setlength{\belowcaptionskip}{-20pt}
	\centering
	\includegraphics[width=4in]{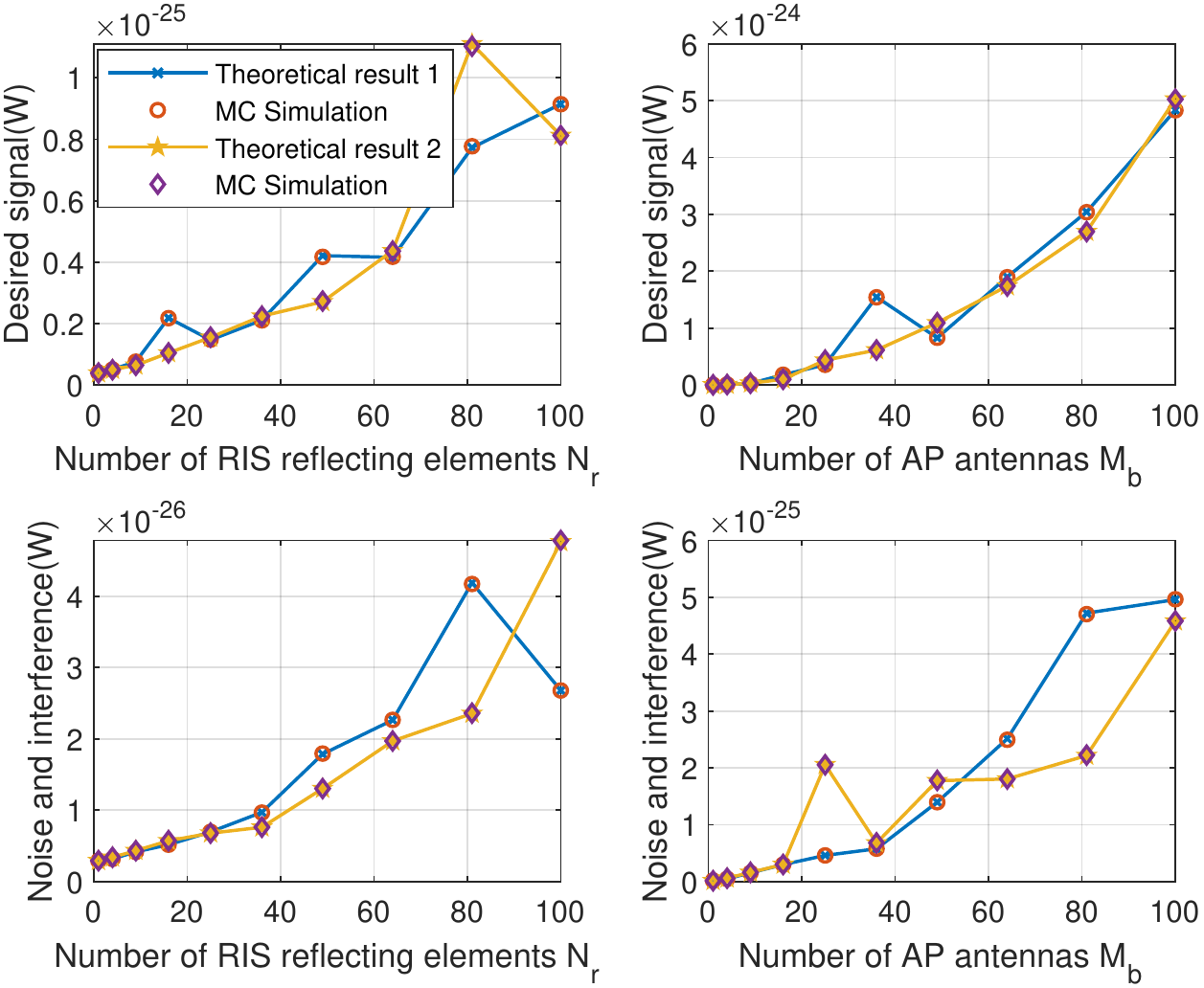}
	\DeclareGraphicsExtensions.
	\caption{Desired signal power and the sum power of interference and noise for user $1$ under random channel realizations.}
	\label{figure1}
\end{figure}
{\color{black}Firstly, we verify the accuracy of the derivations of the closed-form expressions in (\ref{noise}), (\ref{signal power}), and (\ref{interference}). Fig. \ref{figure1} shows the desired signal power $p_{1}\mathbb{E}\left\{\left\|\mathbf{g}_{1}+\mathbf{d}_{1}\right\|^{4}\right\}$, the sum power of multi-user interference $\sum\limits_{i=2}^{4}p_{i} \mathbb{E}\left\{\left|(\mathbf{g}_{1}^{H}+\mathbf{d}_{1}^{H}) (\mathbf{g}_{i}+\mathbf{d}_{i})\right|^{2}\right\}$ and noise $\sigma^{2}\mathbb{E}\left\{\left\|\mathbf{g}_{1}+\mathbf{d}_{1}\right\|^{2}\right\}$ for user $1$ under two independent random channel realizations. The theoretical results of random channel realization are obtained from our derived expressions.  Similarly, based on the same channel realization, we can obtain the results from the MC simulation.  It can be observed that the results obtained from our derived expressions exactly match the MC simulations, which verifies the accuracy of our derivations and the correctness of the expressions.}

{\color{black}Next,} we utilize the GA to optimize the phase shifts of RISs based on the closed-form {\color{black}approximate} expression of the achievable rate in (\ref{rate}), and all of the following MC simulations are presented to compare further the approximations of the rate obtained from the expression. To this end, we respectively solve the optimization Problem (\ref{p1}) and Problem (\ref{p2}) to obtain their optimized phase shifts {\color{black}  $\mathbf{\Phi}^*_{sum}$ and $\mathbf{\Phi}^*_{min}$,} and then calculate two kinds of performance metrics, i.e., {\color{black}the sum rate $\sum_{k=1}^{K} R_k\left(\mathbf{\Phi}^*_{sum}\right)$} and the minimum user rate {\color{black}$\min\limits_{k} R_k\left(\mathbf{\Phi}^*_{min}\right)$}. 

%we evaluate the impact of various system parameters on the uplink achievable rate of the RIS-aided cell-free mMIMO system.

%{\color{black}We refer to the sum user rate obtained by solving Problem (\ref{p1}) as ``sum rate by max-sum'', refer to the minimum user rate obtained by solving Problem (\ref{p1}) as ``min rate by max-sum'', refer to the sum user rate obtained by solving Problem (\ref{p2}) as ``sum rate by max-min'', and refer to the minimum user rate obtained by solving Problem (\ref{p2}) as ``min rate by max-min'', respectively. Besides, we consider the RIS-free cell-free mMIMO system as the standard benchmark system, which means that the number of RISs $N$ is zero and represents the conventional cell-free mMIMO system.
%The random phase shifts-based design of the RISs is also included in the performance evaluation, where the sum user rate and minimum user rate are obtained by averaging $ 10^{4} $ random phase shift realizations.}

{\color{black}
	\subsection{The Optimality and Convergence Analysis of the Proposed GA Method}
	\begin{figure}
		\setlength{\abovecaptionskip}{0pt}
		\setlength{\belowcaptionskip}{-20pt}
		\centering
		\subfigure[\color{black}Convergence behavior of the GA method for different population sizes and RIS elements.] {\includegraphics[width=3in]{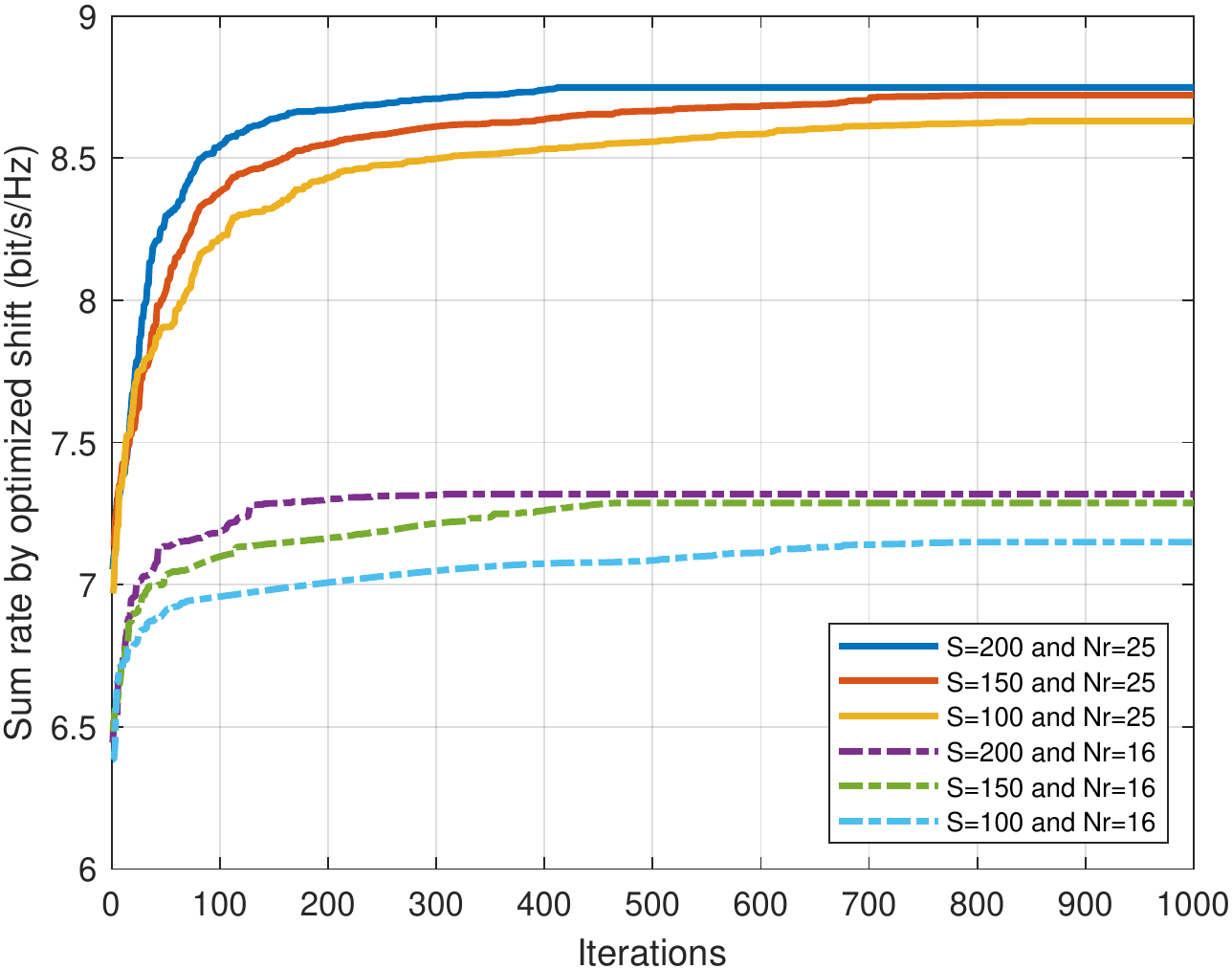}
			\label{figure_convergence}}
		\subfigure[\color{black}Performance comparison of the GA method and the exhaustive search method for $N_{r}=4$.] {\includegraphics[width=3in]{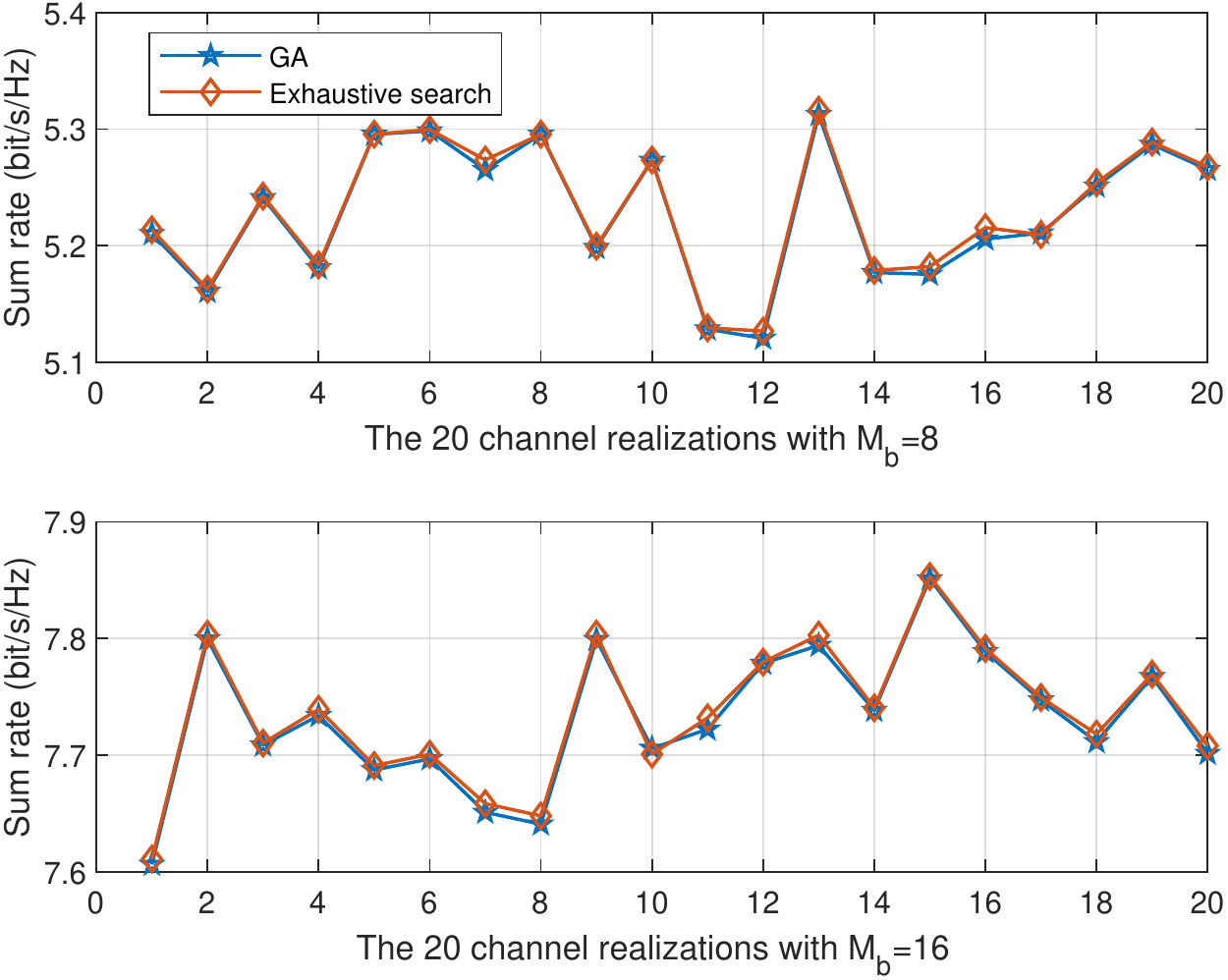}
			\label{figure_optimality}}
		\DeclareGraphicsExtensions.
		\caption{\color{black}The optimality and convergence behaviors of the proposed GA method.}
		\label{figure_opt_con}
	\end{figure}
	In this subsection, based on numerical simulations similar to \cite{6826535,10164189}, we investigate the optimality and convergence behaviors of the GA to verify the effectiveness of the proposed GA method. In Fig. \ref{figure_convergence}, we verify the convergence behavior of the GA for the different values of population size $ S = \left\{100,150,200\right\}$ having the number of RIS elements $N_{r} = \left\{16,25\right\}$, where the sum rate is calculated by the optimized phase shift $\mathbf{\Phi}^*_{sum}$ based on the GA. From this figure, we observe that although the population size $S$ of the GA can affect the convergence speed of this method, the achievable rate is almost unaffected when $S$ is large enough. Furthermore, both the convergence speed and the achievable rate are affected by the number of RIS elements $N_{r}$. The results show that the GA method converges within 1000 iterations, and the maximum number of iterations decreases as the RIS element number $N_{r}$ decreases or the population size $S$ increases. Generally, we cannot guarantee that the GA achieves the optimal design. However, to showcase the effectiveness of the proposed algorithm, we consider comparing it with the exhaustive search method. Constrained by the computational complexity, we consider the case with discrete phase shifts and small $N_r$. Note that the proposed algorithm can be easily extended to the case with discrete phase shifts. The phase shift of the $r$-th reflecting element of the $r$-th RIS with $b$ bits quantization precision can be expressed as ${\theta _{n,r}}\in\left\{0,\frac{2\pi}{2^{b}},...,\left(2^{b}-1\right)\frac{2\pi}{2^{b}}\right\}$, $\forall n,r$. In Fig. \ref{figure_optimality}, we test 20 random channel realizations under different numbers of AP antennas, where ``GA" refers to the sum rate optimized by the GA, and ``exhaustive search" refers to the sum rate obtained by the exhaustive search method \cite{peng2020analysis}. It can be seen that the GA has almost the same rate performance as the globally optimal solution obtained by the exhaustive search method in the considered cases under different channel realizations and AP antennas, which demonstrates the optimality behavior of the proposed GA method.}
%\cite{peng2020analysis}
{\color{black}\subsection{The Interplay Between RIS and Cell-free Massive MIMO}}
In this subsection, we aim to investigate the performance gains of deploying RISs in cell-free mMIMO systems.
% {\color{black}Note that only the continuous phase shift is considered in the following simulations.} 
To achieve a larger system capacity and guarantee user fairness, {\color{black}we consider the sum rate} maximization problem in (\ref{p1}) and the minimum user rate maximization problem in (\ref{p2}) in the following simulations. We consider the cell-free mMIMO systems with the optimized phase shifts-based RISs, where the sum rate calculated by the optimized phase shift $\mathbf{\Phi}^*_{sum}$ is called ``sum rate by max-sum'' or ``sum rate by optimized phase'', the sum rate calculated by the optimized phase shift $\mathbf{\Phi}^*_{min}$ is called ``sum rate by max-min'', the minimum user rate calculated by the optimized phase shift $\mathbf{\Phi}^*_{sum}$ is called ``min rate by max-sum'', and the minimum user rate calculated by the optimized phase shift $\mathbf{\Phi}^*_{min}$ is called ``min rate by max-min" or ``min rate by optimized phase".
{\color{black}For comparison, we adopt the following benchmark systems: 
	\subsubsection{The Cell-free Massive MIMO Systems with RISs under the Random Phase Shifts}
	We consider the random phase shifts-based design of RISs, where the sum rate and minimum user rate obtained by averaging $ 10^{5} $ random phase shift realizations are called ``sum rate by random phase" and ``min rate by random phase", respectively.
	\subsubsection{The RIS-free Cell-free Massive MIMO System}
	We consider the RIS-free cell-free mMIMO system as the benchmark system \cite{9586055}, which means that the number of RISs is zero and represents the conventional cell-free mMIMO system. The sum and minimum user rates of the RIS-free system are called ``sum rate by RIS-free" and ``min rate by RIS-free", respectively.}

\begin{figure}
	\setlength{\abovecaptionskip}{0pt}
	\setlength{\belowcaptionskip}{-20pt}
	\begin{minipage}[t]{0.5\textwidth}
	\centering
	\includegraphics[width=3in]{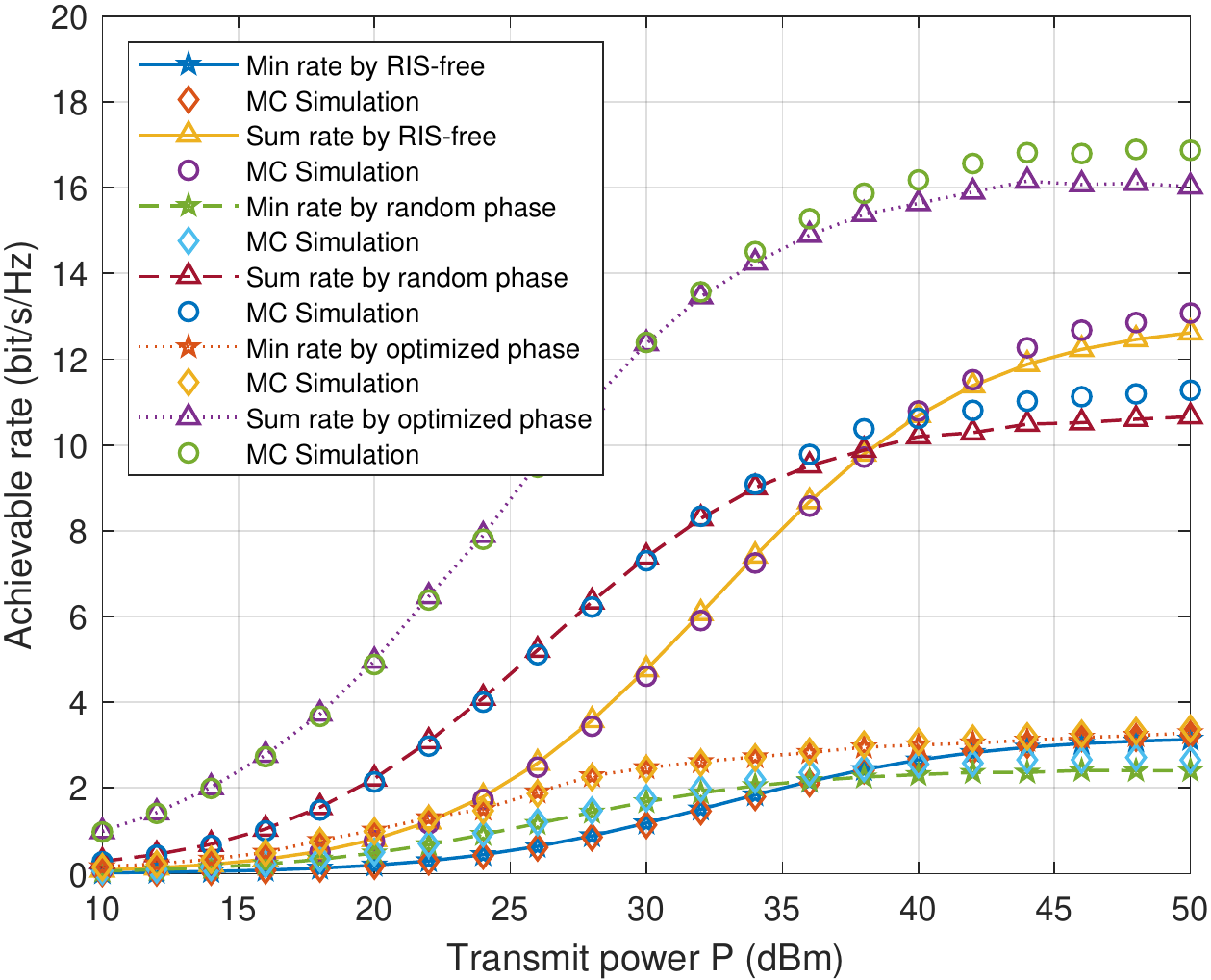}
	\DeclareGraphicsExtensions.
	\caption{Sum rate and minimum user rate versus transmit power $P$.}
	\label{figure2}
    \end{minipage}
	\begin{minipage}[t]{0.5\textwidth}
	\centering
	\includegraphics[width=3in]{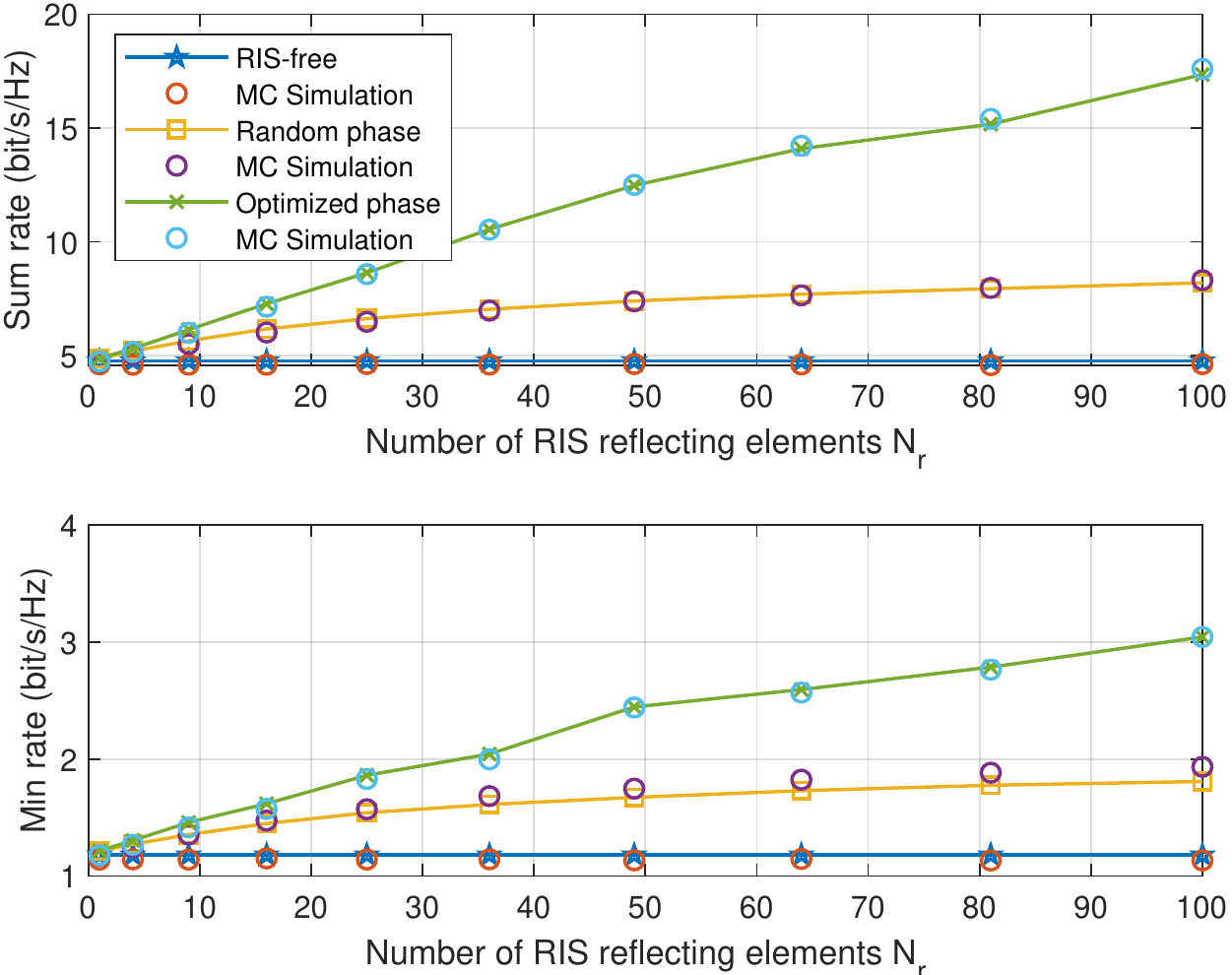}
	\DeclareGraphicsExtensions.
	\caption{Sum rate and minimum user rate versus the number of RIS reflecting elements $N_{r}$.}
	\label{figure3}
	\end{minipage}
\end{figure}
{\color{black}Fig. \ref{figure2} plots the achievable rate versus transmit power $P$. As $P$ increases, there is a small gap between the MC simulation and the analytical result due to the approximate rate expression. In general, the approximated rate expression (\ref{rate}) matches well with the MC simulations, validating the correctness of our derived expressions. Besides, we can see that even with a simple MRC receiver, deployment of the RISs can still effectively improve the rate performance in cell-free mMIMO systems in the low signal-to-noise ratio (SNR) region. }  However, {\color{black} as SNR increases,} the {\color{black}RIS-free cell-free mMIMO systems} will gradually exceed random phase shifts-based RIS systems regarding the {\color{black}achievable rate}. {\color{black}This is because the rate will be limited by multi-user interference in the high SNR region, which aggravates the negative impacts of additional interference caused by RISs. Moreover, with the help of optimized phase shift design, the achievable rate of the RIS-aided system is always higher than that of the RIS-free system for arbitrary SNR, which shows the benefits of deploying RISs and emphasizes the importance of optimizing phase shifts.}
%These results agree with our analysis in Corollary 3 and Corollary 5.

In Fig. \ref{figure3}, we evaluate the achievable {\color{black}rate} as a function of the number of RIS reflecting elements $N_{r}$. {\color{black}The RIS-free system arises for ease of comparison with the RIS-aided cell-free mMIMO system to show the gains of the optimized phase-based RIS.} It can be seen that as the number of RIS reflecting elements $N_{r}$ increases, the optimized phase shifts-based RIS can bring a significant performance improvement to {\color{black}the} cell-free mMIMO systems, which supports the motivation of deploying the RIS in cell-free mMIMO systems. By contrast, the {\color{black}rate} improvement over the random phase shifts-based RIS design is smaller and approaches saturation when $N_{r}\rightarrow\infty$. {\color{black}The result shows that the rate gap caused by the optimized phase shift increases with the increase of RIS elements, which again emphasizes the importance of optimizing phase shifts.}

\begin{figure}
	\setlength{\abovecaptionskip}{0pt}
	\setlength{\belowcaptionskip}{-20pt}
	\centering
	\subfigure[\color{black}Achievable rate of the RIS-free and RIS-aided systems.] {\includegraphics[width=3in]{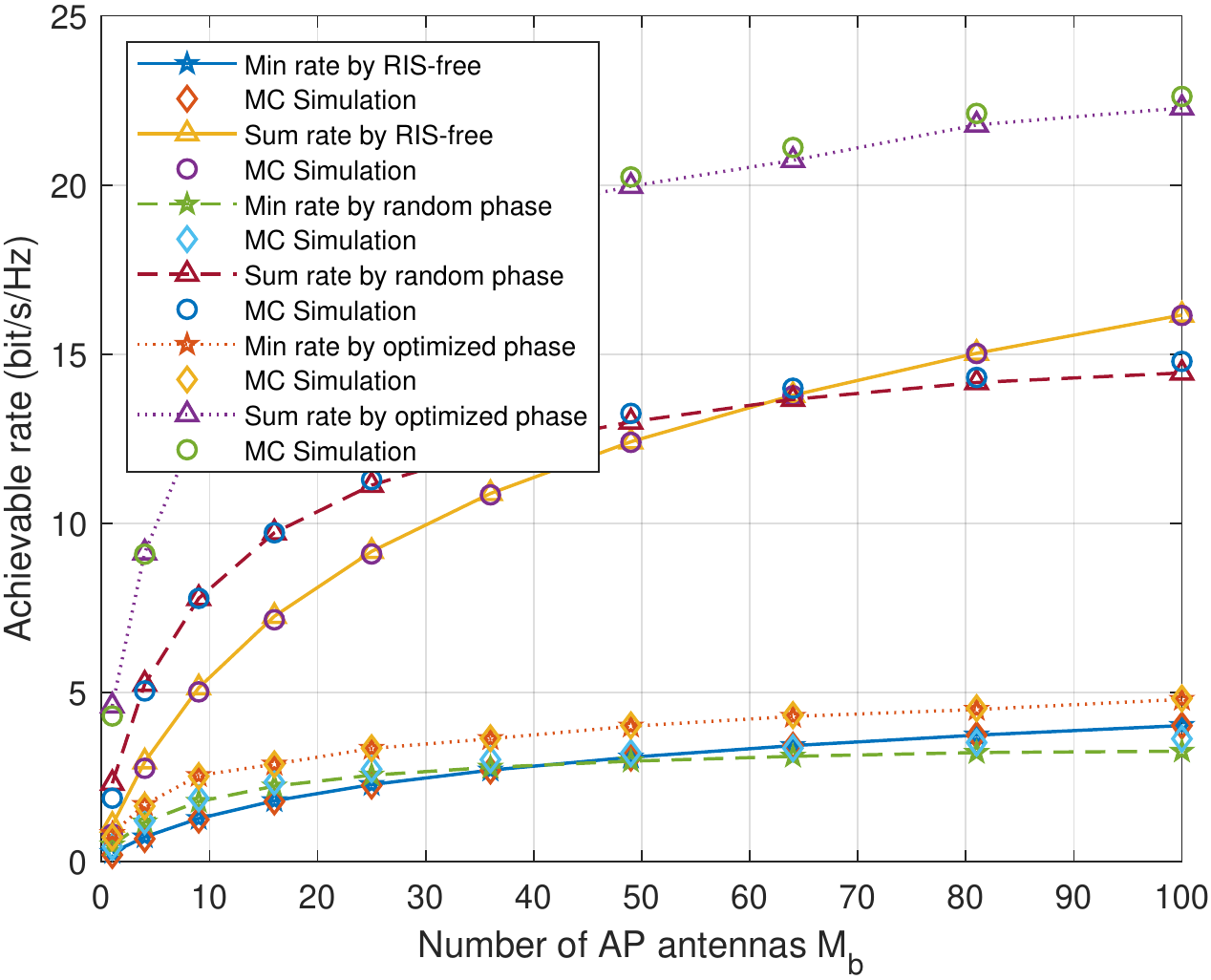}
	\label{figure4_1}}
	\subfigure[\color{black}Achievable rate under the optimized phase shift-based RIS with $N_{r}=25$ and $N_{r}=49$.] {\includegraphics[width=3in]{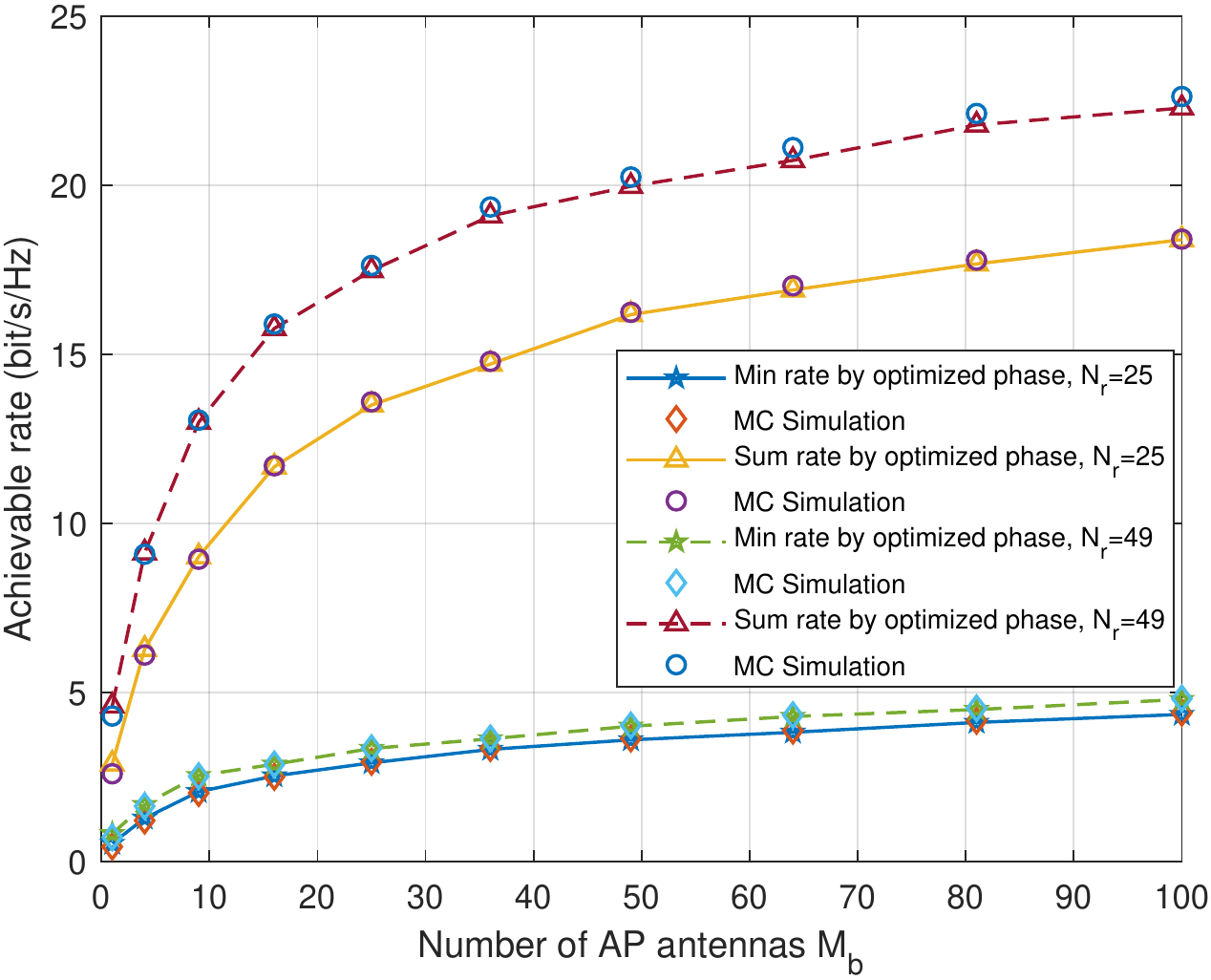}
	\label{figure4_2}}
	\DeclareGraphicsExtensions.
	\caption{Sum rate and minimum user rate versus the number of {\color{black}AP} antennas $M_{b}$.}
	\label{figure4}
\end{figure}

Fig. \ref{figure4} shows the achievable {\color{black}rate} versus the number of {\color{black}AP} antennas $M_{b}$. Specifically, Fig. \ref{figure4_1} compares the achievable rate under different phase shift designs at RISs. 
It can be seen that as $M_{b}$ increases, the {\color{black}rate} could be significantly improved and will approach saturation when $M_{b}\rightarrow\infty$ due to the multi-user interference. Furthermore, when $M_{b}$ is large, {\color{black}the RIS-free} cell-free mMIMO systems will outperform the random phase shifts-based RIS-aided systems. 
Then, in {\color{black}the RIS-free} cell-free mMIMO systems, a large number of {\color{black}AP} antennas are required to serve a large number of users. However, increasing the number of active {\color{black}AP} antennas requires a large-sized array, high power consumption, and high hardware cost. To tackle this problem, we show the achievable rate {\color{black}under the optimized phase shift-based RIS} with $N_{r}=25$ and $N_{r}=49$ in Fig. \ref{figure4_2}. It can be observed {\color{black}that with} the help of the RIS, we can achieve the same throughput as {\color{black}the RIS-free} cell-free mMIMO systems with a smaller number of antennas. In particular, the rate achieved by the system with {\color{black}$25$} {\color{black}AP} antennas and $49$ RIS elements is equal to that achieved by the system with $100$ {\color{black}AP} antennas and $25$ RIS elements. Due to the lower cost and energy consumption of RIS reflecting elements, RIS-aided cell-free mMIMO systems are promising to maintain the network capacity requirement with much-reduced hardware cost and power consumption in future wireless communication systems.
%Since the cost and energy consumption of one RIS element is much lower than that of one BS antenna, we conclude that the integration of RISs in conventional massive MIMO systems is a promising and cost-effective solution for future wireless communication systems.

\begin{figure}
	\setlength{\abovecaptionskip}{0pt}
	\setlength{\belowcaptionskip}{-20pt}
	\begin{minipage}[t]{0.5\textwidth}
		\centering
		\includegraphics[width=3in]{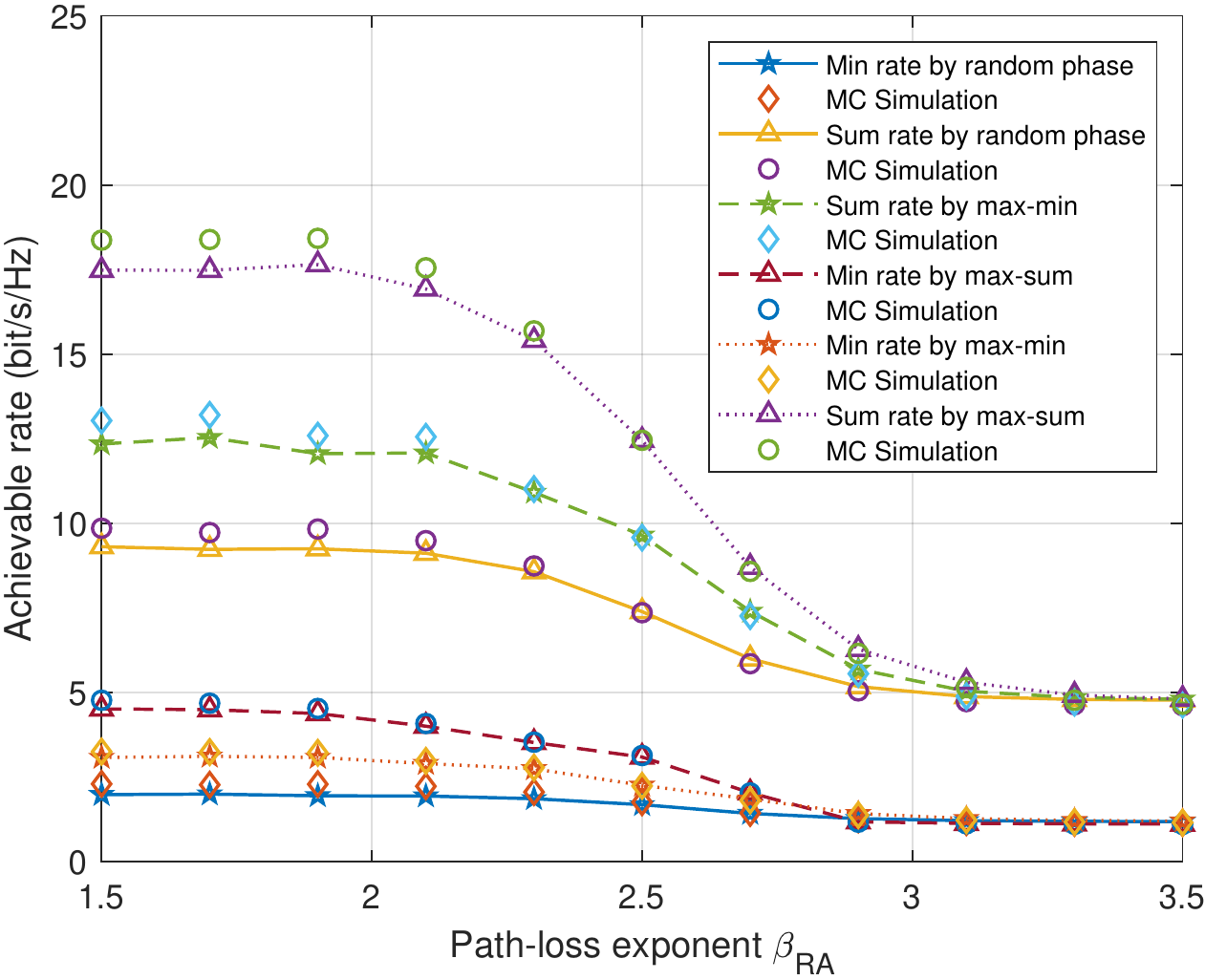}
		\DeclareGraphicsExtensions.
		\caption{Sum rate and minimum rate versus the path-loss exponent {\color{black}$\beta_{RA}$} of RIS-{\color{black}AP} channel.}
		\label{figure5}
	\end{minipage}
	\begin{minipage}[t]{0.5\textwidth}
		\centering
		\includegraphics[width=3in]{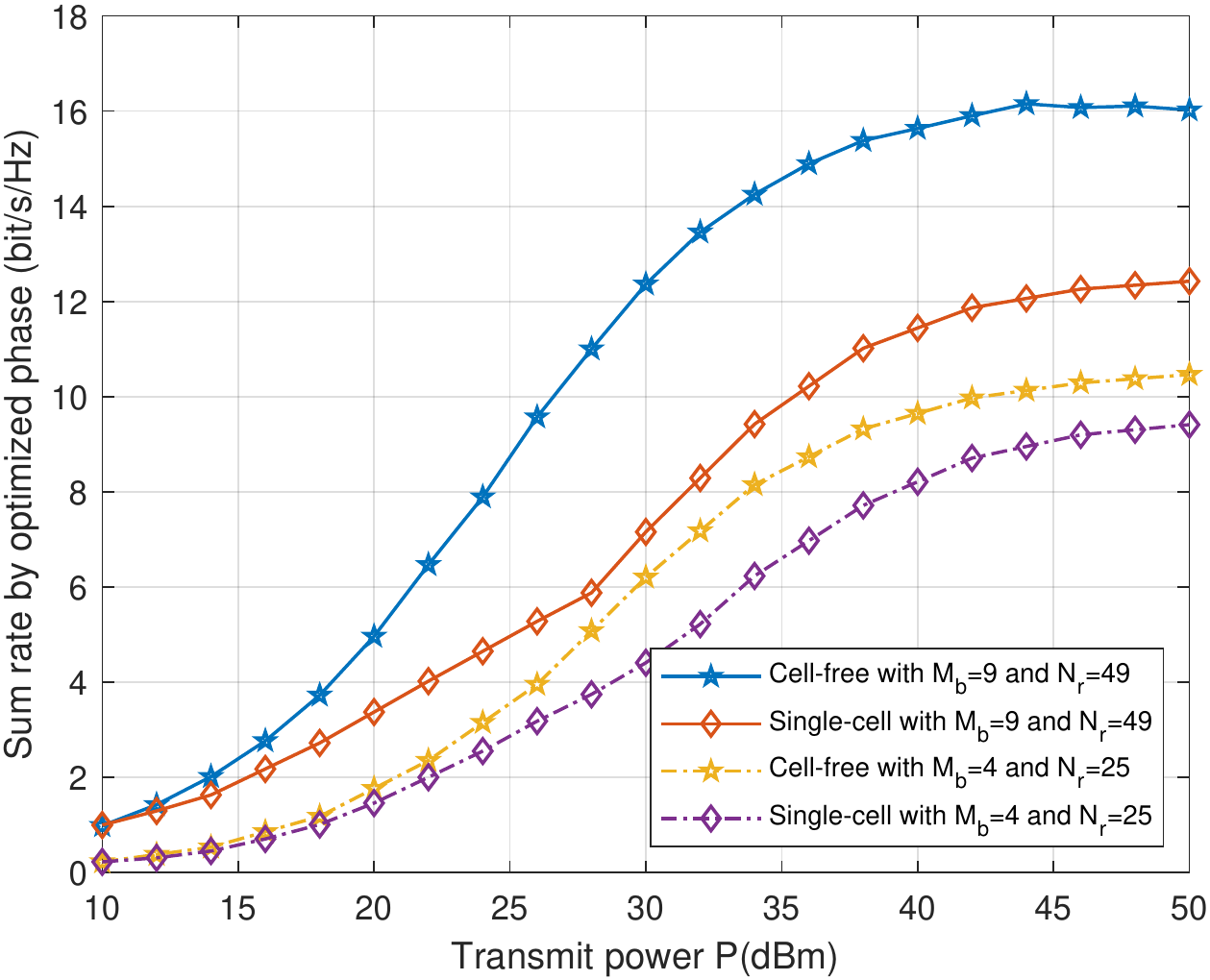}
		\DeclareGraphicsExtensions.
		\caption{\color{black}Sum rate of RIS-aided mMIMO systems in cell-free and single-cell networks.}
		\label{figureCF}
	\end{minipage}
\end{figure}

%\subsection{\color{black}The effect of the path-loss exponent of the RIS-AP link}
The above simulations are carried out for {\color{black}$\beta_{RA}=2.5$} under the assumption that the RIS-{\color{black}AP} channels are less blocked. However, this scenario may not hold in {\color{black}more complex environments.} Hence, it is essential to investigate the impact of path-loss exponent {\color{black}$\beta_{RA}$} on the system performance, {\color{black}as} shown in Fig. \ref{figure5}.

%\begin{figure}
%	\setlength{\abovecaptionskip}{0pt}
%	\setlength{\belowcaptionskip}{-20pt}
%	\centering
%	\includegraphics[width=4in]{fig_pl.pdf}
%	%	\includegraphics[width=4in]{fig4.pdf}
%	\DeclareGraphicsExtensions.
%	\caption{Sum rate and minimum rate versus the path-loss exponent {\color{black}$\beta_{RA}$} of RIS-{\color{black}AP} channel.}
%	\label{figure5}
%\end{figure}
%\begin{figure}
%	\setlength{\abovecaptionskip}{0pt}
%	\setlength{\belowcaptionskip}{-20pt}
%	\centering
%	\includegraphics[width=4in]{fig_CF_P.pdf}
%	%	\includegraphics[width=4in]{fig4.pdf}
%	\DeclareGraphicsExtensions.
%	\caption{\color{black}Sum rate of RIS-aided mMIMO systems in cell-free and single-cell networks.}
%	\label{figureCF}
%\end{figure}
{\color{black}We} can find that both the max-sum problem (\ref{p1}) and max-min problem (\ref{p2}) lead to similarly good performance when {\color{black}$\beta_{RA}$} is small. Secondly, as the path-loss exponent {\color{black}$\beta_{RA}$} increases, if we want to guarantee user fairness, the achievable rate (minimum rate or sum rate) will reduce and ultimately approach the rate obtained by random phase shifts. Based on these findings, we could conclude that if we want to maintain high network capacity and guarantee user fairness, the path-loss exponent should be as small as possible, corresponding to a shorter distance. Therefore, we should carefully select the locations of RISs to find a less blocked scenario associated with a low {\color{black}$\beta_{RA}$}.

{\color{black}\subsection{The Comparison of Cell-free and Single-cell Networks}}
{\color{black}In this subsection,} we evaluate the achievable rate performance in cell-free and single-cell networks. {\color{black}As shown in Fig. \ref{sim_fig}, we consider the single-cell mMIMO system with the optimized phase shifts-based RIS as a new benchmark system \cite{9355404}, where the centralized RIS with $2N_{r}$ elements and the centralized AP with $3M_{b}$ antennas are deployed respectively at $(75\;\text{m},28\;\text{m},5\;\text{m})$ and $(75\;\text{m},-803\;\text{m},8\;\text{m})$. The single-cell network is compared to the cell-free network with the distributed deployment of APs and RISs, where the latter represents the cell-free mMIMO systems with RISs under the optimized phase shifts.}
%the difference between the two systems is the location and number of BSs and RISs. 

{\color{black}Fig. \ref{figureCF} shows the rate performance of RIS-aided mMIMO systems in single-cell and cell-free networks.} We can see that the achievable rate in the cell-free network is higher than that in the single-cell network. This is because different distances from the massive distributed RIS elements to the massive distributed {\color{black}AP} antennas offer extra distance diversity compared {\color{black}to} centralized deployment. 
%Therefore, we can exploit the distributed design of the BS and RIS in the cell-free scenario to improve the rate performance of the system.
%due to the spatial multiplexing gains of both the BS and RIS.
%Furthermore, the distributed RIS design in the cell-free scenario can achieve the interference suppression, which is not realized by a single RIS.
%Then, in the high SNR region, the rate performance gap between these two networks decreases due to the severe multi-user interference. 
Besides, it is found that the performance gap between the two networks is more significant for the larger number of reflecting elements $N_{r}$, which indicates that {\color{black}the RIS-aided cell-free mMIMO system} has great development potential for achieving high system capacity.

{\color{black}\subsection{The Energy Efficiency of RIS-aided Cell-free Massive MIMO Systems}
To investigate the energy efficiency of the RIS-aided cell-free mMIMO system, we first define the energy efficiency metric based on an energy consumption model similar to \cite{huang2019reconfigu}. In general, the total power consumed by an RIS-aided multi-user system mainly includes the transmit power of users, the power consumption of the RIS, and other static hardware power consumption from system operation and maintenance. In particular, for a cell-free network with CPU assistance, we consider the power consumption of the fronthaul used to transfer data between the APs and the CPU, which is proportional to the total user rate \cite{9212395,8097026}. 

Based on the above considerations, we have the total power consumed by the proposed RIS-aided cell-free mMIMO uplink system as follows:
\begin{align}\label{power_consumption}
	P_{\mathrm{total}}= \sum\nolimits_{k=1}^{K} \left( \frac{1}{\xi_k} p_{k} + P_{\mathrm c,k} \right) + \sum\nolimits_{m=1}^{M} \left(M_{b}P_{\mathrm{ap},m}+P_{\mathrm{fh},m}\right) + \sum\nolimits_{n=1}^{N}N_{r} P_{\mathrm{ris},n}(b),
\end{align}
where $0 < \xi_k  \leq 1$ represents the efficiency of the transmit power amplifier deployed at the user $k$, constants $P_{\mathrm{ap},m}$ and $P_{\mathrm c,k}$ are the circuit hardware power consumed by each antenna of AP $m$ and user $k$, respectively. $P_{\mathrm{fh},m}$ is the power consumption of the fronthaul link connecting the CPU and the AP $m$, given by $P_{\mathrm{fh},m}=P_{\mathrm{0},m}+P_{\mathrm{ft},m}W\sum\nolimits_{k=1}^{K}R_{k}$, where $P_{\mathrm{0},m}$ is a fixed power consumption of each fronthaul, $P_{\mathrm{ft},m}$ is the traffic-dependent power (in Watt/bit/s), and $W$ is the system bandwidth. Moreover, the term  $N_{r}P_{\mathrm{ris},n}(b)$ represents the energy consumption of the RIS $n$ with a $b$-bit resolution phase shifter, proportional to the number of RIS reflecting elements $N_{r}$, where $P_{\mathrm{ris},n}(b)$ is the hardware power consumed by each element. 
%In addition, we can observe from \eqref{eq:single} that the RIS operates without consuming any transmit power. As previously stated, the RIS reflectors are ideally passive, thus do not change the amplitude of the impinging signals.
According to the closed-form expression of the achievable rate in (\ref{rate}) and the energy consumption model in (\ref{power_consumption}), the expression of the ergodic energy efficiency is given by
\begin{align}\label{ergodic_EE}
	EE_{\mathrm{ris}} \triangleq  W\frac{\sum\nolimits_{k=1}^{K}R_{k}}{P_{\mathrm{total}}},
\end{align}
%[\mathrm{bit/J}]
where $W$ is again the system bandwidth.

Then, we consider the energy efficiency maximization design of the RIS-aided cell-free mMIMO system by utilizing the GA method, where the phase shifts of RISs are optimized to maximize the energy efficiency $EE_{\mathrm{ris}}$ in (\ref{ergodic_EE}). For comparison, we consider the RIS-free cell-free mMIMO system as the benchmark system \cite{9212395,8097026}, and the corresponding energy efficiency is $EE_{\mathrm{rf}} \triangleq  W \sum\nolimits_{k=1}^{K}R_{k}^{(\mathrm{w})}/{P_{\mathrm{total}}^{(\mathrm{w})}}$, where $P_{\mathrm{total}}^{(\mathrm{w})}= \sum\nolimits_{k=1}^{K} \left( \frac{1}{\xi_k} p_{k} + P_{\mathrm c,k} \right) + \sum\nolimits_{m=1}^{M} \Big(M_{b}P_{\mathrm{ap},m}+P_{\mathrm{0},m}+\\P_{\mathrm{ft},m}W\sum\nolimits_{k=1}^{K}R_{k}^{(\mathrm{w})}\Big)$, and $R_{k}^{(\mathrm{w})}$ is given in Corollary \ref{corollary1}.

\begin{figure}
	\setlength{\abovecaptionskip}{0pt}
	\setlength{\belowcaptionskip}{-20pt}
	\centering
	\includegraphics[width=4in]{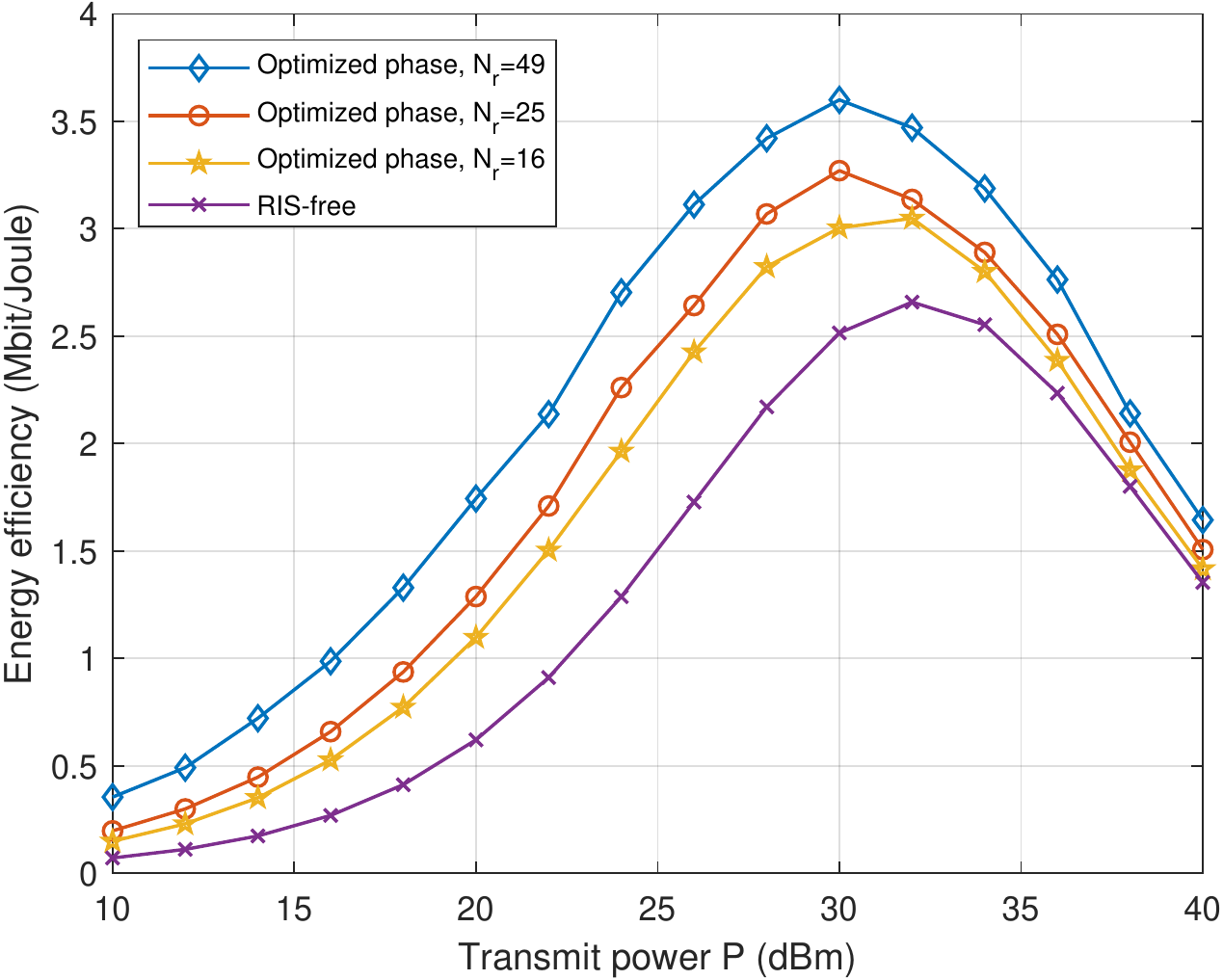}
	\DeclareGraphicsExtensions.
	\caption{The energy efficiency versus transmit power $P$ under different numbers of RIS elements.}
	\label{figure_EE}
\end{figure}

Finally, we present the energy efficiency performance versus transmit power $P$ under different numbers of RIS elements. Specifically, we assume that $ \xi_k = 0.3$, $M_{b}=9$, $W=20$ MHz, $P_{\mathrm{ap},m} =20$ dBm, $P_{\mathrm c,k} = 10$ dBm, $P_{\mathrm{0},m}=23$ dBm, and $P_{\mathrm{ft},m}=24$ dBm/(Gbits/s), $\forall m,k$. In addition, we consider the continuous phase shift case where the RIS reflecting elements have infinite phase resolutions and set $P_{\mathrm{ris},n}(b) =25$ dBm for $ b\to \infty$, $\forall n$. In Fig. \ref{figure_EE}, we can find that there exists a unique transmit power $P_{max}$ that maximizes the energy efficiency of the systems. The energy efficiency increases with transmit power $P$ when $P$ is smaller than the threshold $P_{max}$ but eventually declines rapidly when $P$ is larger than this threshold. This behavior is explained by the reason that the achievable rate of systems reaches saturation as the transmit power $P$ increases, and the corresponding energy efficiency will reach a peak or even decrease due to the increase of total power consumption. Moreover, we can see that the optimized phase shift-based RIS can effectively improve the rate performance of the cell-free mMIMO system. It is worth noting that the rate improvement brought by RIS is more significant with the increase of RIS elements, even though the RIS with infinite phase shift accuracy requires a large circuit hardware power consumption. This observation shows the high energy efficiency of RIS-aided systems, meaning that the RIS is a power-effective solution to support high data rate transmission.

}
\section{Conclusion}\label{section6}
%This paper has investigated a two-timescale design for RIS-aided massive MIMO systems by taking into account the channel estimation errors. An LMMSE estimator has been proposed for obtaining the instantaneous CSI of the $M\times K$ RIS-assisted aggregated channel, whose channel estimates are used by the BS for MRC. We have derived a closed-form expression for the ergodic achievable rate, and have studied the power scaling laws. Based on the derived analytical expressions, we have designed the RIS phase shifts by relying only on statistical CSI, which significantly reduces the signaling overhead and the computational complexity. For the single-user case, we have obtained a closed-form optimal solution. For the general multi-user case, we have adopted a GA for solving the minimum user rate maximization problem. The obtained numerical results show that the transmit power can be reduced proportionally to $1/M$, while maintaining a non-zero rate, as $M\to\infty$ over RIS-BS Rician channels. If the RIS-BS channel is Rayleigh distributed, on the other hand, a non-zero rate can be maintained when the power is scaled proportionally to $1/\sqrt{M}$ as $M\to\infty$ or proportionally to $1/N$ as $N\to\infty$. Finally, we have proved that it is preferable to deploy the RIS close to the users rather than close to the BS.

This paper has investigated and optimized the achievable rate performance in the uplink RIS-aided cell-free {\color{black}mMIMO systems} based on the two-timescale scheme. 
We have considered {\color{black}the general} Rician channel model and applied RISs to provide additional channels to the user in the rich scattering region of {\color{black} cell-free mMIMO systems.} 
{\color{black}The CPU has utilized the MRC technique based on the short-time instantaneous CSI for fully centralized processing}, while the phase shifts of RISs have been designed based on long-time statistical CSI, which could reduce the channel estimation overhead {\color{black}and computational complexity}. 
{\color{black}On this basis, we first derived the closed-form approximate expression of the uplink achievable rate} and provided analytical insights. {\color{black}These valuable insights revealed the impact of various system parameters on the achievable rate and the asymptotic behaviors of the achievable rate, which can serve as clear guidelines for the benefits of the proposed RIS-aided cell-free mMIMO systems.}
Then, we have designed the optimized phase shifts to maximize {\color{black}the user's sum rate and minimum rate} by exploiting the GA method. 
Finally, we have provided the numerical results to validate the effectiveness and the benefits of integrating RISs into {\color{black}the cell-free mMIMO systems. 
Our results have demonstrated the correctness of our derived expressions and verified the proposed GA method's effectiveness by showing the GA's optimality and convergence behaviors. Also, we have investigated the benefits of the optimized phase shift-based RISs in the cell-free mMIMO system, and revealed that the smaller path-loss exponent of the RIS-AP channels is more conducive to simultaneously maintaining high network capacity and guaranteeing user fairness. Compared to the AP antenna with high power consumption and hardware cost, the passive reflecting elements of RISs are promising to provide the same rate increase for cell-free mMIMO systems at a lower cost. To give more beneficial analysis, we have given the closed-form expression of the energy efficiency and presented numerical results to show the high energy efficiency of RIS-aided systems, which implies that the RIS will be power effective. {\color{black}Besides, we have shown the benefits of distributed deployment of APs and RISs for the RIS-aided mMIMO network.}}

\begin{appendices}

\section{}\label{appB}
    As known in Section \ref{section2}, by applying \cite[Lemma 1]{zhang2014ArRank}, the uplink achievable rate expression of user $k$ can be approximated as
\begin{align}\label{Rk}
    {\color{black}R_{k} \approx \mathrm{log_{2}}}\left(1+\frac{{p_k} \mathbb{E}\left\{ \left\|{\bf g}_k+{\bf d}_k\right\|^{4} \right\} }{ \sum\limits_{i=1, i\neq k}^{K}{p_i}\mathbb{E}\left\{\left|({\bf g}_k+{\bf d}_k)^{H} ({\bf g}_i+{\bf d}_i)\right|^{2}\right\}+\sigma^{2}\mathbb{E}\left\{\left\|{\bf g}_k+{\bf d}_k\right\|^{2}\right\}}\right).
\end{align}

    To derive the closed-form expression of (\ref{Rk}), we need to derive $\mathbb{E}\left\{\left|({\bf g}_k+{\bf d}_k)^{H} ({\bf g}_i+{\bf d}_i)\right|^{2}\right\}$,  $\mathbb{E}\left\{\left\|{\bf g}_k+{\bf d}_k\right\|^{4}\right\}$ and $\mathbb{E}\left\{\left\|{\bf g}_k+{\bf d}_k\right\|^{2}\right\}$, respectively. 
    
    To begin with, we can derive the preliminary expression by expanding and simplifying the three mathematical expectation terms. Specially, note that $ {\mathbf{d}}_{k}$, ${\mathbf{d}}_{i}$ and ${\mathbf{g}}_{k}$ are independent of each other, $\forall i \neq k$, and $ {\mathbf{d}}_{k}$ is composed of independent and identically distributed random variables with zero mean. Firstly, the noise term $\mathbb{E}\left\{\left\|{\bf g}_k+{\bf d}_k\right\|^{2}\right\}$ can be written as
\begin{align}\label{noise_term}
	\begin{array}{l}
	\mathbb{E}\left\{\left\|\mathbf{g}_{k}+\mathbf{d}_{k}\right\|^{2}\right\}\\
	=\mathbb{E}\left\{\mathbf{g}_{k}^{H} \mathbf{g}_{k}+\mathbf{g}_{k}^{H} \mathbf{d}_{k}+\mathbf{d}_{k}^{H} \mathbf{g}_{k}+\mathbf{d}_{k}^{H} \mathbf{d}_{k}\right\} \\
	=\mathbb{E}\left\{\mathbf{g}_{k}^{H} \mathbf{g}_{k}+\mathbf{d}_{k}^{H} \mathbf{d}_{k}\right\}=\mathbb{E}\left\{\left\|\mathbf{g}_{k}\right\|^{2}\right\}+\mathbb{E}\left\{\left\|\mathbf{d}_{k}\right\|^{2}\right\} \\
	=\mathbb{E}\left\{\left\|\mathbf{g}_{k}\right\|^{2}\right\}+\sum\limits_{m=1}^{M}\gamma_{m,k} M_{b}.
	\end{array}
\end{align}  

    Next, we can derive the signal term $\mathbb{E}\left\{\left\|{\bf g}_k+{\bf d}_k\right\|^{4}\right\}$ as
\begin{align}\label{E4}
	\begin{array}{l}
	\mathbb{E}\left\{\left\|\mathbf{g}_{k}+\mathbf{d}_{k}\right\|^{4}\right\} \\
	=\mathbb{E}\left\{\Big(\left\|\mathbf{g}_{k}\right\|^{2}+2 \operatorname{Re}\left\{\mathbf{d}_{k}^{H} \mathbf{g}_{k}\right\}+\left\|\mathbf{d}_{k}\right\|^{2}\Big)^{2}\right\} \\
	=\mathbb{E}\left\{\left\|\mathbf{g}_{k}\right\|^{4}\right\}+4 \mathbb{E}\left\{\left(\operatorname{Re}\left\{\mathbf{d}_{k}^{H} \mathbf{g}_{k}\right\}\right)^{2}\right\}
	+\mathbb{E}\left\{\left\|\mathbf{d}_{k}\right\|^{4}\right\}+2 \mathbb{E}\left\{\left\|\mathbf{g}_{k}\right\|^{2}\left\|\mathbf{d}_{k}\right\|^{2}\right\}.
	\end{array}
\end{align}
    Assuming that $ {\left[\mathbf{g}_{k}\right]_{m,b}=\left[\mathbf{g}_{m,k}\right]_{b}} {=v_{m,b}+j w_{m,b}}$ and ${\left[\mathbf{d}_{k}^{H}\right]_{m,b}=\left[\mathbf{d}_{m,k}^{H}\right]_{b}=}{s_{m,b}+j t_{m,b}}$, where both $s_{m,b}$ and $t_{m,b}$ independently follow  $\mathcal{N}\left(0,\frac{\gamma_{m,k}}{2}\right)$, $\forall b$, we have
\begin{align}\label{E4_1}
	\begin{array}{l}
	\mathbb{E}\left\{\left(\operatorname{Re}\left\{\mathbf{d}_{k}^{H} \mathbf{g}_{k}\right\}\right)^{2}\right\} \\
	=\mathbb{E}\left\{\left(\sum\limits_{m=1}^{M} \sum\limits_{b=1}^{M_{b}} s_{m,b} v_{m,b}-t_{m,b} w_{m,b}\right)^{2}\right\} \\
	=\mathbb{E}\left\{\sum\limits_{m=1}^{M} \sum\limits_{b=1}^{M_{b}} \left(s_{m,b} v_{m,b}-t_{m,b} w_{m,b}\right)^{2}\right\}
	\\
	=\sum\limits_{m=1}^{M}\mathbb{E}\left\{ \sum\limits_{b=1}^{M_{b}} \left(s_{m,b} v_{m,b}\right)^{2}+\left(t_{m,b} w_{m,b}\right)^{2}\right\} 
	\\
	=\sum\limits_{m=1}^{M}\frac{\gamma_{m,k}}{2} \mathbb{E}\left\{\sum\limits_{b=1}^{M_{b}} \left(v_{m,b}\right)^{2}+\left(w_{m,b}\right)^{2}\right\} 
	\\
%	\qquad\qquad\qquad\qquad\;\;
	=\sum\limits_{m=1}^{M}\frac{\gamma_{m,k}}{2} \mathbb{E}\left\{\left\|\mathbf{g}_{m,k}\right\|^{2}\right\}.
	\end{array}
\end{align}

    Then, the remaining two terms in (\ref{E4}) can be obtained as
\begin{align}\label{E4_2}
	\begin{array}{l}
	\mathbb{E}\left\{\left\|\mathbf{d}_{k}\right\|^{4}\right\}\\=\mathbb{E}\left\{\left(\sum\limits_{m=1}^{M}\sum\limits_{b=1}^{M_{b}}\left|\left[\mathbf{d}_{k}\right]_{m,b}\right|^{2}\right)^{2}\right\} \\
	=\mathbb{E} \! \left\{\sum\limits_{m=1}^{M}\sum\limits_{b=1}^{M_{b}}\left|\left[\mathbf{d}_{k}\right]_{m,b}\right|^{4}\right\}  \! + \!  \mathbb{E} \!  \left\{	\mathop{\sum\limits_{m_1=1}^{M} \sum\limits_{b_1=1}^{M_{b}}{\sum\limits_{m_2=1}^{M}\sum\limits_{b_2=1}^{M_{b}}} }\limits_{(m_{2},b_{2})\neq(m_{1},b_{1})}
	\!  \!  \left|\left[\mathbf{d}_{k}\right]_{m_1,b_1}\right|^{2}\left|\left[\mathbf{d}_{k}\right]_{m_2,b_2}\right|^{2}\right\} \\
	=\sum\limits_{m=1}^{M}  \gamma_{m,k}^{2}M_{b}+\left(\sum\limits_{m=1}^{M} \gamma_{m,k}M_{b}\right)^{2},
	\end{array}
\end{align}
and
    \begin{align}\label{E4_3}
	\mathbb{E}\left\{\left\|\mathbf{g}_{k}\right\|^{2} \! \left\|\mathbf{d}_{k}\right\|^{2}\right\} \! = \!  \mathbb{E}\left\{\left\|\mathbf{g}_{k}\right\|^{2}\right\}  \!  \mathbb{E}\left\{\left\|\mathbf{d}_{k}\right\|^{2}\right\} \! = \! \sum_{m=1}^{M}\gamma_{m,k}M_{b}  \mathbb{E}\left\{\left\|\mathbf{g}_{k}\right\|^{2}\right\} \! .
    \end{align}

Substituting (\ref{E4_1}), (\ref{E4_2}) and (\ref{E4_3}) into (\ref{E4}), we obtain the expression of signal term as follows
\begin{align}\label{signal_term}
	\begin{array}{l}
	\mathbb{E}\left\{\left\|\mathbf{g}_{k}+\mathbf{d}_{k}\right\|^{4}\right\} \\
	=\mathbb{E}\left\{\left\|\mathbf{g}_{k}\right\|^{4}\right\}+\sum\limits_{m=1}^{M}\gamma_{m,k}\Big(M_{b}\big(2\mathbb{E}\left\{\left\|\mathbf{g}_{k}\right\|^{2}\right\}+\gamma_{m,k}\big)+2\mathbb{E}\left\{\left\|\mathbf{g}_{m,k}\right\|^{2}\right\}\Big)+\left(\sum\limits_{m=1}^{M} \gamma_{m,k}M_{b}\right)^{2}.
	\end{array}
\end{align}
    Finally, the interference term can be expanded as
\begin{align}\label{interference_term}
	\begin{array}{l}
		\mathbb{E}\left\{\left|\left(\mathbf{g}_{k}^{H}+\mathbf{d}_{k}^{H}\right)\left(\mathbf{g}_{i}+\mathbf{d}_{i}\right)\right|^{2}\right\} \\
		=\mathbb{E}\left\{\left|\mathbf{g}_{k}^{H} \mathbf{g}_{i}+\mathbf{g}_{k}^{H} \mathbf{d}_{i}+\mathbf{d}_{k}^{H} \mathbf{g}_{i}+\mathbf{d}_{k}^{H} \mathbf{d}_{i}\right|^{2}\right\} \\
		=\mathbb{E}\! \left\{\left|\mathbf{g}_{k}^{H} \mathbf{g}_{i}\right|^{2}\! \right\}   \!+\!   \mathbb{E}\! \left\{\left|\mathbf{g}_{k}^{H} \mathbf{d}_{i}\right|^{2}\! \right\}
		\!+\!   \mathbb{E}\! \left\{\left|\mathbf{d}_{k}^{H} \mathbf{g}_{i}\right|^{2}\! \right\}    \!+\!   \mathbb{E}\! \left\{\left|\mathbf{d}_{k}^{H} \mathbf{d}_{i}\right|^{2}\! \right\},
	\end{array}
\end{align}
and
\begin{align}\label{interference_term_re3}
	\begin{array}{l}
	\mathbb{E}\left\{\left|\mathbf{g}_{k}^{H} \mathbf{d}_{i}\right|^{2}\right\}
	=\mathbb{E}\left\{\mathbf{g}_{k}^{H} \mathbb{E}\left\{\mathbf{d}_{i} \mathbf{d}_{i}^{H}\right\} \mathbf{g}_{k}\right\} 
	=\sum\limits_{m=1}^{M}\gamma_{m,i} \mathbb{E}\left\{\left\|\mathbf{g}_{m,k}\right\|^{2}\right\},\\
	\mathbb{E}\left\{\left|\mathbf{d}_{k}^{H} \mathbf{g}_{i}\right|^{2}\right\} 
	=\mathbb{E}\left\{\mathbf{g}_{i}^{H} \mathbb{E}\left\{\mathbf{d}_{k} \mathbf{d}_{k}^{H}\right\} \mathbf{g}_{i}\right\} 
	=\sum\limits_{m=1}^{M}\gamma_{m,k} \mathbb{E}\left\{\left\|\mathbf{g}_{m,i}\right\|^{2}\right\},\\
	\mathbb{E}\left\{\left|\mathbf{d}_{k}^{H} \mathbf{d}_{i}\right|^{2}\right\} 
	=\mathbb{E}\left\{\mathbf{d}_{k}^{H} \mathbb{E}\left\{\mathbf{d}_{i} \mathbf{d}_{i}^{H}\right\} \mathbf{d}_{k}\right\} 
	=\sum\limits_{m=1}^{M} \gamma_{m,i} \gamma_{m,k} M_{b}.
	\end{array}
\end{align}

	Therefore, combining (\ref{noise_term}), (\ref{signal_term}) and (\ref{interference_term}) with (\ref{Rk}) and after some simplifications, we can gain the preliminary expression of (\ref{Rk}) with $ \mathbb{E}\left\{\left\|\mathbf{g}_{k}\right\|^{2}\right\} $, $ \mathbb{E}\left\{\left\|\mathbf{g}_{k}\right\|^{4}\right\} $, and $\mathbb{E}\left\{\left|\mathbf{g}_{k}^{H} \mathbf{g}_{i}\right|^{2}\right\}$. 

To derive the complete closed-form expressions of the terms above, we present some definitions and properties which will be utilized in the following derivation. According to the definitions of Rician channels in (\ref{rician1}), (\ref{rician2}) and (\ref{rician3}), we can rewrite the cascaded channels ${\bf g}_{k}$ for user $k$ and ${\bf g}_{i}$ for user $i$ as follows
\begin{align}\label{g_k}
	\mathbf{g}_{k} = \sum\limits_{n=1}^{N} \mathbf{g}_{n,k}= \sum\limits_{n=1}^{N}\left[\mathbf{g}^{H}_{1,n,k},...,\mathbf{g}^{H}_{m,n,k},...,\mathbf{g}^{H}_{M,n,k}\right]^{H},
\end{align}
\begin{align}\label{g_i}
	\mathbf{g}_{i} = \sum\limits_{n=1}^{N} \mathbf{g}_{n,i}= \sum\limits_{n=1}^{N}\left[\mathbf{g}^{H}_{1,n,i},...,\mathbf{g}^{H}_{m,n,i},...,\mathbf{g}^{H}_{M,n,i}\right]^{H},
\end{align}

\begin{align}\label{g_mnk}
	\begin{array}{l}
		\mathbf{g}_{m,n,k}=\!\mathbf{Z}_{m,n} {\bf \Phi}_{n} \mathbf{h}_{n,k}=\!\sqrt{\frac{\beta_{m,n}\alpha_{n,k}}{(\delta_{m,n}+1)\left(\varepsilon_{n,k}+1\right)}} (\underbrace{\sqrt{\delta_{m,n} \varepsilon_{n,k}} \overline{\mathbf{Z}}_{m,n} 	{\bf \Phi}_{n} \overline{\mathbf{h}}_{n,k}}_{\mathbf{g}_{m,n,k}^{1}} 
		+\underbrace{\sqrt{\delta_{m,n}} \overline{\mathbf{Z}}_{m,n} 	{\bf \Phi}_{n} \tilde{\mathbf{h}}_{n,k}}_{\mathbf{g}_{m,n,k}^{2}}
		\\+
		\underbrace{\sqrt{\varepsilon_{n,k}} \tilde{\mathbf{Z}}_{m,n} 	{\bf \Phi}_{n} \overline{\mathbf{h}}_{n,k}}_{\mathbf{g}_{m,n,k}^{3}}+\underbrace{\tilde{\mathbf{Z}}_{m,n} {\bf \Phi}_{n} \tilde{\mathbf{h}}_{n,k}}_{\mathbf{g}_{m,n,k}^{4}}),
	\end{array}
\end{align}
\begin{align}\label{g_mni}
	\begin{array}{l}
		\mathbf{g}_{m,n,i}=\!\mathbf{Z}_{m,n} {\bf \Phi}_{n} \mathbf{h}_{n,i}=\!\sqrt{\frac{\beta_{m,n}\alpha_{n,i}}{(\delta_{m,n}+1)\left(\varepsilon_{n,i}+1\right)}} (\underbrace{\sqrt{\delta_{m,n} \varepsilon_{n,i}} \overline{\mathbf{Z}}_{m,n} 	{\bf \Phi}_{n} \overline{\mathbf{h}}_{n,i}}_{\mathbf{g}_{m,n,i}^{1}} 
		+\underbrace{\sqrt{\delta_{m,n}} \overline{\mathbf{Z}}_{m,n} 	{\bf \Phi}_{n} \tilde{\mathbf{h}}_{n,i}}_{\mathbf{g}_{m,n,i}^{2}}
		\\+
		\underbrace{\sqrt{\varepsilon_{n,i}} \tilde{\mathbf{Z}}_{m,n} 	{\bf \Phi}_{n} \overline{\mathbf{h}}_{n,i}}_{\mathbf{g}_{m,n,i}^{3}}+\underbrace{\tilde{\mathbf{Z}}_{m,n} {\bf \Phi}_{n} \tilde{\mathbf{h}}_{n,i}}_{\mathbf{g}_{m,n,i}^{4}}).
	\end{array}
\end{align}

Note that $ \tilde{\mathbf{Z}}_{m,n}$, $\tilde{\mathbf{h}}_{n,k}$ and $\tilde{\mathbf{h}}_{n,i}$ are independent of each other, and $ \tilde{\mathbf{Z}}_{m,n}$, $\tilde{\mathbf{h}}_{n,k}$ and $\tilde{\mathbf{h}}_{n,i}$ are composed of independent and identically distributed random variables following $\mathcal{CN}\left(0,1\right)$. Therefore, for arbitrary $x$, $y$ and $t$, we have
\begin{align}\label{property1}
	\begin{array}{l}
		\mathbb{E}\left\{\left[\tilde{\mathbf{Z}}_{m,n}\right]_{x,y}\right\}=\mathbb{E}\left\{\left[\tilde{\mathbf{h}}_{n,k}\right]_{t}\right\}=\mathbb{E}\left\{\left[\tilde{\mathbf{h}}_{n,i}\right]_{t}\right\}=0, \\
		\mathbb{E}\left\{\left[\tilde{\mathbf{Z}}_{m,n}\right]_{x,y} \left[\tilde{\mathbf{h}}_{n,k}\right]_{t} \left[\tilde{\mathbf{h}}_{n,i}\right]_{t}\right\} =\mathbb{E}\left\{\left[\tilde{\mathbf{Z}}_{m,n}\right]_{x,y} \right\}  \mathbb{E}\left\{ \left[\tilde{\mathbf{h}}_{n,k}\right]_{t}\right\}  \mathbb{E}\left\{\left[\tilde{\mathbf{h}}_{n,i}\right]_{t}\right\}=0,\\
		\mathbb{E}\left\{\left[\tilde{\mathbf{h}}_{n,k}\right]_{t_{1}}\left[
		\tilde{\mathbf{h}}_{n,k}\right]^*_{t_{2}}\right\}=\mathbb{E}\left\{\left[\tilde{\mathbf{h}}_{n,k}\right]_{t_{1}}\right\}\mathbb{E}\left\{\left[\tilde{\mathbf{h}}_{n,k}\right]^*_{t_{2}}\right\}=0, \forall t_{1} \neq t_{2},\\
		\mathbb{E}\left\{\left|\left[\tilde{\mathbf{h}}_{n,k}\right]_{t}\right|^{2}\right\}=\mathbb{E}\left\{\left|\left[\tilde{\mathbf{h}}_{n,i}\right]_{t}\right|^{2}\right\}=\mathbb{E}\left\{\left|\left[\tilde{\mathbf{Z}}_{m,n}\right]_{x,y}\right|^{2}\right\}=1,
	\end{array}
\end{align}
where $\left[{\mathbf{Z}}_{m,n}\right]_{x,y}$ denotes the $(x,y)$-th entry of matrix ${\mathbf{Z}}_{m,n}$ and $\left[\mathbf{h}_{n,k}\right]_{t} $ represents the $t$-th element of column vector ${\bf h}_{n,k}$.

Next, we will derive $\mathbb{E}\left\{\left\|\mathbf{g}_{k}\right\|^{2}\right\}$, $\mathbb{E}\left\{\left\|\mathbf{g}_{k}\right\|^{4}\right\}$ and $\mathbb{E}\left\{\left|\mathbf{g}_{k}^{H} \mathbf{g}_{i}\right|^{2}\right\}$, respectively.
\subsection{Derivation of $\mathbb{E}\left\{\left\|\mathbf{g}_{k}\right\|^{2}\right\}$}\label{subsection1}
Based on the definition in (\ref{g_k}) and (\ref{g_mnk}), $\mathbb{E}\left\{\left\|\mathbf{g}_{k}\right\|^{2}\right\}$ can be expressed as
\begin{align}
	\begin{array}{l}
		\mathbb{E}\left\{\left\|\mathbf{g}_{k}\right\|^{2}\right\} \\= \mathbb{E}\left\{\mathbf{g}^H_{k}\mathbf{g}_{k}   \right\}  \\=\sum\limits_{m=1}^{M}\mathbb{E}\left\{\left\|\mathbf{g}_{m,k}\right\|^{2}\right\}\\=\sum\limits_{m=1}^{M}\mathbb{E}\left\{\mathbf{g}^H_{m,k}\mathbf{g}_{m,k}\right\} \\= \sum\limits_{m=1}^{M}\mathbb{E} \left\{ \sum\limits_{n_{1}=1}^{N}\mathbf{g}^H_{m,n_{1},k} \sum\limits_{n_{2}=1}^{N}\mathbf{g}_{m,n_{2},k} \right\} \\ 
		=\sum\limits_{m=1}^{M}\sum\limits_{n=1}^{N}\mathbb{E} \left\{\mathbf{g}^H_{m,n,k}\mathbf{g}_{m,n,k} \right\} + \sum\limits_{m=1}^{M}\sum\limits_{n_{1}=1}^{N}\sum\limits_{n_{2}=1 \atop n_{2}\neq n_{1}}^{N}\mathbb{E} \left\{\mathbf{g}^H_{m,n_{1},k}\mathbf{g}_{m,n_{2},k} \right\} \\
		=\sum\limits_{m=1}^{M}\sum\limits_{n=1}^{N}c_{m,n,k}\mathbb{E}\left\{\sum\limits_{\omega=1}^{4}\left(\mathbf{g}_{m,n,k}^{\omega}\right)^{H} \sum\limits_{\psi=1}^{4} \mathbf{g}_{m,n,k}^{\psi}\right\} \\ \quad +\sum\limits_{m=1}^{M}\sum\limits_{n_{1}=1}^{N}\sum\limits_{n_{2}=1 \atop n_{2}\neq n_{1}}^{N} \sqrt{c_{m,n_{1},k}c_{m,n_{2},k}}\mathbb{E}\left\{\sum\limits_{\omega=1}^{4}\left(\mathbf{g}_{m,n_{1},k}^{\omega}\right)^{H} \sum\limits_{\psi=1}^{4} \mathbf{g}_{m,n_{2},k}^{\psi}\right\}.
		%=\frac{\beta \alpha_{k}}{(\delta+1)\left(\varepsilon_{k}+1\right)} \mathbb{E}\left\{\sum_{\omega=1}^{4}\left(\mathbf{g}_{k}^{\omega}\right)^{H} \sum_{\psi=1}^{4} \mathbf{g}_{k}^{\psi}\right\}.
	\end{array}
\end{align}

Based on (\ref{property1}), we have
\begin{align}
	\begin{array}{l}
		\mathbb{E}\left\{\left(\mathbf{g}_{m,n,k}^{\omega}\right)^{H} \mathbf{g}_{m,n,k}^{\psi}\right\}=0, \forall \omega \neq \psi. %\\
		%\mathbb{E}\left\{\left(\mathbf{g}_{k}^{\omega}\right)^{H} \mathbf{g}_{i}^{\psi}\right\}=0, \omega \neq \psi \cup \omega=2 \cup \omega=4,
	\end{array}
\end{align}

Therefore, we have
\begin{align}\label{lemma1kk}
	\begin{array}{l}
		\mathbb{E}\left\{\mathbf{g}_{m,n,k}^{H} \mathbf{g}_{m,n,k}\right\}\\
		=c_{m,n,k}  \mathbb{E}\left\{\sum\limits_{\omega=1}^{4}\left(\mathbf{g}_{m,n,k}^{\omega}\right)^{H}  \mathbf{g}_{m,n,k}^{\omega}\right\} \\
		=c_{m,n,k}\Bigg(\delta_{m,n} \varepsilon_{n,k}\left\|\overline{\mathbf{Z}}_{m,n} {\bf \Phi}_{n}	 \overline{\mathbf{h}}_{n,k}\right\|^{2}+ \delta_{m,n}\mathbb{E}\left\{\left\|\overline{\mathbf{Z}}_{m,n} {\bf \Phi}_{n}	 \tilde{\mathbf{h}}_{n,k}\right\|^{2}\right\} \\
		\quad\quad+\varepsilon_{n,k} \mathbb{E}\left\{\left\|\tilde{\mathbf{Z}}_{m,n}{\bf \Phi}_{n}	 \overline{\mathbf{h}}_{n,k}\right\|^{2}\right\}+\mathbb{E}\left\{\left\|\tilde{\mathbf{Z}}_{m,n} {\bf \Phi}_{n}	 \tilde{\mathbf{h}}_{n,k}\right\|^{2}\right\}\Bigg) \\ 
		{\mathop  = \limits^{\left( a \right)} }c_{m,n,k}\left(\delta_{m,n} \varepsilon_{n,k} M_{b}\left|f_{m,n,k}({\bf \Phi}	)\right|^{2}+\delta_{m,n} M_{b} N_{r}+\varepsilon_{n,k} M_{b} N_{r}+M_{b} N_{r}\right)\\
		=M_{b}c_{m,n,k}\left(\delta_{m,n} \varepsilon_{n,k}\left|f_{m,n,k}({\bf \Phi}	)\right|^{2}+\left(\delta_{m,n}+\varepsilon_{n,k}+1\right) N_{r}\right),
	\end{array}
\end{align}
%{\bf \Phi}	
%{\mathop  = \limits^{\left( a \right)} }
where $(a)$ follows by exploiting the identities
\begin{align}
	\begin{array}{l}
		\left\|\overline{\mathbf{Z}}_{m,n} {\bf \Phi}_{n}\overline{\mathbf{h}}_{n,k}\right\|^{2}=\left\|\mathbf{a}_{M_{b}}(m,n)\right\|^{2}\left\|\mathbf{a}_{N_{r}}^{H}(m,n) {\bf \Phi}_{n} \overline{\mathbf{h}}_{n,k}\right\|^{2}=M_{b}\left|f_{m,n,k}({\bf \Phi}	)\right|^{2}, \\
		\mathrm{Tr}\big\{\overline{\mathbf{Z}}_{m,n}^{H}\overline{\mathbf{Z}}_{m,n}\big\}=\mathrm{Tr}\big\{\overline{\mathbf{Z}}_{m,n}\overline{\mathbf{Z}}_{m,n}^{H}\big\}=M_{b}N_{r},\\
		\mathbb{E}\left\{\tilde{\mathbf{h}}_{n,k} \tilde{\mathbf{h}}_{n,k}^{H}\right\}=\mathbf{I}_{N_{r}}, {\bf\Phi}_{n}{\bf\Phi}_{n}^H={\bf I}_{N_{r}},
		\mathbb{E}\left\{\tilde{\mathbf{h}}_{n,k}^{H} \tilde{\mathbf{h}}_{n,k}\right\}=\overline{\mathbf{h}}_{n,k}^{H} \overline{\mathbf{h}}_{n,k}=N_{r}, \\
		\mathbb{E}\left\{\tilde{\mathbf{Z}}_{m,n}^{H} \tilde{\mathbf{Z}}_{m,n}\right\}=M_{b} \mathbf{I}_{N_{r}}, 
		\mathbb{E}\left\{ \tilde{\mathbf{Z}}_{m,n}\tilde{\mathbf{Z}}_{m,n}^{H}\right\}=N_{r} \mathbf{I}_{M_{b}}.
	\end{array}
\end{align}

Besides, we have 
\begin{align}
	\begin{array}{l}
		\mathbb{E}\left\{\mathbf{g}_{m,n_{1},k}^{H} \mathbf{g}_{m,n_{2},k}\right\}\\
		=\sqrt{c_{m,n_{1},k}c_{m,n_{2},k}} \mathbb{E}\left\{\sum\limits_{\omega=1}^{4}\left(\mathbf{g}_{m,n_{1},k}^{\omega}\right)^{H}\sum\limits_{\psi=1}^{4}\mathbf{g}_{m,n_{2},k}^{\psi}\right\} \\
		=\sqrt{c_{m,n_{1},k}c_{m,n_{2},k}} \Bigg(\sqrt{\delta_{m,n_{1}}\delta_{m,n_{2}}\varepsilon_{n_{1},k}\varepsilon_{n_{2},k}} {\,}\overline{\mathbf{h}}_{n_{1},k}^{H}{\bf \Phi}_{n_{1}}^{H}\overline{\mathbf{Z}}_{m,n_{1}}^{H}\overline{\mathbf{Z}}_{m,n_{2}} {\bf \Phi}_{n_{2}}\overline{\mathbf{h}}_{n_{2},k}+ \sqrt{\delta_{m,n_{1}}\delta_{m,n_{1}}} \\ \mathbf{E}\Big\{\tilde{\mathbf{h}}_{n_{1},k}^{H}{\bf\Phi}_{n_{1}}^{H}\overline{\mathbf{Z}}_{m,n_{1}}^{H}\overline{\mathbf{Z}}_{m,n_{2}}{\bf\Phi}_{n_{2}}\tilde{\mathbf{h}}_{n_{2},k}\Big\}+ \sqrt{\varepsilon_{n_{1},k}\varepsilon_{n_{2},k}}\mathbf{E}\Big\{\overline{\mathbf{h}}_{n_{1},k}^{H}{\bf\Phi}_{n_{1}}^{H} \tilde{\mathbf{Z}}_{m,n_{1}}^{H}   \tilde{\mathbf{Z}}_{m,n_{2}}{\bf\Phi}_{n_{2}}\overline{\mathbf{h}}_{n_{2},k}\Big\}\\+  \mathbf{E}\Big\{\tilde{\mathbf{h}}_{n_{1},k}^{H}{\bf\Phi}_{n_{1}}^{H} \tilde{\mathbf{Z}}_{m,n_{1}}^{H}   \tilde{\mathbf{Z}}_{m,n_{2}}{\bf\Phi}_{n_{2}}\tilde{\mathbf{h}}_{n_{2},k}\Big\}\Bigg) \\
		{\mathop = \limits^{\left(b \right)} }\sqrt{c_{m,n_{1},k}c_{m,n_{2},k} \delta_{m,n_{1}}\delta_{m,n_{2}} \varepsilon_{n_{1},k}\varepsilon_{n_{2},k}}
		{\,}\overline{\mathbf{h}}_{n_{1},k}^{H}{\bf \Phi}_{n_{1}}^{H}\overline{\mathbf{Z}}_{m,n_{1}}^{H}\overline{\mathbf{Z}}_{m,n_{2}} {\bf \Phi}_{n_{2}}\overline{\mathbf{h}}_{n_{2},k} \\
		=\sqrt{c_{m,n_{1},k}c_{m,n_{2},k} \delta_{m,n_{1}}\delta_{m,n_{2}} \varepsilon_{n_{1},k}\varepsilon_{n_{2},k}} f_{m,n_{1},k}^{H}({\bf \Phi})f_{m,n_{2},k}({\bf \Phi}) {\bf a}^{H}_{M_{b}}(m,n_{1}){\bf a}_{M_{b}}(m,n_{2}),
	\end{array}
\end{align}
where $(b)$ applies (\ref{property1}) and exploits the independence of $\tilde{\mathbf{Z}}_{m,n_{1}}$, $\tilde{\mathbf{Z}}_{m,n_{2}}$, $\tilde{\mathbf{h}}_{n_{1},k}$ and $\tilde{\mathbf{h}}_{n_{2},k}$, $ \forall n_{1} \neq n_{2}$.

In general, $\mathbb{E}\left\{\left\|\mathbf{g}_{k}\right\|^{2}\right\}$ can be expressed as 
\begin{align}\label{g_k2}
	\begin{array}{l}
		\mathbb{E}\left\{\left\|\mathbf{g}_{k}\right\|^{2}\right\} \\ =\sum\limits_{m=1}^{M}	\mathbb{E}\left\{\left\|\mathbf{g}_{m,k}\right\|^{2}\right\}\\
		=\sum\limits_{m=1}^{M}\bigg(\sum\limits_{n_{1}=1}^{N}\sum\limits_{n_{2}=1}^{N} \sqrt{c_{m,n_{1},k}c_{m,n_{2},k} \delta_{m,n_{1}}\delta_{m,n_{2}} \varepsilon_{n_{1},k}\varepsilon_{n_{2},k}} f_{m,n_{1},k}^{H}({\bf \Phi})f_{m,n_{2},k}({\bf \Phi}) \\
		\qquad{\bf a}^{H}_{M_{b}}(m,n_{1}){\bf a}_{M_{b}}(m,n_{2})
		+\sum\limits_{n=1}^{N} c_{m,n,k}M_{b}N_{r}\left(\delta_{m,n}+\varepsilon_{n,k}+1\right)\bigg).
	\end{array}
\end{align}
Note that the expression of $ \mathbb{E}\left\{\left\|\mathbf{g}_{m,k}\right\|^{2}\right\}$ can also be obtained from the above derivation. Finally, we can complete the calculation of the noise term $\mathbb{E}\left\{\left\|{\bf g}_k+{\bf d}_k\right\|^{2}\right\}$ based on (\ref{noise_term}) and (\ref{g_k2}).

\subsection{Derivation of $\mathbb{E}\left\{\left|\mathbf{g}_{k}^{H} \mathbf{g}_{i}\right|^{2}\right\}$}\label{subsection2}

Before the derivation, we first provide some important properties as follows
\begin{align}\label{ZYZ}
	\begin{array}{l}
		\mathbb{E}\left\{\tilde{\mathbf{Z}}_{m,n} \mathbf{Y} \tilde{\mathbf{Z}}_{m,n}\right\}=	\mathbb{E}\left\{\operatorname{Re}\left\{\tilde{\mathbf{Z}}_{m,n} \mathbf{Y} \tilde{\mathbf{Z}}_{m,n}\right\}\right\}= \mathbf{0},\\
		\mathbb{E}\left\{\tilde{\mathbf{h}}_{n,k}^{H} \mathbf{y}_{1} \tilde{\mathbf{h}}_{n,k}^{H}\mathbf{y}_{2}\right\}=\mathbb{E}\left\{\operatorname{Re}\left\{\tilde{\mathbf{h}}_{n,k}^{H} \mathbf{y}_{1} \tilde{\mathbf{h}}_{n,k}^{H}\mathbf{y}_{2}\right\}\right\}= 0,
	\end{array}
\end{align}
where $ \mathbf{Y} \in \mathbb{C}^{N_{r} \times M_{b}} $ is an arbitrary deterministic matrix, and $ \mathbf{y}_{1} , \mathbf{y}_{2} \in \mathbb{C}^{N_{r} \times 1}$ are the arbitrary deterministic vectors. This conclusion can be proved by firstly considering the case of dimension of one and then generalizing it to high dimensions through mathematical induction.

Besides, by exploiting the moment properties of normal distribution, for an deterministic matrice     $ \mathbf{W} \in \mathbb{C}^{N_{r} \times N_{r}} $, we have 
\begin{align}\label{ZWZ}
	\begin{array}{l}
		\mathbb{E}\left\{\tilde{\mathbf{Z}}_{m,n} \mathbf{W}\tilde{\mathbf{Z}}_{m,n}^{H}\right\}=\operatorname{Tr}\{\mathbf{W}\} \mathbf{I}_{M_{b}},\\
		\mathbb{E}\left\{\tilde{\mathbf{h}}_{n,k}^{H} \mathbf{W}\tilde{\mathbf{h}}_{n,k}\right\}=\operatorname{Tr}\{\mathbf{W}\},
	\end{array}
\end{align}
and
\begin{align}\label{ZZWZZ}
	\begin{array}{l}
		\mathbb{E}\left\{\tilde{\mathbf{Z}}_{m,n}^{H}\tilde{\mathbf{Z}}_{m,n} \mathbf{W}\tilde{\mathbf{Z}}_{m,n}^{H}\tilde{\mathbf{Z}}_{m,n}\right\}=M_{b}^{2}\mathbf{W}+M_{b}\operatorname{Tr}\{\mathbf{W}\}\mathbf{I}_{N_{r}},\\
		\mathbb{E}\left\{\tilde{\mathbf{h}}_{n,k}\tilde{\mathbf{h}}_{n,k}^{H} \mathbf{W}\tilde{\mathbf{h}}_{n,k}\tilde{\mathbf{h}}_{n,k}^{H}\right\}=\mathbf{W}+\operatorname{Tr}\{\mathbf{W}\}\mathbf{I}_{N_{r}}.
	\end{array}
\end{align}

Unlike the scenario without RIS, since the cascaded links between the BSs and different users contain the same RIS-BS channel $\mathbf{Z}$, $ \mathbf{g}_i $ is no longer independent of $\mathbf{g}_k $. Recalling (\ref{g_k}) $\sim$ (\ref{g_mni}), when deriving $\mathbb{E}\left\{\left|\mathbf{g}_{k}^{H} \mathbf{g}_{i}\right|^{2}\right\}$, we can ignore the zero-mean terms based on (\ref{property1}) and (\ref{ZYZ}). Then we have
\begin{align}\label{gkgi_2}
	\begin{array}{l}
		\mathbb{E}\left\{\left|\mathbf{g}_{k}^{H} \mathbf{g}_{i}\right|^{2}\right\}=	\mathbb{E}\left\{\left|\sum_{m=1}^{M}\mathbf{g}_{m,k}^{H} \mathbf{g}_{m,i}\right|^{2}\right\}\\= \mathbb{E}\left\{\left|\sum_{m=1}^{M}\left(\sum_{n_{1}=1}^{N}\mathbf{g}_{m,n_{1},k}^{H}\sum_{n_{2}=1}^{N} \mathbf{g}_{m,n_{2},i}\right)\right|^{2}\right\}\\
		=\mathbb{E}\left\{\sum\limits_{m_{1}=1}^{M}\left(\sum\limits_{n_{1}=1}^{N}\mathbf{g}_{m_{1},n_{1},k}^{H}\sum\limits_{n_{2}=1}^{N} \mathbf{g}_{m_{1},n_{2},i}\right) \sum\limits_{m_{2}=1}^{M}\left(\sum\limits_{n_{3}=1}^{N}\mathbf{g}_{m_{2},n_{3},i}^{H}\sum\limits_{n_{4}=1}^{N}\mathbf{g}_{m_{2},n_{4},k}\right)\right\}\\		={\sum\limits_{m_{1}=1}^{M}\sum\limits_{m_{2}=1}^{M}\sum\limits_{n_{1}=1}^{N}\sum\limits_{n_{2}=1}^{N}\sum\limits_{n_{3}=1}^{N}\sum\limits_{n_{4}=1}^{N}} \mathbb{E}\Big\{\mathbf{g}_{m_{1},n_{1},k}^{H} \mathbf{g}_{m_{1},n_{2},i} \mathbf{g}_{m_{2},n_{3},i}^{H}\mathbf{g}_{m_{2},n_{4},k}\Big\}\\
		={\sum\limits_{m_{1}=1}^{M}\sum\limits_{m_{2}=1}^{M}\sum\limits_{n_{1}=1}^{N}\sum\limits_{n_{2}=1}^{N}\sum\limits_{n_{3}=1}^{N}\sum\limits_{n_{4}=1}^{N}}
		\sqrt{c_{m_{1},n_{1},k} c_{m_{1},n_{2},i} c_{m_{2},n_{3},i} c_{m_{2},n_{4},k}}
		\\  \quad
		\mathbb{E}\Bigg\{ \sum\limits_{\omega_{1}=1}^{4}
		\Big(\mathbf{g}_{m_{1},n_{1},k}^{\omega_{1}}\Big)^{H}\sum\limits_{\psi_{1}=1}^{4} \mathbf{g}_{m_{1},n_{2},i}^{\psi_{1}}
		\sum\limits_{\psi_{2}=1}^{4}
		\Big(\mathbf{g}_{m_{2},n_{3},i}^{\psi_{2}}\Big)^{H}
		\sum\limits_{\omega_{2}=1}^{4}
		\mathbf{g}_{m_{2},n_{4},k}^{\omega_{2}}
		\Bigg\} \\
		={\sum\limits_{m_{1}=1}^{M}\sum\limits_{m_{2}=1}^{M}\sum\limits_{n_{1}=1}^{N}\sum\limits_{n_{2}=1}^{N}\sum\limits_{n_{3}=1}^{N}\sum\limits_{n_{4}=1}^{N}}
		\sqrt{c_{m_{1},n_{1},k} c_{m_{1},n_{2},i} c_{m_{2},n_{3},i} c_{m_{2},n_{4},k}}
		\\ \qquad\qquad\qquad
		\mathbb{E}\Bigg\{ \sum\limits_{\omega=1}^{4} \sum\limits_{\psi=1}^{4} 
		\Big(\mathbf{g}_{m_{1},n_{1},k}^{\omega}\Big)^{H}
		\mathbf{g}_{m_{1},n_{2},i}^{\psi}
		\Big(\mathbf{g}_{m_{2},n_{3},i}^{\psi}\Big)^{H}
		\mathbf{g}_{m_{2},n_{4},k}^{\omega}
		\\+	2\mathrm{Re}\bigg\{
		\Big(\mathbf{g}_{m_{1},n_{1},k}^{1}\Big)^{H}\mathbf{g}_{m_{1},n_{2},i}^{1}
		\Big(\mathbf{g}_{m_{2},n_{3},i}^{3}\Big)^{H}
		\mathbf{g}_{m_{2},n_{4},k}^{3}
		+
		\Big(\mathbf{g}_{m_{1},n_{1},k}^{1}\Big)^{H}\mathbf{g}_{m_{1},n_{2},i}^{2}
		\Big(\mathbf{g}_{m_{2},n_{3},i}^{4}\Big)^{H}
		\mathbf{g}_{m_{2},n_{4},k}^{3}
		\\+
		\Big(\mathbf{g}_{m_{1},n_{1},k}^{2}\Big)^{H}\mathbf{g}_{m_{1},n_{2},i}^{1} 
		\Big(\mathbf{g}_{m_{2},n_{3},i}^{3}\Big)^{H}
		\mathbf{g}_{m_{2},n_{4},k}^{4}
		+
		\Big(\mathbf{g}_{m_{1},n_{1},k}^{2}\Big)^{H}\mathbf{g}_{m_{1},n_{2},i}^{2}
		\Big(\mathbf{g}_{m_{2},n_{3},i}^{4}\Big)^{H}
		\mathbf{g}_{m_{2},n_{4},k}^{4}\bigg\}\Bigg\}.	
	\end{array}
\end{align}
%	\mathbb{E}\left\{     \right\}

Next, we will derive the above terms in (\ref{gkgi_2}) one by one. The derivation utilizes the properties in (\ref{ZWZ}) and (\ref{ZZWZZ}), and the independence between $\tilde{\mathbf{Z}}_{m_{1},n_{1}}$, $\tilde{\mathbf{Z}}_{m_{2},n_{2}}$, $\tilde{\mathbf{h}}_{n_{1},k}$, $\tilde{\mathbf{h}}_{n_{2},i}$, $\tilde{\mathbf{h}}_{n_{3},i}$ and $\tilde{\mathbf{h}}_{n_{4},k}$, where $ (m_{1},n_{1}) \neq (m_{2},n_{2})$, $ n_{2} \neq n_{3}$, $ n_{1} \neq n_{4}$. To facilitate the derivation, we omit the notation of accumulation $ \sum\limits_{m_{1}=1}^{M}\sum\limits_{m_{2}=1}^{M}\sum\limits_{n_{1}=1}^{N}\sum\limits_{n_{2}=1}^{N}\sum\limits_{n_{3}=1}^{N}\sum\limits_{n_{4}=1}^{N} $ and the common coefficient $\sqrt{c_{m_{1},n_{1},k} c_{m_{1},n_{2},i} c_{m_{2},n_{3},i} c_{m_{2},n_{4},k}}$, which will vary with the parameters $m_{1}$, $m_{2}$, $n_{1}$, $n_{2}$, $n_{3}$, and $n_{4}$ in the next derivation.

First, we focus on the terms of $\mathbb{E}\Bigg\{  
\Big(\mathbf{g}_{m_{1},n_{1},k}^{\omega}\Big)^{H}\mathbf{g}_{m_{1},n_{2},i}^{\psi}
\Big(\mathbf{g}_{m_{2},n_{3},i}^{\psi}\Big)^{H}
\mathbf{g}_{m_{2},n_{4},k}^{\omega}
\Bigg\}, 1\leq\omega,\psi\leq4$. 

When $\omega=1$ and $\psi=1$, we have
\begin{align}\label{gkgi2_begin}
	\begin{aligned}
		&\mathbb{E}\Bigg\{  
		\Big(\mathbf{g}_{m_{1},n_{1},k}^{1}\Big)^{H}\mathbf{g}_{m_{1},n_{2},i}^{1}
		\Big(\mathbf{g}_{m_{2},n_{3},i}^{1}\Big)^{H}
		\mathbf{g}_{m_{2},n_{4},k}^{1}
		\Bigg\} \\ 
		&=\sqrt{\delta_{m_{1},n_{1}}\delta_{m_{1},n_{2}}\delta_{m_{2},n_{3}}\delta_{m_{2},n_{4}} \varepsilon_{n_{1},k}\varepsilon_{n_{2},i} \varepsilon_{n_{3},i}\varepsilon_{n_{4},k}} \\ &\qquad{\overline{\mathbf{h}}_{n_{1},k}^{H}  {\bf\Phi}^{H}_{n_{1}} \overline{\mathbf{Z}}_{m_{1},n_{1}}^{H} \overline{\mathbf{Z}}_{m_{1},n_{2}} {\bf\Phi}_{n_{2}}\overline{\mathbf{h}}_{n_{2},i} \overline{\mathbf{h}}_{n_{3},i}^{H} {\bf\Phi}^{H}_{n_{3}} \overline{\mathbf{Z}}_{m_{2},n_{3}}^{H} \overline{\mathbf{Z}}_{m_{2},n_{4}} {\bf\Phi}_{n_{4}}\overline{\mathbf{h}}_{n_{4},k}}\\
		&=\sqrt{\delta_{m_{1},n_{1}}\delta_{m_{1},n_{2}}\delta_{m_{2},n_{3}}\delta_{m_{2},n_{4}} \varepsilon_{n_{1},k}\varepsilon_{n_{2},i} \varepsilon_{n_{3},i}\varepsilon_{n_{4},k}}  
		f_{m_{1},n_{1},k}^{H}({\bf \Phi})
		f_{m_{1},n_{2},i}({\bf \Phi}) \\	
		&\qquad f_{m_{2},n_{3},i}^{H}({\bf \Phi}) f_{m_{2},n_{4},k}({\bf \Phi}) 
		{\bf a}^{H}_{M_{b}}(m_{1},n_{1})
		{\bf a}_{M_{b}}(m_{1},n_{2})
		{\bf a}^{H}_{M_{b}}(m_{2},n_{3})
		{\bf a}_{M_{b}}(m_{2},n_{4}).
	\end{aligned}
\end{align}

When $\omega=1$ and $\psi=2$, we have
\begin{align}
	\begin{array}{l}
		\mathbb{E}\Bigg\{  
		\Big(\mathbf{g}_{m_{1},n_{1},k}^{1}\Big)^{H}\mathbf{g}_{m_{1},n_{2},i}^{2}
		\Big(\mathbf{g}_{m_{2},n_{3},i}^{2}\Big)^{H}
		\mathbf{g}_{m_{2},n_{4},k}^{1}
		\Bigg\} \\ %\triangleq
		{\mathop = \limits^{\left(c \right)} } \sqrt{\delta_{m_{1},n_{1}}\delta_{m_{1},n_{2}}\delta_{m_{2},n_{2}}\delta_{m_{2},n_{4}} \varepsilon_{n_{1},k}\varepsilon_{n_{4},k}} \\ \qquad{\overline{\mathbf{h}}_{n_{1},k}^{H}  {\bf\Phi}^{H}_{n_{1}} \overline{\mathbf{Z}}_{m_{1},n_{1}}^{H} \overline{\mathbf{Z}}_{m_{1},n_{2}} {\bf\Phi}_{n_{2}}\mathbb{E} \Big\{\tilde{\mathbf{h}}_{n_{2},i}\tilde{\mathbf{h}}_{n_{2},i}^{H} \Big\} {\bf\Phi}^{H}_{n_{2}} \overline{\mathbf{Z}}_{m_{2},n_{2}}^{H} \overline{\mathbf{Z}}_{m_{2},n_{4}} {\bf\Phi}_{n_{4}}\overline{\mathbf{h}}_{n_{4},k}}\\
		=\sqrt{\delta_{m_{1},n_{1}}\delta_{m_{1},n_{2}}\delta_{m_{2},n_{2}}\delta_{m_{2},n_{4}} \varepsilon_{n_{1},k}\varepsilon_{n_{4},k}}
		f_{m_{1},n_{1},k}^{H}({\bf \Phi}) f_{m_{2},n_{4},k}({\bf \Phi}) 
		\\
		\qquad{\bf a}^{H}_{M_{b}}(m_{1},n_{1})
		{\bf a}_{M_{b}}(m_{1},n_{2})
		{\bf a}^{H}_{N_{r}}(m_{1},n_{2})
		{\bf a}_{N_{r}}(m_{2},n_{2})
		{\bf a}^{H}_{M_{b}}(m_{2},n_{2})
		{\bf a}_{M_{b}}(m_{2},n_{4}),
	\end{array}
\end{align}
where $(c)$ represents $n_{2}=n_{3}$, which removes the zero terms based on the independence and the zero-mean properties of $\tilde{\mathbf{h}}_{n_{2},i}$ and $\tilde{\mathbf{h}}_{n_{3},i}$ with $n_{2} \neq n_{3}$. Note that as $(c)$ occurs in the derivation, the accumulation notation of this term becomes ${\sum\limits_{m_{1}=1}^{M}\sum\limits_{m_{2}=1}^{M}\sum\limits_{n_{1}=1}^{N}\sum\limits_{n_{2}=1}^{N} \sum\limits_{n_{4}=1}^{N}}$, the omitted coefficient of this term becomes $\sqrt{c_{m_{1},n_{1},k} c_{m_{1},n_{2},i} c_{m_{2},n_{2},i} c_{m_{2},n_{4},k}}$, and the following derivations are the same.  

When $\omega=1$ and $\psi=3$, we have
\begin{align}
\begin{array}{l}
		\mathbb{E}\Bigg\{  
		\Big(\mathbf{g}_{m_{1},n_{1},k}^{1}\Big)^{H}\mathbf{g}_{m_{1},n_{2},i}^{3}
		\Big(\mathbf{g}_{m_{2},n_{3},i}^{3}\Big)^{H}
		\mathbf{g}_{m_{2},n_{4},k}^{1}
		\Bigg\} \\
		{\mathop = \limits^{\substack{(d)}}} \sqrt{\delta_{m_{1},n_{1}}\delta_{m_{1},n_{4}} \varepsilon_{n_{1},k}\varepsilon_{n_{4},k}} \varepsilon_{n_{2},i}\\ \qquad{\overline{\mathbf{h}}_{n_{1},k}^{H}  {\bf\Phi}^{H}_{n_{1}} \overline{\mathbf{Z}}_{m_{1},n_{1}}^{H} \mathbb{E}\Big\{\tilde{\mathbf{Z}}_{m_{1},n_{2}}{\bf\Phi}_{n_{2}}\overline{\mathbf{h}}_{n_{2},i}\overline{\mathbf{h}}_{n_{2},i}^{H} {\bf\Phi}^{H}_{n_{2}}  \tilde{\mathbf{Z}}_{m_{1},n_{2}}^{H}\Big\} \overline{\mathbf{Z}}_{m_{1},n_{4}} {\bf\Phi}_{n_{4}}\overline{\mathbf{h}}_{n_{4},k}}\\
		=\sqrt{\delta_{m_{1},n_{1}}\delta_{m_{1},n_{4}} \varepsilon_{n_{1},k}\varepsilon_{n_{4},k}} \varepsilon_{n_{2},i}
		{\overline{\mathbf{h}}_{n_{1},k}^{H}  {\bf\Phi}^{H}_{n_{1}} \overline{\mathbf{Z}}_{m_{1},n_{1}}^{H} \mathbb{E}\Big\{\tilde{\mathbf{Z}}_{m_{1},n_{2}} \tilde{\mathbf{Z}}_{m_{1},n_{2}}^{H}\Big\} \overline{\mathbf{Z}}_{m_{1},n_{4}} {\bf\Phi}_{n_{4}}\overline{\mathbf{h}}_{n_{4},k}}\\
		=\sqrt{\delta_{m_{1},n_{1}}\delta_{m_{1},n_{4}} \varepsilon_{n_{1},k}\varepsilon_{n_{4},k}} \varepsilon_{n_{2},i}N_{r}
		{\overline{\mathbf{h}}_{n_{1},k}^{H}  {\bf\Phi}^{H}_{n_{1}} \overline{\mathbf{Z}}_{m_{1},n_{1}}^{H}  \overline{\mathbf{Z}}_{m_{1},n_{4}} {\bf\Phi}_{n_{4}}\overline{\mathbf{h}}_{n_{4},k}}\\
		=\sqrt{\delta_{m_{1},n_{1}}\delta_{m_{1},n_{4}} \varepsilon_{n_{1},k}\varepsilon_{n_{4},k}} \varepsilon_{n_{2},i} N_{r}
		{f_{m_{1},n_{1},k}^{H}({\bf \Phi}) 
			f_{m_{1},n_{4},k}({\bf \Phi}) 
			{\bf a}^{H}_{M_{b}}(m_{1},n_{1})    
			{\bf a}_{M_{b}}(m_{1},n_{4})},
	\end{array}
\end{align}
where $(d)$ represents $m_{1}=m_{2}$, $n_{2}=n_{3}$ and removes the zero terms.

When $\omega=1$ and $\psi=4$, we arrive at
\begin{align}
	\begin{array}{l}
		\mathbb{E}\Bigg\{  
		\Big(\mathbf{g}_{m_{1},n_{1},k}^{1}\Big)^{H}\mathbf{g}_{m_{1},n_{2},i}^{4}
		\Big(\mathbf{g}_{m_{2},n_{3},i}^{4}\Big)^{H}
		\mathbf{g}_{m_{2},n_{4},k}^{1}
		\Bigg\} \\
		{\mathop = \limits^{\substack{(d)}}}
		\sqrt{\delta_{m_{1},n_{1}}\delta_{m_{1},n_{4}} \varepsilon_{n_{1},k}\varepsilon_{n_{4},k}} 
		\\ %{\xlongequal[m_{1}=m_{2}]{n_{2}=n_{3}}}
		\qquad{\mathbb{E}\Big\{\overline{\mathbf{h}}_{n_{1},k}^{H}  {\bf\Phi}^{H}_{n_{1}} \overline{\mathbf{Z}}_{m_{1},n_{1}}^{H} \tilde{\mathbf{Z}}_{m_{1},n_{2}} {\bf\Phi}_{n_{2}} \tilde{\mathbf{h}}_{n_{2},i} \tilde{\mathbf{h}}_{n_{2},i}^{H} {\bf\Phi}^{H}_{n_{2}}  \tilde{\mathbf{Z}}_{m_{1},n_{2}}^{H} \overline{\mathbf{Z}}_{m_{1},n_{4}} {\bf\Phi}_{n_{4}}\overline{\mathbf{h}}_{n_{4},k}\Big\}}\\
		= 
		\sqrt{\delta_{m_{1},n_{1}}\delta_{m_{1},n_{4}} \varepsilon_{n_{1},k}\varepsilon_{n_{4},k}} 
		\\ %{\xlongequal[m_{1}=m_{2}]{n_{2}=n_{3}}}
		\qquad{\overline{\mathbf{h}}_{n_{1},k}^{H}  {\bf\Phi}^{H}_{n_{1}} \overline{\mathbf{Z}}_{m_{1},n_{1}}^{H} \mathbb{E}\Big\{\tilde{\mathbf{Z}}_{m_{1},n_{2}} {\bf\Phi}_{n_{2}} \mathbb{E}\Big\{\tilde{\mathbf{h}}_{n_{2},i} \tilde{\mathbf{h}}_{n_{2},i}^{H}\Big\} {\bf\Phi}^{H}_{n_{2}}  \tilde{\mathbf{Z}}_{m_{1},n_{2}}^{H}\Big\} \overline{\mathbf{Z}}_{m_{1},n_{4}} {\bf\Phi}_{n_{4}}\overline{\mathbf{h}}_{n_{4},k}}\\
		=\sqrt{\delta_{m_{1},n_{1}}\delta_{m_{1},n_{4}} \varepsilon_{n_{1},k}\varepsilon_{n_{4},k}}  
		{\overline{\mathbf{h}}_{n_{1},k}^{H}  {\bf\Phi}^{H}_{n_{1}} \overline{\mathbf{Z}}_{m_{1},n_{1}}^{H} \mathbb{E}\Big\{\tilde{\mathbf{Z}}_{m_{1},n_{2}} \tilde{\mathbf{Z}}_{m_{1},n_{2}}^{H}\Big\} \overline{\mathbf{Z}}_{m_{1},n_{4}} {\bf\Phi}_{n_{4}}\overline{\mathbf{h}}_{n_{4},k}}\\
		=\sqrt{\delta_{m_{1},n_{1}}\delta_{m_{1},n_{4}} \varepsilon_{n_{1},k}\varepsilon_{n_{4},k}} N_{r}
		{f_{m_{1},n_{1},k}^{H}({\bf \Phi}) 
			f_{m_{1},n_{4},k}({\bf \Phi}) 
			{\bf a}^{H}_{M_{b}}(m_{1},n_{1})    
			{\bf a}_{M_{b}}(m_{1},n_{4})}.
	\end{array}
\end{align}

Secondly, we consider the terms with $\omega=2$. When $\psi=1$, we have
\begin{align}
	\begin{array}{l}
		\mathbb{E}\Bigg\{  
		\Big(\mathbf{g}_{m_{1},n_{1},k}^{2}\Big)^{H}\mathbf{g}_{m_{1},n_{2},i}^{1}
		\Big(\mathbf{g}_{m_{2},n_{3},i}^{1}\Big)^{H}
		\mathbf{g}_{m_{2},n_{4},k}^{2}
		\Bigg\} \\
		\overset{\substack{(e)}}{=}
		\sqrt{\delta_{m_{1},n_{1}}\delta_{m_{1},n_{2}} \delta_{m_{2},n_{3}}\delta_{m_{2},n_{1}} \varepsilon_{n_{2},i}\varepsilon_{n_{3},i}} {\mathbb{E}\Big\{\tilde{\mathbf{h}}_{n_{1},k}^{H}  {\bf\Phi}^{H}_{n_{1}} \overline{\mathbf{Z}}_{m_{1},n_{1}}^{H} \overline{\mathbf{Z}}_{m_{1},n_{2}} {\bf\Phi}_{n_{2}} \overline{\mathbf{h}}_{n_{2},i} \overline{\mathbf{h}}_{n_{3},i}^{H} {\bf\Phi}^{H}_{n_{3}}  \overline{\mathbf{Z}}_{m_{2},n_{3}}^{H} \overline{\mathbf{Z}}_{m_{2},n_{1}} {\bf\Phi}_{n_{1}}\tilde{\mathbf{h}}_{n_{1},k}\Big\}}   \\
		=\sqrt{\delta_{m_{1},n_{1}}\delta_{m_{1},n_{2}} \delta_{m_{2},n_{3}}\delta_{m_{2},n_{1}} \varepsilon_{n_{2},i}\varepsilon_{n_{3},i}} {\overline{\mathbf{h}}_{n_{3},i}^{H}  {\bf\Phi}^{H}_{n_{3}} \overline{\mathbf{Z}}_{m_{2},n_{3}}^{H} \overline{\mathbf{Z}}_{m_{2},n_{1}} {\bf\Phi}_{n_{1}} \mathbb{E}\Big\{\tilde{\mathbf{h}}_{n_{1},k} \tilde{\mathbf{h}}_{n_{1},k}^{H}\Big\} {\bf\Phi}^{H}_{n_{1}}  \overline{\mathbf{Z}}_{m_{1},n_{1}}^{H} \overline{\mathbf{Z}}_{m_{1},n_{2}} {\bf\Phi}_{n_{2}}\overline{\mathbf{h}}_{n_{2},i}}\\
		=\sqrt{\delta_{m_{1},n_{1}}\delta_{m_{1},n_{2}} \delta_{m_{2},n_{3}}\delta_{m_{2},n_{1}} \varepsilon_{n_{2},i}\varepsilon_{n_{3},i}} 
		f_{m_{2},n_{3},i}^{H}({\bf \Phi}) 
		f_{m_{1},n_{2},i}({\bf \Phi}) 
		\\
		\qquad{{\bf a}^{H}_{M_{b}}(m_{2},n_{3})    
			{\bf a}_{M_{b}}(m_{2},n_{1})
			{\bf a}^{H}_{N_{r}}(m_{2},n_{1})    
			{\bf a}_{N_{r}}(m_{1},n_{1})	
			{\bf a}^{H}_{M_{b}}(m_{1},n_{1})    
			{\bf a}_{M_{b}}(m_{1},n_{2})},
	\end{array}
\end{align}
where $(e)$ represents $n_{1}=n_{4}$ and removes the zero terms.

When $\psi=2$, we have
\begin{align}
	\begin{array}{l}
		%2 2 2 2
		\mathbb{E}\Bigg\{  
		\Big(\mathbf{g}_{m_{1},n_{1},k}^{2}\Big)^{H}\mathbf{g}_{m_{1},n_{2},i}^{2}
		\Big(\mathbf{g}_{m_{2},n_{3},i}^{2}\Big)^{H}
		\mathbf{g}_{m_{2},n_{4},k}^{2}
		\Bigg\} \\
		\overset{\substack{(f)}}{=}
		\sqrt{\delta_{m_{1},n_{1}}\delta_{m_{1},n_{2}} \delta_{m_{2},n_{1}}\delta_{m_{2},n_{2}}} 
		\\ 
		\qquad{\mathbb{E}\Big\{\tilde{\mathbf{h}}_{n_{1},k}^{H}  {\bf\Phi}^{H}_{n_{1}} \overline{\mathbf{Z}}_{m_{1},n_{1}}^{H} \overline{\mathbf{Z}}_{m_{1},n_{2}} {\bf\Phi}_{n_{2}} \mathbb{E}\Big\{\tilde{\mathbf{h}}_{n_{2},i} \tilde{\mathbf{h}}_{n_{2},i}^{H}\Big\} {\bf\Phi}^{H}_{n_{2}}  \overline{\mathbf{Z}}_{m_{2},n_{2}}^{H} \overline{\mathbf{Z}}_{m_{2},n_{1}} {\bf\Phi}_{n_{1}}\tilde{\mathbf{h}}_{n_{1},k}\Big\}}   \\
		=\sqrt{\delta_{m_{1},n_{1}}\delta_{m_{1},n_{2}} \delta_{m_{2},n_{1}}\delta_{m_{2},n_{2}}} 
		{\mathbb{E}\Big\{\tilde{\mathbf{h}}_{n_{1},k}^{H}{\bf\Phi}^{H}_{n_{1}} \overline{\mathbf{Z}}_{m_{1},n_{1}}^{H} \overline{\mathbf{Z}}_{m_{1},n_{2}}  \overline{\mathbf{Z}}_{m_{2},n_{2}}^{H} \overline{\mathbf{Z}}_{m_{2},n_{1}} {\bf\Phi}_{n_{1}}\tilde{\mathbf{h}}_{n_{1},k}\Big\}}   \\
		=
		\sqrt{\delta_{m_{1},n_{1}}\delta_{m_{1},n_{2}} \delta_{m_{2},n_{1}}\delta_{m_{2},n_{2}}} 
		{\mathrm{Tr}\Big\{ \overline{\mathbf{Z}}_{m_{2},n_{2}}^{H} \overline{\mathbf{Z}}_{m_{2},n_{1}} \mathbb{E}\Big\{{\bf\Phi}_{n_{1}}\tilde{\mathbf{h}}_{n_{1},k} \tilde{\mathbf{h}}_{n_{1},k}^{H} {\bf\Phi}^{H}_{n_{1}}\Big\} \overline{\mathbf{Z}}_{m_{1},n_{1}}^{H} \overline{\mathbf{Z}}_{m_{1},n_{2}} \Big\}}   
		\\
		=\sqrt{\delta_{m_{1},n_{1}}\delta_{m_{1},n_{2}} \delta_{m_{2},n_{1}}\delta_{m_{2},n_{2}}} 
		{\mathrm{Tr}\Big\{ \overline{\mathbf{Z}}_{m_{2},n_{2}}^{H} \overline{\mathbf{Z}}_{m_{2},n_{1}}  \overline{\mathbf{Z}}_{m_{1},n_{1}}^{H} \overline{\mathbf{Z}}_{m_{1},n_{2}} \Big\}}   
		\\
\end{array}
\end{align}
where $(f)$ represents $n_{1}=n_{4}$, $n_{2}=n_{3}$ and removes the zero terms.

When $\psi=3$, based on (\ref{ZWZ}), we have
\begin{align}
	\begin{array}{l}  
		%  2 3 3 2
		\mathbb{E}\Bigg\{  
		\Big(\mathbf{g}_{m_{1},n_{1},k}^{2}\Big)^{H}\mathbf{g}_{m_{1},n_{2},i}^{3}
		\Big(\mathbf{g}_{m_{2},n_{3},i}^{3}\Big)^{H}
		\mathbf{g}_{m_{2},n_{4},k}^{2}
		\Bigg\} \\
		\overset{\substack{(g)}}{=} \delta_{m_{1},n_{1}}\varepsilon_{n_{2},i}
		{\mathbb{E}\Big\{\tilde{\mathbf{h}}_{n_{1},k}^{H}{\bf\Phi}^{H}_{n_{1}} \overline{\mathbf{Z}}_{m_{1},n_{1}}^{H} \mathbb{E}\Big\{\tilde{\mathbf{Z}}_{m_{1},n_{2}} {\bf\Phi}_{n_{2}} \overline{\mathbf{h}}_{n_{2},i}\overline{\mathbf{h}}_{n_{2},i}^{H}{\bf\Phi}^{H}_{n_{2}}  \tilde{\mathbf{Z}}_{m_{1},n_{2}}^{H}\Big\}  \overline{\mathbf{Z}}_{m_{1},n_{1}} {\bf\Phi}_{n_{1}}\tilde{\mathbf{h}}_{n_{1},k}\Big\}}   \\
		=\delta_{m_{1},n_{1}}\varepsilon_{n_{2},i} 
		{\mathbb{E}\Big\{\tilde{\mathbf{h}}_{n_{1},k}^{H}{\bf\Phi}^{H}_{n_{1}} \overline{\mathbf{Z}}_{m_{1},n_{1}}^{H}
			\mathrm{Tr}\Big\{ {\bf\Phi}_{n_{2}} \overline{\mathbf{h}}_{n_{2},i}\overline{\mathbf{h}}_{n_{2},i}^{H}{\bf\Phi}^{H}_{n_{2}} \Big\}
			\overline{\mathbf{Z}}_{m_{1},n_{1}} {\bf\Phi}_{n_{1}}\tilde{\mathbf{h}}_{n_{1},k}\Big\}}   \\
		=\delta_{m_{1},n_{1}}\varepsilon_{n_{2},i}N_{r}
		{\mathrm{Tr}\Big\{ {\bf\Phi}^{H}_{n_{1}} \overline{\mathbf{Z}}_{m_{1},n_{1}}^{H}
			\overline{\mathbf{Z}}_{m_{1},n_{1}} {\bf\Phi}_{n_{1}}\Big\}}   
		\\
		=\delta_{m_{1},n_{1}}\varepsilon_{n_{2},i}N_{r}
		{\mathrm{Tr}\Big\{  \overline{\mathbf{Z}}_{m_{1},n_{1}}^{H}
			\overline{\mathbf{Z}}_{m_{1},n_{1}} \Big\}}   
		\\
		=\delta_{m_{1},n_{1}}\varepsilon_{n_{2},i}M_{b}N_{r}^{2},
	\end{array}
\end{align}
where $(g)$ represents $m_{1}=m_{2}$, $n_{1}=n_{4}$, $n_{2}=n_{3}$ and removes the zero terms.

When $\psi=4$, we arrive at
\begin{align}
	\begin{array}{l}
		%   2  4  4  2 
		\mathbb{E}\Bigg\{  
		\Big(\mathbf{g}_{m_{1},n_{1},k}^{2}\Big)^{H}\mathbf{g}_{m_{1},n_{2},i}^{4}
		\Big(\mathbf{g}_{m_{2},n_{3},i}^{4}\Big)^{H}
		\mathbf{g}_{m_{2},n_{4},k}^{2}
		\Bigg\} \\
		\overset{\substack{(g)}}{=} \delta_{m_{1},n_{1}} 
		{\mathbb{E}\Big\{\tilde{\mathbf{h}}_{n_{1},k}^{H}{\bf\Phi}^{H}_{n_{1}} \overline{\mathbf{Z}}_{m_{1},n_{1}}^{H} \mathbb{E}\Big\{\tilde{\mathbf{Z}}_{m_{1},n_{2}} {\bf\Phi}_{n_{2}} \mathbb{E}\Big\{\tilde{\mathbf{h}}_{n_{2},i} \tilde{\mathbf{h}}_{n_{2},i}^{H}\Big\} {\bf\Phi}^{H}_{n_{2}}  \tilde{\mathbf{Z}}_{m_{1},n_{2}}^{H}\Big\}  \overline{\mathbf{Z}}_{m_{1},n_{1}} {\bf\Phi}_{n_{1}}\tilde{\mathbf{h}}_{n_{1},k}\Big\}}   \\
		=\delta_{m_{1},n_{1}}N_{r} 
		{\mathbb{E}\Big\{\tilde{\mathbf{h}}_{n_{1},k}^{H}{\bf\Phi}^{H}_{n_{1}} \overline{\mathbf{Z}}_{m_{1},n_{1}}^{H}  \overline{\mathbf{Z}}_{m_{1},n_{1}} {\bf\Phi}_{n_{1}}\tilde{\mathbf{h}}_{n_{1},k}\Big\}}   \\
		=\delta_{m_{1},n_{1}}N_{r} 
		{\mathrm{Tr}\Big\{ {\bf\Phi}^{H}_{n_{1}} \overline{\mathbf{Z}}_{m_{1},n_{1}}^{H}
			\overline{\mathbf{Z}}_{m_{1},n_{1}} {\bf\Phi}_{n_{1}}\Big\}}   \\
		=\delta_{m_{1},n_{1}}M_{b}N_{r}^{2} .
	\end{array}
\end{align}

Thirdly, we consider the terms with $\omega= 3$. When $\psi= 1$, based on (\ref{ZWZ}), we have
\begin{align}
	\begin{array}{l}
		%   3  1  1  3 
		\mathbb{E}\Bigg\{  
		\Big(\mathbf{g}_{m_{1},n_{1},k}^{3}\Big)^{H}\mathbf{g}_{m_{1},n_{2},i}^{1}
		\Big(\mathbf{g}_{m_{2},n_{3},i}^{1}\Big)^{H}
		\mathbf{g}_{m_{2},n_{4},k}^{3}
		\Bigg\} \\
		\overset{\substack{(h)}}{=}\varepsilon_{n_{1},k}
		\sqrt{\delta_{m_{1},n_{2}}\delta_{m_{1},n_{3}} \varepsilon_{n_{2},i}\varepsilon_{n_{3},i}}
		\\
		\qquad{\overline{\mathbf{h}}_{n_{1},k}^{H}{\bf\Phi}^{H}_{n_{1}} \mathbb{E}\Big\{\tilde{\mathbf{Z}}_{m_{1},n_{1}}^{H} \overline{\mathbf{Z}}_{m_{1},n_{2}} {\bf\Phi}_{n_{2}} \overline{\mathbf{h}}_{n_{2},i} \overline{\mathbf{h}}_{n_{3},i}^{H} {\bf\Phi}^{H}_{n_{3}}  \overline{\mathbf{Z}}_{m_{1},n_{3}}^{H} \tilde{\mathbf{Z}}_{m_{1},n_{1}}\Big\}  {\bf\Phi}_{n_{1}}\overline{\mathbf{h}}_{n_{1},k}}\\
		=\varepsilon_{n_{1},k}
		\sqrt{\delta_{m_{1},n_{2}}\delta_{m_{1},n_{3}} \varepsilon_{n_{2},i}\varepsilon_{n_{3},i}}
		\;{\overline{\mathbf{h}}_{n_{1},k}^{H}{\bf\Phi}^{H}_{n_{1}}{\bf\Phi}_{n_{1}}\overline{\mathbf{h}}_{n_{1},k} \mathrm{Tr}\Big\{\overline{\mathbf{Z}}_{m_{1},n_{2}} {\bf\Phi}_{n_{2}} \overline{\mathbf{h}}_{n_{2},i} \overline{\mathbf{h}}_{n_{3},i}^{H} {\bf\Phi}^{H}_{n_{3}}  \overline{\mathbf{Z}}_{m_{1},n_{3}}^{H}\Big\} }
		\\
		=\varepsilon_{n_{1},k} 
		\sqrt{\delta_{m_{1},n_{2}}\delta_{m_{1},n_{3}} \varepsilon_{n_{2},i}\varepsilon_{n_{3},i}}N_{r}
		{\mathrm{Tr}\Big\{\overline{\mathbf{Z}}_{m_{1},n_{2}} {\bf\Phi}_{n_{2}} \overline{\mathbf{h}}_{n_{2},i} \overline{\mathbf{h}}_{n_{3},i}^{H} {\bf\Phi}^{H}_{n_{3}}  \overline{\mathbf{Z}}_{m_{1},n_{3}}^{H}\Big\} }
		\\
		=\varepsilon_{n_{1},k}
		\sqrt{\delta_{m_{1},n_{2}}\delta_{m_{1},n_{3}} \varepsilon_{n_{2},i}\varepsilon_{n_{3},i}}N_{r} 
		\;{ \overline{\mathbf{h}}_{n_{3},i}^{H} {\bf\Phi}^{H}_{n_{3}}  \overline{\mathbf{Z}}_{m_{1},n_{3}}^{H} \overline{\mathbf{Z}}_{m_{1},n_{2}} {\bf\Phi}_{n_{2}} \overline{\mathbf{h}}_{n_{2},i} }
		\\
\end{array}
\end{align}
where $(h)$ represents $m_{1}=m_{2}$, $n_{1}=n_{4}$ and removes the zero terms.

When $\psi= 2$, we have
\begin{align}
	\begin{array}{l}
		%   3  2  2  3 
		\mathbb{E}\Bigg\{  
		\Big(\mathbf{g}_{m_{1},n_{1},k}^{3}\Big)^{H}\mathbf{g}_{m_{1},n_{2},i}^{2}
		\Big(\mathbf{g}_{m_{2},n_{3},i}^{2}\Big)^{H}
		\mathbf{g}_{m_{2},n_{4},k}^{3}
		\Bigg\} \\
		\overset{\substack{(g)}}{=}
		\delta_{m_{1},n_{2}}\varepsilon_{n_{1},k}
		{\overline{\mathbf{h}}_{n_{1},k}^{H}{\bf\Phi}^{H}_{n_{1}} \mathbb{E}\Big\{\tilde{\mathbf{Z}}_{m_{1},n_{1}}^{H} \overline{\mathbf{Z}}_{m_{1},n_{2}} {\bf\Phi}_{n_{2}} \mathbb{E}\Big\{\tilde{\mathbf{h}}_{n_{2},i} \tilde{\mathbf{h}}_{n_{2},i}^{H}\Big\} {\bf\Phi}^{H}_{n_{2}}  \overline{\mathbf{Z}}_{m_{1},n_{2}}^{H} \tilde{\mathbf{Z}}_{m_{1},n_{1}}\Big\}  {\bf\Phi}_{n_{1}}\overline{\mathbf{h}}_{n_{1},k}}\\
		=\delta_{m_{1},n_{2}}\varepsilon_{n_{1},k}
		{\overline{\mathbf{h}}_{n_{1},k}^{H}{\bf\Phi}^{H}_{n_{1}} \mathbb{E}\Big\{\tilde{\mathbf{Z}}_{m_{1},n_{1}}^{H} \overline{\mathbf{Z}}_{m_{1},n_{2}}  \overline{\mathbf{Z}}_{m_{1},n_{2}}^{H} \tilde{\mathbf{Z}}_{m_{1},n_{1}}\Big\}  {\bf\Phi}_{n_{1}}\overline{\mathbf{h}}_{n_{1},k}}\\
		=\delta_{m_{1},n_{2}}\varepsilon_{n_{1},k}
		{\overline{\mathbf{h}}_{n_{1},k}^{H}{\bf\Phi}^{H}_{n_{1}}{\bf\Phi}_{n_{1}}\overline{\mathbf{h}}_{n_{1},k}}\mathrm{Tr}\Big\{ \overline{\mathbf{Z}}_{m_{1},n_{2}}  \overline{\mathbf{Z}}_{m_{1},n_{2}}^{H}\Big\} 
		\\
		=\delta_{m_{1},n_{2}}\varepsilon_{n_{1},k}M_{b} N_{r}^{2}.
	\end{array}
\end{align}

When $\psi= 3$, the term is the addition of the following two parts
\\
part 1: 
\begin{align}
	\begin{array}{l}
		%   3  3  3  3 (1)
		\Bigg(\mathbb{E}\Bigg\{  
		\Big(\mathbf{g}_{m_{1},n_{1},k}^{3}\Big)^{H}\mathbf{g}_{m_{1},n_{2},i}^{3}
		\Big(\mathbf{g}_{m_{2},n_{3},i}^{3}\Big)^{H}
		\mathbf{g}_{m_{2},n_{4},k}^{3}
		\Bigg\}\Bigg)_{1} \\
		\overset{\substack{(g)}}{=}
		\varepsilon_{n_{1},k}\varepsilon_{n_{2},i}
		{\overline{\mathbf{h}}_{n_{1},k}^{H}{\bf\Phi}^{H}_{n_{1}} \mathbb{E}\Big\{\tilde{\mathbf{Z}}_{m_{1},n_{1}}^{H} \mathbb{E}\Big\{\tilde{\mathbf{Z}}_{m_{1},n_{2}} {\bf\Phi}_{n_{2}} \overline{\mathbf{h}}_{n_{2},i} \overline{\mathbf{h}}_{n_{2},i}^{H} {\bf\Phi}^{H}_{n_{2}}  \tilde{\mathbf{Z}}_{m_{1},n_{2}}^{H}\Big\}  \tilde{\mathbf{Z}}_{m_{1},n_{1}}\Big\}  {\bf\Phi}_{n_{1}}\overline{\mathbf{h}}_{n_{1},k}}\\
		=\varepsilon_{n_{1},k}\varepsilon_{n_{2},i}
		{\overline{\mathbf{h}}_{n_{1},k}^{H}{\bf\Phi}^{H}_{n_{1}} \mathbb{E}\Big\{\tilde{\mathbf{Z}}_{m_{1},n_{1}}^{H}  \tilde{\mathbf{Z}}_{m_{1},n_{1}}\Big\}  {\bf\Phi}_{n_{1}}\overline{\mathbf{h}}_{n_{1},k} \mathrm{Tr}\Big\{ {\bf\Phi}_{n_{2}} \overline{\mathbf{h}}_{n_{2},i} \overline{\mathbf{h}}_{n_{2},i}^{H} {\bf\Phi}^{H}_{n_{2}}  \Big\} }
		\\
		%	=\varepsilon_{n_{1},k}\varepsilon_{n_{2},i}N_{r}
		%	{\overline{\mathbf{h}}_{n_{1},k}^{H}{\bf\Phi}^{H}_{n_{1}} \mathbb{E}\Big\{\tilde{\mathbf{Z}}_{m_{1},n_{1}}^{H}  \tilde{\mathbf{Z}}_{m_{1},n_{1}}\Big\}  {\bf\Phi}_{n_{1}}\overline{\mathbf{h}}_{n_{1},k}}\\
		=\varepsilon_{n_{1},k}\varepsilon_{n_{2},i}M_{b} N_{r}^{2},
	\end{array}
\end{align}
part 2: 
\begin{align}
	\begin{array}{l}  %   3  3  3  3 (2)
		\Bigg(\mathbb{E}\Bigg\{  
		\Big(\mathbf{g}_{m_{1},n_{1},k}^{3}\Big)^{H}\mathbf{g}_{m_{1},n_{2},i}^{3}
		\Big(\mathbf{g}_{m_{2},n_{3},i}^{3}\Big)^{H}
		\mathbf{g}_{m_{2},n_{4},k}^{3}
		\Bigg\}	\Bigg)_{2} \\
		\overset{\substack{(i)}}{=}
		\sqrt{\varepsilon_{n_{1},k}\varepsilon_{n_{1},i} \varepsilon_{n_{3},i}\varepsilon_{n_{3},k}}
		{\overline{\mathbf{h}}_{n_{1},k}^{H}{\bf\Phi}^{H}_{n_{1}} \mathbb{E}\Big\{\tilde{\mathbf{Z}}_{m_{1},n_{1}}^{H} \tilde{\mathbf{Z}}_{m_{1},n_{1}}\Big\} {\bf\Phi}_{n_{1}} \overline{\mathbf{h}}_{n_{1},i} \overline{\mathbf{h}}_{n_{3},i}^{H} {\bf\Phi}^{H}_{n_{3}}  \mathbb{E}\Big\{\tilde{\mathbf{Z}}_{m_{2},n_{3}}^{H}  \tilde{\mathbf{Z}}_{m_{2},n_{3}}\Big\}  {\bf\Phi}_{n_{3}}\overline{\mathbf{h}}_{n_{3},k}}\\
		= \sqrt{\varepsilon_{n_{1},k}\varepsilon_{n_{1},i}\varepsilon_{n_{3},i}\varepsilon_{n_{3},k}}M_{b}^{2}
		{\overline{\mathbf{h}}_{n_{1},k}^{H}{\bf\Phi}^{H}_{n_{1}}  {\bf\Phi}_{n_{1}} \overline{\mathbf{h}}_{n_{1},i} \overline{\mathbf{h}}_{n_{3},i}^{H} {\bf\Phi}^{H}_{n_{3}}   {\bf\Phi}_{n_{3}}\overline{\mathbf{h}}_{n_{3},k}}\\
		= \sqrt{\varepsilon_{n_{1},k}\varepsilon_{n_{1},i}\varepsilon_{n_{3},i}\varepsilon_{n_{3},k}}M_{b}^{2}
		{\overline{\mathbf{h}}_{n_{1},k}^{H} \overline{\mathbf{h}}_{n_{1},i} \overline{\mathbf{h}}_{n_{3},i}^{H} \overline{\mathbf{h}}_{n_{3},k}},
	\end{array}
\end{align}
where $(i)$ represents $n_{1}=n_{2}$, $n_{3}=n_{4}$ and removes the zero terms.  

When $\psi= 4$, the term is the addition of the following two parts
\\
part 1:
\begin{align}
	\begin{array}{l}
		%   3  4  4  3 (1)
		\Bigg(\mathbb{E}\Bigg\{  
		\Big(\mathbf{g}_{m_{1},n_{1},k}^{3}\Big)^{H}\mathbf{g}_{m_{1},n_{2},i}^{4}
		\Big(\mathbf{g}_{m_{2},n_{3},i}^{4}\Big)^{H}
		\mathbf{g}_{m_{2},n_{4},k}^{3}
		\Bigg\} \Bigg)_{1}\\
		\overset{\substack{(g)}}{=}
		\varepsilon_{n_{1},k}
		{\overline{\mathbf{h}}_{n_{1},k}^{H}{\bf\Phi}^{H}_{n_{1}} \mathbb{E}\Big\{\tilde{\mathbf{Z}}_{m_{1},n_{1}}^{H} \mathbb{E}\Big\{\tilde{\mathbf{Z}}_{m_{1},n_{2}} {\bf\Phi}_{n_{2}} \mathbb{E}\Big\{\tilde{\mathbf{h}}_{n_{2},i} \tilde{\mathbf{h}}_{n_{2},i}^{H}\Big\} {\bf\Phi}^{H}_{n_{2}}  \tilde{\mathbf{Z}}_{m_{1},n_{2}}^{H}\Big\}  \tilde{\mathbf{Z}}_{m_{1},n_{1}}\Big\}  {\bf\Phi}_{n_{1}}\overline{\mathbf{h}}_{n_{1},k}}\\
		=\varepsilon_{n_{1},k}
		{\overline{\mathbf{h}}_{n_{1},k}^{H}{\bf\Phi}^{H}_{n_{1}} \mathbb{E}\Big\{\tilde{\mathbf{Z}}_{m_{1},n_{1}}^{H} \mathbb{E}\Big\{\tilde{\mathbf{Z}}_{m_{1},n_{2}} \tilde{\mathbf{Z}}_{m_{1},n_{2}}^{H}\Big\}  \tilde{\mathbf{Z}}_{m_{1},n_{1}}\Big\}  {\bf\Phi}_{n_{1}}\overline{\mathbf{h}}_{n_{1},k}}\\
		=\varepsilon_{n_{1},k}N_{r} 
		{\overline{\mathbf{h}}_{n_{1},k}^{H}{\bf\Phi}^{H}_{n_{1}} \mathbb{E}\Big\{\tilde{\mathbf{Z}}_{m_{1},n_{1}}^{H}  \tilde{\mathbf{Z}}_{m_{1},n_{1}}\Big\}  {\bf\Phi}_{n_{1}}\overline{\mathbf{h}}_{n_{1},k}}\\
		=\varepsilon_{n_{1},k}M_{b} N_{r}^{2} ,
	\end{array}
\end{align}
part 2:
\begin{align}
	\begin{array}{l}
		%   3  4  4  3 (2)
		\Bigg(\mathbb{E}\Bigg\{  
		\Big(\mathbf{g}_{m_{1},n_{1},k}^{3}\Big)^{H}\mathbf{g}_{m_{1},n_{2},i}^{4}
		\Big(\mathbf{g}_{m_{2},n_{3},i}^{4}\Big)^{H}
		\mathbf{g}_{m_{2},n_{4},k}^{3}
		\Bigg\} \Bigg)_{2}\\
		\overset{\substack{(j)}}{=}
		\varepsilon_{n_{1},k}
		{\overline{\mathbf{h}}_{n_{1},k}^{H}{\bf\Phi}^{H}_{n_{1}} \mathbb{E}\Big\{\tilde{\mathbf{Z}}_{m_{1},n_{1}}^{H} \tilde{\mathbf{Z}}_{m_{1},n_{1}}\Big\} {\bf\Phi}_{n_{1}} \mathbb{E}\Big\{\tilde{\mathbf{h}}_{n_{1},i} \tilde{\mathbf{h}}_{n_{1},i}^{H}\Big\} {\bf\Phi}^{H}_{n_{1}}  \mathbb{E}\Big\{\tilde{\mathbf{Z}}_{m_{2},n_{1}}^{H}\tilde{\mathbf{Z}}_{m_{2},n_{1}}\Big\}  {\bf\Phi}_{n_{1}}\overline{\mathbf{h}}_{n_{1},k}}\\
		=\varepsilon_{n_{1},k}M_{b}^{2} 
		{\overline{\mathbf{h}}_{n_{1},k}^{H}{\bf\Phi}^{H}_{n_{1}}  {\bf\Phi}_{n_{1}}  {\bf\Phi}^{H}_{n_{1}}   {\bf\Phi}_{n_{1}}\overline{\mathbf{h}}_{n_{1},k}}\\
		=\varepsilon_{n_{1},k}M_{b}^{2} N_{r} ,
	\end{array}
\end{align}
where $(j)$ represents $n_{1}=n_{2}=n_{3}=n_{4}$ and removes the zero terms.

Fourthly, we consider the terms with $\omega= 4$. When $\psi= 1$, we have
\begin{align}
	\begin{array}{l}
		%   4  1  1  4 
		\mathbb{E}\Bigg\{  
		\Big(\mathbf{g}_{m_{1},n_{1},k}^{4}\Big)^{H}\mathbf{g}_{m_{1},n_{2},i}^{1}
		\Big(\mathbf{g}_{m_{2},n_{3},i}^{1}\Big)^{H}
		\mathbf{g}_{m_{2},n_{4},k}^{4}
		\Bigg\} \\
		\overset{\substack{(h)}}{=}
		\sqrt{\delta_{m_{1},n_{2}}\delta_{m_{1},n_{3}} \varepsilon_{n_{2},i}\varepsilon_{n_{3},i}}
		\\
		\qquad{\mathbb{E}\Big\{\tilde{\mathbf{h}}_{n_{1},k}^{H}{\bf\Phi}^{H}_{n_{1}} \mathbb{E}\Big\{\tilde{\mathbf{Z}}_{m_{1},n_{1}}^{H} \overline{\mathbf{Z}}_{m_{1},n_{2}} {\bf\Phi}_{n_{2}} \overline{\mathbf{h}}_{n_{2},i} \overline{\mathbf{h}}_{n_{3},i}^{H}{\bf\Phi}^{H}_{n_{3}}  \overline{\mathbf{Z}}_{m_{1},n_{3}}^{H}\tilde{\mathbf{Z}}_{m_{1},n_{1}}\Big\}  {\bf\Phi}_{n_{1}}\tilde{\mathbf{h}}_{n_{1},k}\Big\} }
		\\
		=\sqrt{\delta_{m_{1},n_{2}}\delta_{m_{1},n_{3}} \varepsilon_{n_{2},i}\varepsilon_{n_{3},i}}
		{\mathbb{E}\Big\{\tilde{\mathbf{h}}_{n_{1},k}^{H}{\bf\Phi}^{H}_{n_{1}}  {\bf\Phi}_{n_{1}}\tilde{\mathbf{h}}_{n_{1},k}\Big\} \mathrm{Tr}\Big\{\overline{\mathbf{Z}}_{m_{1},n_{2}} {\bf\Phi}_{n_{2}} \overline{\mathbf{h}}_{n_{2},i} \overline{\mathbf{h}}_{n_{3},i}^{H}{\bf\Phi}^{H}_{n_{3}}  \overline{\mathbf{Z}}_{m_{1},n_{3}}^{H}\Big\} }
		\\
		=\sqrt{\delta_{m_{1},n_{2}}\delta_{m_{1},n_{3}} \varepsilon_{n_{2},i}\varepsilon_{n_{3},i}}N_{r}
		{f_{m_{1},n_{3},i}^{H}({\bf \Phi}) f_{m_{1},n_{2},i}({\bf \Phi})  
			{\bf a}^{H}_{M_{b}}(m_{1},n_{3}) 
			{\bf a}_{M_{b}}(m_{1},n_{2})  }.
	\end{array}
\end{align}

When $\psi= 2$, we have
\begin{align}
	\begin{array}{l}
		%   4  2  2  4 
		\mathbb{E}\Bigg\{  
		\Big(\mathbf{g}_{m_{1},n_{1},k}^{4}\Big)^{H}\mathbf{g}_{m_{1},n_{2},i}^{2}
		\Big(\mathbf{g}_{m_{2},n_{3},i}^{2}\Big)^{H}
		\mathbf{g}_{m_{2},n_{4},k}^{4}
		\Bigg\} \\
		\overset{\substack{(g)}}{=}
		\delta_{m_{1},n_{2}}
		{\mathbb{E}\Big\{\tilde{\mathbf{h}}_{n_{1},k}^{H}{\bf\Phi}^{H}_{n_{1}} \mathbb{E}\Big\{\tilde{\mathbf{Z}}_{m_{1},n_{1}}^{H} \overline{\mathbf{Z}}_{m_{1},n_{2}} {\bf\Phi}_{n_{2}} \mathbb{E}\Big\{\tilde{\mathbf{h}}_{n_{2},i} \tilde{\mathbf{h}}_{n_{2},i}^{H}\Big\}{\bf\Phi}^{H}_{n_{2}}  \overline{\mathbf{Z}}_{m_{1},n_{2}}^{H}\tilde{\mathbf{Z}}_{m_{1},n_{1}}\Big\}  {\bf\Phi}_{n_{1}}\tilde{\mathbf{h}}_{n_{1},k}\Big\} }
		\\
		=\delta_{m_{1},n_{2}}
		{\mathbb{E}\Big\{\tilde{\mathbf{h}}_{n_{1},k}^{H}{\bf\Phi}^{H}_{n_{1}} \mathbb{E}\Big\{\tilde{\mathbf{Z}}_{m_{1},n_{1}}^{H} \overline{\mathbf{Z}}_{m_{1},n_{2}}  \overline{\mathbf{Z}}_{m_{1},n_{2}}^{H}\tilde{\mathbf{Z}}_{m_{1},n_{1}}\Big\}  {\bf\Phi}_{n_{1}}\tilde{\mathbf{h}}_{n_{1},k}\Big\} }
		\\
		=\delta_{m_{1},n_{2}}
		{\mathbb{E}\Big\{\tilde{\mathbf{h}}_{n_{1},k}^{H}{\bf\Phi}^{H}_{n_{1}}   {\bf\Phi}_{n_{1}}\tilde{\mathbf{h}}_{n_{1},k}\Big\}\mathrm{Tr}\Big\{\overline{\mathbf{Z}}_{m_{1},n_{2}}\overline{\mathbf{Z}}_{m_{1},n_{2}}^{H}\Big\} }
		\\
		=\delta_{m_{1},n_{2}}M_{b}N_{r}^{2}.
	\end{array}
\end{align}

When $\psi= 3$, the term is the addition of the following two parts
\\part 1:
\begin{align}
	\begin{array}{l}
		%   4  3  3  4  (1)
		\Bigg(\mathbb{E}\Bigg\{  
		\Big(\mathbf{g}_{m_{1},n_{1},k}^{4}\Big)^{H}\mathbf{g}_{m_{1},n_{2},i}^{3}
		\Big(\mathbf{g}_{m_{2},n_{3},i}^{3}\Big)^{H}
		\mathbf{g}_{m_{2},n_{4},k}^{4}
		\Bigg\} \Bigg)_{1}\\
		\overset{\substack{(g)}}{=}
		\varepsilon_{n_{2},i}
		{\mathbb{E}\Big\{\tilde{\mathbf{h}}_{n_{1},k}^{H}{\bf\Phi}^{H}_{n_{1}} \mathbb{E}\Big\{\tilde{\mathbf{Z}}_{m_{1},n_{1}}^{H} \mathbb{E}\Big\{\tilde{\mathbf{Z}}_{m_{1},n_{2}} {\bf\Phi}_{n_{2}} \overline{\mathbf{h}}_{n_{2},i} \overline{\mathbf{h}}_{n_{2},i}^{H}{\bf\Phi}^{H}_{n_{2}}  \tilde{\mathbf{Z}}_{m_{1},n_{2}}^{H}\Big\}\tilde{\mathbf{Z}}_{m_{1},n_{1}}\Big\}  {\bf\Phi}_{n_{1}}\tilde{\mathbf{h}}_{n_{1},k}\Big\} }
		\\
		=\varepsilon_{n_{2},i}
		{\mathbb{E}\Big\{\tilde{\mathbf{h}}_{n_{1},k}^{H}{\bf\Phi}^{H}_{n_{1}} \mathbb{E}\Big\{\tilde{\mathbf{Z}}_{m_{1},n_{1}}^{H} \tilde{\mathbf{Z}}_{m_{1},n_{1}}\Big\}  {\bf\Phi}_{n_{1}}\tilde{\mathbf{h}}_{n_{1},k}\Big\}\mathrm{Tr}\Big\{{\bf\Phi}_{n_{2}} \overline{\mathbf{h}}_{n_{2},i} \overline{\mathbf{h}}_{n_{2},i}^{H}{\bf\Phi}^{H}_{n_{2}}\Big\} }
		\\
		=\varepsilon_{n_{2},i}M_{b}N_{r}^{2},
	\end{array}
\end{align}
part 2:
\begin{align}
	\begin{array}{l}
		%   4  3  3  4  (2)
		\Bigg(\mathbb{E}\Bigg\{  
		\Big(\mathbf{g}_{m_{1},n_{1},k}^{4}\Big)^{H}\mathbf{g}_{m_{1},n_{2},i}^{3}
		\Big(\mathbf{g}_{m_{2},n_{3},i}^{3}\Big)^{H}
		\mathbf{g}_{m_{2},n_{4},k}^{4}
		\Bigg\} \Bigg)_{2}\\
		\overset{\substack{(j)}}{=}
		\varepsilon_{n_{1},i}
		{\mathbb{E}\Big\{\tilde{\mathbf{h}}_{n_{1},k}^{H}{\bf\Phi}^{H}_{n_{1}} \mathbb{E}\Big\{\tilde{\mathbf{Z}}_{m_{1},n_{1}}^{H} \tilde{\mathbf{Z}}_{m_{1},n_{1}}\Big\} {\bf\Phi}_{n_{1}} \overline{\mathbf{h}}_{n_{1},i} \overline{\mathbf{h}}_{n_{1},i}^{H}{\bf\Phi}^{H}_{n_{1}}  \mathbb{E}\Big\{\tilde{\mathbf{Z}}_{m_{2},n_{1}}^{H}\tilde{\mathbf{Z}}_{m_{2},n_{1}}\Big\}  {\bf\Phi}_{n_{1}}\tilde{\mathbf{h}}_{n_{1},k}\Big\} }
		\\
		=\varepsilon_{n_{1},i}M_{b}^{2} 
		{\mathbb{E}\Big\{\tilde{\mathbf{h}}_{n_{1},k}^{H}{\bf\Phi}^{H}_{n_{1}}  {\bf\Phi}_{n_{1}} \overline{\mathbf{h}}_{n_{1},i} \overline{\mathbf{h}}_{n_{1},i}^{H}{\bf\Phi}^{H}_{n_{1}}   {\bf\Phi}_{n_{1}}\tilde{\mathbf{h}}_{n_{1},k}\Big\} }
		\\
		=\varepsilon_{n_{1},i}M_{b}^{2} 
		{\overline{\mathbf{h}}_{n_{1},i}^{H}   \mathbb{E}\Big\{\tilde{\mathbf{h}}_{n_{1},k} \tilde{\mathbf{h}}_{n_{1},k}^{H}\Big\} \overline{\mathbf{h}}_{n_{1},i} }
		\\
		=\varepsilon_{n_{1},i}M_{b}^{2} N_{r} .
	\end{array}
\end{align}

When $\psi= 4$, the term is the addition of the following two parts
\\part 1:
\begin{align}
	\begin{array}{l}
		%   4  4  4  4  (1)  
		\Bigg(\mathbb{E}\Bigg\{  
		\Big(\mathbf{g}_{m_{1},n_{1},k}^{4}\Big)^{H}\mathbf{g}_{m_{1},n_{2},i}^{4}
		\Big(\mathbf{g}_{m_{2},n_{3},i}^{4}\Big)^{H}
		\mathbf{g}_{m_{2},n_{4},k}^{4}
		\Bigg\} \Bigg)_{1}\\
		\overset{\substack{(g)}}{=}
		{\mathbb{E}\Big\{\tilde{\mathbf{h}}_{n_{1},k}^{H}{\bf\Phi}^{H}_{n_{1}} \mathbb{E}\Big\{\tilde{\mathbf{Z}}_{m_{1},n_{1}}^{H}\mathbb{E}\Big\{\tilde{\mathbf{Z}}_{m_{1},n_{2}} {\bf\Phi}_{n_{2}} \mathbb{E}\Big\{\tilde{\mathbf{h}}_{n_{2},i} \tilde{\mathbf{h}}_{n_{2},i}^{H}\Big\} {\bf\Phi}^{H}_{n_{2}}  \tilde{\mathbf{Z}}_{m_{1},n_{2}}^{H}\Big\} \tilde{\mathbf{Z}}_{m_{1},n_{1}}\Big\}  {\bf\Phi}_{n_{1}}\tilde{\mathbf{h}}_{n_{1},k}\Big\} }
		\\
		={\mathbb{E}\Big\{\tilde{\mathbf{h}}_{n_{1},k}^{H}{\bf\Phi}^{H}_{n_{1}} \mathbb{E}\Big\{\tilde{\mathbf{Z}}_{m_{1},n_{1}}^{H}\mathbb{E}\Big\{\tilde{\mathbf{Z}}_{m_{1},n_{2}}   \tilde{\mathbf{Z}}_{m_{1},n_{2}}^{H}\Big\} \tilde{\mathbf{Z}}_{m_{1},n_{1}}\Big\}  {\bf\Phi}_{n_{1}}\tilde{\mathbf{h}}_{n_{1},k}\Big\} }
		\\
		=M_{b} N_{r} {\mathbb{E}\Big\{\tilde{\mathbf{h}}_{n_{1},k}^{H}{\bf\Phi}^{H}_{n_{1}}  {\bf\Phi}_{n_{1}}\tilde{\mathbf{h}}_{n_{1},k}\Big\} }
		\\
		=M_{b} N_{r}^{2},
	\end{array}
\end{align}
part 2:
\begin{align}
	\begin{array}{l}
		%   4  4  4  4  (2)
		\Bigg(\mathbb{E}\Bigg\{  
		\Big(\mathbf{g}_{m_{1},n_{1},k}^{4}\Big)^{H}\mathbf{g}_{m_{1},n_{2},i}^{4}
		\Big(\mathbf{g}_{m_{2},n_{3},i}^{4}\Big)^{H}
		\mathbf{g}_{m_{2},n_{4},k}^{4}
		\Bigg\} \Bigg)_{2}\\
		\overset{\substack{(j)}}{=}
		{\mathbb{E}\Big\{\tilde{\mathbf{h}}_{n_{1},k}^{H}{\bf\Phi}^{H}_{n_{1}} \mathbb{E}\Big\{\tilde{\mathbf{Z}}_{m_{1},n_{1}}^{H} \tilde{\mathbf{Z}}_{m_{1},n_{1}}\Big\} {\bf\Phi}_{n_{1}} \mathbb{E}\Big\{\tilde{\mathbf{h}}_{n_{1},i} \tilde{\mathbf{h}}_{n_{1},i}^{H}\Big\} {\bf\Phi}^{H}_{n_{1}}  \mathbb{E}\Big\{\tilde{\mathbf{Z}}_{m_{2},n_{1}}^{H}\tilde{\mathbf{Z}}_{m_{2},n_{1}}\Big\}  {\bf\Phi}_{n_{1}}\tilde{\mathbf{h}}_{n_{1},k}\Big\} }
		\\
		=M_{b}^{2} {\mathbb{E}\Big\{\tilde{\mathbf{h}}_{n_{1},k}^{H}\mathbb{E}\Big\{\tilde{\mathbf{h}}_{n_{1},i} \tilde{\mathbf{h}}_{n_{1},i}^{H}\Big\} \tilde{\mathbf{h}}_{n_{1},k}\Big\} }
		\\
		=M_{b}^{2} N_{r}.
	\end{array}
\end{align}

Similar to the above derivations, we derive the remaining four parts in (\ref{gkgi_2}) one by one.

The first term is derived as
\begin{align}
	\begin{array}{l}
		%   1  1  3  3  
		2\mathbb{E}\Bigg\{\mathrm{Re}\Bigg\{  
		\Big(\mathbf{g}_{m_{1},n_{1},k}^{1}\Big)^{H}\mathbf{g}_{m_{1},n_{2},i}^{1}
		\Big(\mathbf{g}_{m_{2},n_{3},i}^{3}\Big)^{H}
		\mathbf{g}_{m_{2},n_{4},k}^{3}	\Bigg\}
		\Bigg\} \\
		\overset{\substack{(k)}}{=}
		2{\sqrt{\delta_{m_{1},n_{1}}\delta_{m_{1},n_{2}} \varepsilon_{n_{1},k}\varepsilon_{n_{2},i} \varepsilon_{n_{3},i}\varepsilon_{n_{3},k}}}
		\\ \qquad{\mathrm{Re}\Big\{\overline{\mathbf{h}}_{n_{1},k}^{H}{\bf\Phi}^{H}_{n_{1}} \overline{\mathbf{Z}}_{m_{1},n_{1}}^{H} \overline{\mathbf{Z}}_{m_{1},n_{2}} {\bf\Phi}_{n_{2}}\overline{\mathbf{h}}_{n_{2},i} \overline{\mathbf{h}}_{n_{3},i}^{H} {\bf\Phi}^{H}_{n_{3}}  \mathbb{E}\Big\{\tilde{\mathbf{Z}}_{m_{2},n_{3}}^{H}\tilde{\mathbf{Z}}_{m_{2},n_{3}}\Big\}  {\bf\Phi}_{n_{3}} \overline{\mathbf{h}}_{n_{3},k}\Big\} }
		\\
		=2{\sqrt{\delta_{m_{1},n_{1}}\delta_{m_{1},n_{2}} \varepsilon_{n_{1},k}\varepsilon_{n_{2},i} \varepsilon_{n_{3},i}\varepsilon_{n_{3},k}}}
		{M_{b}\mathrm{Re}\Big\{\overline{\mathbf{h}}_{n_{1},k}^{H}{\bf\Phi}^{H}_{n_{1}} \overline{\mathbf{Z}}_{m_{1},n_{1}}^{H} \overline{\mathbf{Z}}_{m_{1},n_{2}} {\bf\Phi}_{n_{2}}\overline{\mathbf{h}}_{n_{2},i} \overline{\mathbf{h}}_{n_{3},i}^{H}  \overline{\mathbf{h}}_{n_{3},k}\Big\} }
		\\
		=2{\sqrt{\delta_{m_{1},n_{1}}\delta_{m_{1},n_{2}} \varepsilon_{n_{1},k}\varepsilon_{n_{2},i} \varepsilon_{n_{3},i}\varepsilon_{n_{3},k}}}
		M_{b} 
		\\
		\qquad\mathrm{Re}\Big\{ 
		f_{m_{1},n_{1},k}^{H}({\bf \Phi}) f_{m_{1},n_{2},i}({\bf \Phi}) 
		{\bf a}^{H}_{M_{b}}(m_{1},n_{1})	
		{\bf a}_{M_{b}}(m_{1},n_{2}) \overline{\mathbf{h}}_{n_{3},i}^{H}  \overline{\mathbf{h}}_{n_{3},k} \Big\},
	\end{array}
\end{align}
where $(k)$ represents $n_{3}=n_{4}$ and removes the zero terms.

The second term is 
\begin{align}
	\begin{array}{l}
		%   1  2  4  3  
		2\mathbb{E}\Bigg\{\mathrm{Re}\Bigg\{  
		\Big(\mathbf{g}_{m_{1},n_{1},k}^{1}\Big)^{H}\mathbf{g}_{m_{1},n_{2},i}^{2}
		\Big(\mathbf{g}_{m_{2},n_{3},i}^{4}\Big)^{H}
		\mathbf{g}_{m_{2},n_{4},k}^{3}	\Bigg\}
		\Bigg\} \\
		\overset{\substack{(l)}}{=}
		2{\sqrt{\delta_{m_{1},n_{1}}\delta_{m_{1},n_{2}}  \varepsilon_{n_{1},k}\varepsilon_{n_{2},k}}}
		\\ \qquad{\mathrm{Re}\Big\{\overline{\mathbf{h}}_{n_{1},k}^{H}{\bf\Phi}^{H}_{n_{1}} \overline{\mathbf{Z}}_{m_{1},n_{1}}^{H} \overline{\mathbf{Z}}_{m_{1},n_{2}} {\bf\Phi}_{n_{2}}\mathbb{E}\Big\{ \tilde{\mathbf{h}}_{n_{2},i} \tilde{\mathbf{h}}_{n_{2},i}^{H}\Big\}  {\bf\Phi}^{H}_{n_{2}}  \mathbb{E}\Big\{\tilde{\mathbf{Z}}_{m_{2},n_{2}}^{H}\tilde{\mathbf{Z}}_{m_{2},n_{2}}\Big\}  {\bf\Phi}_{n_{2}} \overline{\mathbf{h}}_{n_{2},k}\Big\} }
		\\
		=2{\sqrt{\delta_{m_{1},n_{1}}\delta_{m_{1},n_{2}}\varepsilon_{n_{1},k}\varepsilon_{n_{2},k}}M_{b}}
		{\mathrm{Re}\Big\{\overline{\mathbf{h}}_{n_{1},k}^{H}{\bf\Phi}^{H}_{n_{1}} \overline{\mathbf{Z}}_{m_{1},n_{1}}^{H} \overline{\mathbf{Z}}_{m_{1},n_{2}} {\bf\Phi}_{n_{2}} \overline{\mathbf{h}}_{n_{2},k}\Big\} }
		\\
		=2{\sqrt{\delta_{m_{1},n_{1}}\delta_{m_{1},n_{2}}\varepsilon_{n_{1},k}\varepsilon_{n_{2},k}}M_{b}}
		{\mathrm{Re}\Big\{	
			f_{m_{1},n_{1},k}^{H}({\bf \Phi}) f_{m_{1},n_{2},k}({\bf \Phi}) 
			{\bf a}^{H}_{M_{b}}(m_{1},n_{1})	
			{\bf a}_{M_{b}}(m_{1},n_{2})\Big\} },
	\end{array}
\end{align}
where $(l)$ represents $n_{2}=n_{3}=n_{4}$ and removes the zero terms.

The third term is 
\begin{align}
	\begin{array}{l}
		%   2  1  3  4  
		2\mathbb{E}\Bigg\{\mathrm{Re}\Bigg\{  
		\Big(\mathbf{g}_{m_{1},n_{1},k}^{2}\Big)^{H}\mathbf{g}_{m_{1},n_{2},i}^{1}
		\Big(\mathbf{g}_{m_{2},n_{3},i}^{3}\Big)^{H}
		\mathbf{g}_{m_{2},n_{4},k}^{4
		}	\Bigg\}\Bigg\} 
		\\
		\overset{\substack{(m)}}{=}
		2{\sqrt{\delta_{m_{1},n_{1}}\delta_{m_{1},n_{2}}  \varepsilon_{n_{1},i}\varepsilon_{n_{2},i}}}
		\\ 
		\qquad{\mathbb{E}\Big\{\mathrm{Re}\Big\{ \tilde{\mathbf{h}}_{n_{1},k}^{H}{\bf\Phi}^{H}_{n_{1}} \overline{\mathbf{Z}}_{m_{1},n_{1}}^{H} \overline{\mathbf{Z}}_{m_{1},n_{2}} {\bf\Phi}_{n_{2}} \overline{\mathbf{h}}_{n_{2},i} 
			\overline{\mathbf{h}}_{n_{1},i}^{H}  {\bf\Phi}^{H}_{n_{1}}  \mathbb{E}\Big\{\tilde{\mathbf{Z}}_{m_{2},n_{1}}^{H}\tilde{\mathbf{Z}}_{m_{2},n_{1}}\Big\}  {\bf\Phi}_{n_{1}} \tilde{\mathbf{h}}_{n_{1},k}\Big\}\Big\}  }
		\\
		=2{\sqrt{\delta_{m_{1},n_{1}}\delta_{m_{1},n_{2}}\varepsilon_{n_{1},i}\varepsilon_{n_{2},i}}M_{b}}
		{\mathbb{E}\Big\{\mathrm{Re}\Big\{ \tilde{\mathbf{h}}_{n_{1},k}^{H}{\bf\Phi}^{H}_{n_{1}} \overline{\mathbf{Z}}_{m_{1},n_{1}}^{H} \overline{\mathbf{Z}}_{m_{1},n_{2}} {\bf\Phi}_{n_{2}} \overline{\mathbf{h}}_{n_{2},i} 
			\overline{\mathbf{h}}_{n_{1},i}^{H}   \tilde{\mathbf{h}}_{n_{1},k}\Big\}\Big\}  }
		\\
		=2{\sqrt{\delta_{m_{1},n_{1}}\delta_{m_{1},n_{2}}\varepsilon_{n_{1},i}\varepsilon_{n_{2},i}}M_{b}}
		{\mathrm{Re}\Big\{\mathrm{Tr}\Big\{  \overline{\mathbf{Z}}_{m_{1},n_{2}} {\bf\Phi}_{n_{2}} \overline{\mathbf{h}}_{n_{2},i} 
			\overline{\mathbf{h}}_{n_{1},i}^{H}   \mathbb{E}\Big\{ \tilde{\mathbf{h}}_{n_{1},k} \tilde{\mathbf{h}}_{n_{1},k}^{H}\Big\} {\bf\Phi}^{H}_{n_{1}} \overline{\mathbf{Z}}_{m_{1},n_{1}}^{H}\Big\}\Big\}  }
		\\
\end{array}
\end{align}
where $(m)$ represents $n_{1}=n_{3}=n_{4}$ and removes the zero terms.

The fourth term is 
\begin{align}\label{gkgi2_end}
	\begin{array}{l}
		%   2  2  4  4  
		2\mathbb{E}\Bigg\{\mathrm{Re}\Bigg\{  
		\Big(\mathbf{g}_{m_{1},n_{1},k}^{2}\Big)^{H}\mathbf{g}_{m_{1},n_{2},i}^{2}
		\Big(\mathbf{g}_{m_{2},n_{3},i}^{4}\Big)^{H}
		\mathbf{g}_{m_{2},n_{4},k}^{4
		}	\Bigg\}
		\Bigg\} \\
		\overset{\substack{(j)}}{=}
		2{\delta_{m_{1},n_{1}}} {\mathbb{E}\Big\{\mathrm{Re}\Big\{ \tilde{\mathbf{h}}_{n_{1},k}^{H}{\bf\Phi}^{H}_{n_{1}} \overline{\mathbf{Z}}_{m_{1},n_{1}}^{H} \overline{\mathbf{Z}}_{m_{1},n_{1}} {\bf\Phi}_{n_{1}}   \mathbb{E}\Big\{\tilde{\mathbf{h}}_{n_{1},i} 
			\tilde{\mathbf{h}}_{n_{1},i}^{H} \Big\} {\bf\Phi}^{H}_{n_{1}}  \mathbb{E}\Big\{\tilde{\mathbf{Z}}_{m_{2},n_{1}}^{H}\tilde{\mathbf{Z}}_{m_{2},n_{1}}\Big\}  {\bf\Phi}_{n_{1}} \tilde{\mathbf{h}}_{n_{1},k}\Big\}\Big\}  }
		\\
		=2{\delta_{m_{1},n_{1}}}
		{\mathbb{E}\Big\{\mathrm{Re}\Big\{ \tilde{\mathbf{h}}_{n_{1},k}^{H}{\bf\Phi}^{H}_{n_{1}} \overline{\mathbf{Z}}_{m_{1},n_{1}}^{H} \overline{\mathbf{Z}}_{m_{1},n_{1}}   {\bf\Phi}_{n_{1}} \tilde{\mathbf{h}}_{n_{1},k}\Big\}\Big\}  }
		\\
		=2{\delta_{m_{1},n_{1}}} M_{b}
		{\mathrm{Re}\Big\{\mathrm{Tr}\Big\{  \overline{\mathbf{Z}}_{m_{1},n_{1}}^{H} \overline{\mathbf{Z}}_{m_{1},n_{1}} {\bf\Phi}_{n_{1}}   \mathbb{E}\Big\{\tilde{\mathbf{h}}_{n_{1},k} 
			\tilde{\mathbf{h}}_{n_{1},k}^{H} \Big\} {\bf\Phi}^{H}_{n_{1}}  \Big\}\Big\}  }
		\\
		=2{\delta_{m_{1},n_{1}}} M_{b}
		{\mathrm{Re}\Big\{\mathrm{Tr}\Big\{  \overline{\mathbf{Z}}_{m_{1},n_{1}}^{H} \overline{\mathbf{Z}}_{m_{1},n_{1}}   \Big\}\Big\}  }
		\\
		=2{\delta_{m_{1},n_{1}}} M_{b}^{2} N_{r}.
	\end{array}
\end{align}

We have completed the calculation of the remaining four parts. After some simplifications,  $\mathbb{E}\left\{\left|\mathbf{g}_{k}^{H} \mathbf{g}_{i}\right|^{2}\right\}$ can be obtained by substituting (\ref{gkgi2_begin}) $\sim$ (\ref{gkgi2_end}) into (\ref{gkgi_2}). With the aid of (\ref{interference_term}) and (\ref{interference_term_re3}), we can complete the calculation of the interference term $\mathbb{E}\left\{\left|({\bf g}_k+{\bf d}_k)^{H} ({\bf g}_i+{\bf d}_i)\right|^{2}\right\}$.
\subsection{Derivation of $\mathbb{E}\left\{\left\|\mathbf{g}_{k}\right\|^{4}\right\}$}\label{subsection3}
Similarly, based on the derivation of the above subsection and (\ref{g_k}) $\sim$ (\ref{g_mni}), when deriving $\mathbb{E}\left\{\left\|\mathbf{g}_{k}\right\|^{4}\right\}$, we can ignore the zero-mean terms based on (\ref{property1}) and (\ref{ZYZ}). Then we have
\begin{align}\label{gk_4}
	\begin{array}{l}
		\mathbb{E}\left\{\left\|\mathbf{g}_{k}\right\|^{4}\right\}=	\mathbb{E}\left\{\left|\sum_{m=1}^{M}\mathbf{g}_{m,k}^{H} \mathbf{g}_{m,k}\right|^{2}\right\}= \mathbb{E}\left\{\left|\sum_{m=1}^{M}\left(\sum_{n_{1}=1}^{N}\mathbf{g}_{m,n_{1},k}^{H}\sum_{n_{2}=1}^{N} \mathbf{g}_{m,n_{2},k}\right)\right|^{2}\right\}\\
		=\mathbb{E}\left\{\sum\limits_{m_{1}=1}^{M}\left(\sum\limits_{n_{1}=1}^{N}\mathbf{g}_{m_{1},n_{1},k}^{H}\sum\limits_{n_{2}=1}^{N} \mathbf{g}_{m_{1},n_{2},k}\right) \sum\limits_{m_{2}=1}^{M}\left(\sum\limits_{n_{3}=1}^{N}\mathbf{g}_{m_{2},n_{3},k}^{H}\sum\limits_{n_{4}=1}^{N}\mathbf{g}_{m_{2},n_{4},k}\right)\right\}\\		={\sum\limits_{m_{1}=1}^{M}\sum\limits_{m_{2}=1}^{M}\sum\limits_{n_{1}=1}^{N}\sum\limits_{n_{2}=1}^{N}\sum\limits_{n_{3}=1}^{N}\sum\limits_{n_{4}=1}^{N}} \mathbb{E}\Big\{\mathbf{g}_{m_{1},n_{1},k}^{H} \mathbf{g}_{m_{1},n_{2},k} \mathbf{g}_{m_{2},n_{3},k}^{H}\mathbf{g}_{m_{2},n_{4},k}\Big\}\\
		={\sum\limits_{m_{1}=1}^{M}\sum\limits_{m_{2}=1}^{M}\sum\limits_{n_{1}=1}^{N}\sum\limits_{n_{2}=1}^{N}\sum\limits_{n_{3}=1}^{N}\sum\limits_{n_{4}=1}^{N}}
		\sqrt{c_{m_{1},n_{1},k} c_{m_{1},n_{2},k} c_{m_{2},n_{3},k} c_{m_{2},n_{4},k}}
		\\  \quad
		\mathbb{E}\Bigg\{ \sum\limits_{\omega_{1}=1}^{4}
		\Big(\mathbf{g}_{m_{1},n_{1},k}^{\omega_{1}}\Big)^{H}\sum\limits_{\psi_{1}=1}^{4} \mathbf{g}_{m_{1},n_{2},k}^{\psi_{1}}
		\sum\limits_{\psi_{2}=1}^{4}
		\Big(\mathbf{g}_{m_{2},n_{3},k}^{\psi_{2}}\Big)^{H}
		\sum\limits_{\omega_{2}=1}^{4}
		\mathbf{g}_{m_{2},n_{4},k}^{\omega_{2}}
		\Bigg\} \\
		={\sum\limits_{m_{1}=1}^{M}\sum\limits_{m_{2}=1}^{M}\sum\limits_{n_{1}=1}^{N}\sum\limits_{n_{2}=1}^{N}\sum\limits_{n_{3}=1}^{N}\sum\limits_{n_{4}=1}^{N}}
		\sqrt{c_{m_{1},n_{1},k} c_{m_{1},n_{2},k} c_{m_{2},n_{3},k} c_{m_{2},n_{4},k}}
		\\ \quad\quad\quad  
		\mathbb{E}\Bigg\{ \sum\limits_{\omega=1}^{4} \sum\limits_{\psi=1}^{4} 
		\Big(\mathbf{g}_{m_{1},n_{1},k}^{\omega}\Big)^{H}\mathbf{g}_{m_{1},n_{2},k}^{\psi}
		\Big(\mathbf{g}_{m_{2},n_{3},k}^{\psi}\Big)^{H}
		\mathbf{g}_{m_{2},n_{4},k}^{\omega} 	\\+2\mathrm{Re}\bigg\{
		\Big(\mathbf{g}_{m_{1},n_{1},k}^{1}\Big)^{H}\mathbf{g}_{m_{1},n_{2},k}^{1}
		\Big(\mathbf{g}_{m_{2},n_{3},k}^{2}\Big)^{H}
		\mathbf{g}_{m_{2},n_{4},k}^{2}
		+
		\Big(\mathbf{g}_{m_{1},n_{1},k}^{1}\Big)^{H}\mathbf{g}_{m_{1},n_{2},k}^{1}
		\Big(\mathbf{g}_{m_{2},n_{3},k}^{3}\Big)^{H}
		\mathbf{g}_{m_{2},n_{4},k}^{3}
		\\ 
		+
		\Big(\mathbf{g}_{m_{1},n_{1},k}^{1}\Big)^{H}\mathbf{g}_{m_{1},n_{2},k}^{1}
		\Big(\mathbf{g}_{m_{2},n_{3},k}^{4}\Big)^{H}
		\mathbf{g}_{m_{2},n_{4},k}^{4}
		+
		\Big(\mathbf{g}_{m_{1},n_{1},k}^{1}\Big)^{H}\mathbf{g}_{m_{1},n_{2},k}^{2}
		\Big(\mathbf{g}_{m_{2},n_{3},k}^{4}\Big)^{H}
		\mathbf{g}_{m_{2},n_{4},k}^{3}
		\\ 
		+
		\Big(\mathbf{g}_{m_{1},n_{1},k}^{1}\Big)^{H}\mathbf{g}_{m_{1},n_{2},k}^{3}
		\Big(\mathbf{g}_{m_{2},n_{3},k}^{4}\Big)^{H}
		\mathbf{g}_{m_{2},n_{4},k}^{2}
		+
		\Big(\mathbf{g}_{m_{1},n_{1},k}^{2}\Big)^{H}\mathbf{g}_{m_{1},n_{2},k}^{1}
		\Big(\mathbf{g}_{m_{2},n_{3},k}^{3}\Big)^{H}
		\mathbf{g}_{m_{2},n_{4},k}^{4}
		\\ 
		+
		\Big(\mathbf{g}_{m_{1},n_{1},k}^{2}\Big)^{H}\mathbf{g}_{m_{1},n_{2},k}^{2}
		\Big(\mathbf{g}_{m_{2},n_{3},k}^{3}\Big)^{H}
		\mathbf{g}_{m_{2},n_{4},k}^{3}
		+
		\Big(\mathbf{g}_{m_{1},n_{1},k}^{2}\Big)^{H}\mathbf{g}_{m_{1},n_{2},k}^{2}
		\Big(\mathbf{g}_{m_{2},n_{3},k}^{4}\Big)^{H}
		\mathbf{g}_{m_{2},n_{4},k}^{4}	
		\\ +
		\Big(\mathbf{g}_{m_{1},n_{1},k}^{3}\Big)^{H}\mathbf{g}_{m_{1},n_{2},k}^{1}
		\Big(\mathbf{g}_{m_{2},n_{3},k}^{2}\Big)^{H}
		\mathbf{g}_{m_{2},n_{4},k}^{4}
		+
		\Big(\mathbf{g}_{m_{1},n_{1},k}^{3}\Big)^{H}\mathbf{g}_{m_{1},n_{2},k}^{3}
		\Big(\mathbf{g}_{m_{2},n_{3},k}^{4}\Big)^{H}
		\mathbf{g}_{m_{2},n_{4},k}^{4}\bigg\}\Bigg\}.
	\end{array}
\end{align}

Next, the above terms in (\ref{gk_4}) will be derived in sequence. The derivation utilizes the properties in (\ref{ZWZ}) and (\ref{ZZWZZ}), and the independence between $\tilde{\mathbf{Z}}_{m_{1},n_{1}}$, $\tilde{\mathbf{Z}}_{m_{2},n_{2}}$, $\tilde{\mathbf{h}}_{n_{1},k}$, $\tilde{\mathbf{h}}_{n_{2},k}$, $\tilde{\mathbf{h}}_{n_{3},k}$ and $\tilde{\mathbf{h}}_{n_{4},k}$. To facilitate the derivation, we omit the notation of accumulation $ \sum\limits_{m_{1}=1}^{M}\sum\limits_{m_{2}=1}^{M}\sum\limits_{n_{1}=1}^{N}\sum\limits_{n_{2}=1}^{N}\sum\limits_{n_{3}=1}^{N}\sum\limits_{n_{4}=1}^{N} $ and the common coefficient $\sqrt{c_{m_{1},n_{1},k} c_{m_{1},n_{2},k} c_{m_{2},n_{3},k} c_{m_{2},n_{4},k}}$, which will vary with these accumulation parameters $m_{1}$, $m_{2}$, $n_{1}$, $n_{2}$, $n_{3}$, and $n_{4}$ in the next derivation.

Next, we first calculate the terms of $\mathbb{E}\Bigg\{  
\Big(\mathbf{g}_{m_{1},n_{1},k}^{\omega}\Big)^{H}\mathbf{g}_{m_{1},n_{2},k}^{\psi}
\Big(\mathbf{g}_{m_{2},n_{3},k}^{\psi}\Big)^{H}
\mathbf{g}_{m_{2},n_{4},k}^{\omega}
\Bigg\}$, $1\leq\omega,\psi\leq4$. 

Firstly, we consider the terms with $\omega=1$. When $\psi=1$, we have
\begin{align}\label{gk4_begin}
	\begin{aligned}
		&\mathbb{E}\Bigg\{  
		\Big(\mathbf{g}_{m_{1},n_{1},k}^{1}\Big)^{H}\mathbf{g}_{m_{1},n_{2},k}^{1}
		\Big(\mathbf{g}_{m_{2},n_{3},k}^{1}\Big)^{H}
		\mathbf{g}_{m_{2},n_{4},k}^{1}
		\Bigg\} \\ 
		&=\sqrt{\delta_{m_{1},n_{1}}\delta_{m_{1},n_{2}}\delta_{m_{2},n_{3}}\delta_{m_{2},n_{4}} \varepsilon_{n_{1},k}\varepsilon_{n_{2},k} \varepsilon_{n_{3},k}\varepsilon_{n_{4},k}} \\ &\qquad{\overline{\mathbf{h}}_{n_{1},k}^{H}  {\bf\Phi}^{H}_{n_{1}} \overline{\mathbf{Z}}_{m_{1},n_{1}}^{H} \overline{\mathbf{Z}}_{m_{1},n_{2}} {\bf\Phi}_{n_{2}}\overline{\mathbf{h}}_{n_{2},k} \overline{\mathbf{h}}_{n_{3},k}^{H} {\bf\Phi}^{H}_{n_{3}} \overline{\mathbf{Z}}_{m_{2},n_{3}}^{H} \overline{\mathbf{Z}}_{m_{2},n_{4}} {\bf\Phi}_{n_{4}}\overline{\mathbf{h}}_{n_{4},k}}\\
		&=\sqrt{\delta_{m_{1},n_{1}}\delta_{m_{1},n_{2}}\delta_{m_{2},n_{3}}\delta_{m_{2},n_{4}} \varepsilon_{n_{1},k}\varepsilon_{n_{2},k} \varepsilon_{n_{3},k}\varepsilon_{n_{4},k}}  
		f_{m_{1},n_{1},k}^{H}({\bf \Phi})
		f_{m_{1},n_{2},k}({\bf \Phi}) \\	
		&\qquad f_{m_{2},n_{3},k}^{H}({\bf \Phi}) f_{m_{2},n_{4},k}({\bf \Phi}) 
		{\bf a}^{H}_{M_{b}}(m_{1},n_{1})
		{\bf a}_{M_{b}}(m_{1},n_{2})
		{\bf a}^{H}_{M_{b}}(m_{2},n_{3})
		{\bf a}_{M_{b}}(m_{2},n_{4}).
	\end{aligned}
\end{align}

When $\psi=2$, we have
\begin{align}
	\begin{array}{l}
		%1   2    2    1
		\mathbb{E}\Bigg\{  
		\Big(\mathbf{g}_{m_{1},n_{1},k}^{1}\Big)^{H}\mathbf{g}_{m_{1},n_{2},k}^{2}
		\Big(\mathbf{g}_{m_{2},n_{3},k}^{2}\Big)^{H}
		\mathbf{g}_{m_{2},n_{4},k}^{1}
		\Bigg\} \\ %\triangleq
		{\mathop = \limits^{(c)}} \sqrt{\delta_{m_{1},n_{1}}\delta_{m_{1},n_{2}}\delta_{m_{2},n_{2}}\delta_{m_{2},n_{4}} \varepsilon_{n_{1},k}\varepsilon_{n_{4},k}} \\ \qquad{\overline{\mathbf{h}}_{n_{1},k}^{H}  {\bf\Phi}^{H}_{n_{1}} \overline{\mathbf{Z}}_{m_{1},n_{1}}^{H} \overline{\mathbf{Z}}_{m_{1},n_{2}} {\bf\Phi}_{n_{2}}\mathbb{E} \Big\{\tilde{\mathbf{h}}_{n_{2},k}\tilde{\mathbf{h}}_{n_{2},k}^{H} \Big\} {\bf\Phi}^{H}_{n_{2}} \overline{\mathbf{Z}}_{m_{2},n_{2}}^{H} \overline{\mathbf{Z}}_{m_{2},n_{4}} {\bf\Phi}_{n_{4}}\overline{\mathbf{h}}_{n_{4},k}}\\
		=\sqrt{\delta_{m_{1},n_{1}}\delta_{m_{1},n_{2}}\delta_{m_{2},n_{2}}\delta_{m_{2},n_{4}} \varepsilon_{n_{1},k}\varepsilon_{n_{4},k}}
		f_{m_{1},n_{1},k}^{H}({\bf \Phi}) f_{m_{2},n_{4},k}({\bf \Phi}) 
		\\
		\qquad{\bf a}^{H}_{M_{b}}(m_{1},n_{1})
		{\bf a}_{M_{b}}(m_{1},n_{2})
		{\bf a}^{H}_{N_{r}}(m_{1},n_{2})
		{\bf a}_{N_{r}}(m_{2},n_{2})
		{\bf a}^{H}_{M_{b}}(m_{2},n_{2})
		{\bf a}_{M_{b}}(m_{2},n_{4}).
	\end{array}
\end{align}

When $\psi=3$, we have
\begin{align}
	\begin{array}{l}
		% 1   3    3     1
		\mathbb{E}\Bigg\{  
		\Big(\mathbf{g}_{m_{1},n_{1},k}^{1}\Big)^{H}\mathbf{g}_{m_{1},n_{2},k}^{3}
		\Big(\mathbf{g}_{m_{2},n_{3},k}^{3}\Big)^{H}
		\mathbf{g}_{m_{2},n_{4},k}^{1}
		\Bigg\} \\
		{\mathop = \limits^{\substack{(d)}}} \sqrt{\delta_{m_{1},n_{1}}\delta_{m_{1},n_{4}} \varepsilon_{n_{1},k}\varepsilon_{n_{4},k}} \varepsilon_{n_{2},k}\\ \qquad{\overline{\mathbf{h}}_{n_{1},k}^{H}  {\bf\Phi}^{H}_{n_{1}} \overline{\mathbf{Z}}_{m_{1},n_{1}}^{H} \mathbb{E}\Big\{\tilde{\mathbf{Z}}_{m_{1},n_{2}}{\bf\Phi}_{n_{2}}\overline{\mathbf{h}}_{n_{2},k}\overline{\mathbf{h}}_{n_{2},k}^{H} {\bf\Phi}^{H}_{n_{2}}  \tilde{\mathbf{Z}}_{m_{1},n_{2}}^{H}\Big\} \overline{\mathbf{Z}}_{m_{1},n_{4}} {\bf\Phi}_{n_{4}}\overline{\mathbf{h}}_{n_{4},k}}\\
		=\sqrt{\delta_{m_{1},n_{1}}\delta_{m_{1},n_{4}} \varepsilon_{n_{1},k}\varepsilon_{n_{4},k}} \varepsilon_{n_{2},k}
		{\overline{\mathbf{h}}_{n_{1},k}^{H}  {\bf\Phi}^{H}_{n_{1}} \overline{\mathbf{Z}}_{m_{1},n_{1}}^{H} \mathbb{E}\Big\{\tilde{\mathbf{Z}}_{m_{1},n_{2}} \tilde{\mathbf{Z}}_{m_{1},n_{2}}^{H}\Big\} \overline{\mathbf{Z}}_{m_{1},n_{4}} {\bf\Phi}_{n_{4}}\overline{\mathbf{h}}_{n_{4},k}}\\
		=\sqrt{\delta_{m_{1},n_{1}}\delta_{m_{1},n_{4}} \varepsilon_{n_{1},k}\varepsilon_{n_{4},k}} \varepsilon_{n_{2},k}N_{r}
		{\overline{\mathbf{h}}_{n_{1},k}^{H}  {\bf\Phi}^{H}_{n_{1}} \overline{\mathbf{Z}}_{m_{1},n_{1}}^{H}  \overline{\mathbf{Z}}_{m_{1},n_{4}} {\bf\Phi}_{n_{4}}\overline{\mathbf{h}}_{n_{4},k}}\\
		=\sqrt{\delta_{m_{1},n_{1}}\delta_{m_{1},n_{4}} \varepsilon_{n_{1},k}\varepsilon_{n_{4},k}} \varepsilon_{n_{2},i} N_{r}
		{f_{m_{1},n_{1},k}^{H}({\bf \Phi}) 
			f_{m_{1},n_{4},k}({\bf \Phi}) 
			{\bf a}^{H}_{M_{b}}(m_{1},n_{1})    
			{\bf a}_{M_{b}}(m_{1},n_{4})}.
	\end{array}
\end{align}

When $\psi=4$, we arrive at
\begin{align}
	\begin{array}{l}
		%1   4    4     1
		\mathbb{E}\Bigg\{  
		\Big(\mathbf{g}_{m_{1},n_{1},k}^{1}\Big)^{H}\mathbf{g}_{m_{1},n_{2},k}^{4}
		\Big(\mathbf{g}_{m_{2},n_{3},k}^{4}\Big)^{H}
		\mathbf{g}_{m_{2},n_{4},k}^{1}
		\Bigg\} \\
		{\mathop = \limits^{\substack{(d)}}}
		\sqrt{\delta_{m_{1},n_{1}}\delta_{m_{1},n_{4}} \varepsilon_{n_{1},k}\varepsilon_{n_{4},k}} 
		\\ %{\xlongequal[m_{1}=m_{2}]{n_{2}=n_{3}}}
		\qquad{\overline{\mathbf{h}}_{n_{1},k}^{H}  {\bf\Phi}^{H}_{n_{1}} \overline{\mathbf{Z}}_{m_{1},n_{1}}^{H} \mathbb{E}\Big\{\tilde{\mathbf{Z}}_{m_{1},n_{2}} {\bf\Phi}_{n_{2}} \mathbb{E}\Big\{\tilde{\mathbf{h}}_{n_{2},k} \tilde{\mathbf{h}}_{n_{2},k}^{H}\Big\} {\bf\Phi}^{H}_{n_{2}}  \tilde{\mathbf{Z}}_{m_{1},n_{2}}^{H}\Big\} \overline{\mathbf{Z}}_{m_{1},n_{4}} {\bf\Phi}_{n_{4}}\overline{\mathbf{h}}_{n_{4},k}}\\
		=\sqrt{\delta_{m_{1},n_{1}}\delta_{m_{1},n_{4}} \varepsilon_{n_{1},k}\varepsilon_{n_{4},k}}  
		\;{\overline{\mathbf{h}}_{n_{1},k}^{H}  {\bf\Phi}^{H}_{n_{1}} \overline{\mathbf{Z}}_{m_{1},n_{1}}^{H} \mathbb{E}\Big\{\tilde{\mathbf{Z}}_{m_{1},n_{2}} \tilde{\mathbf{Z}}_{m_{1},n_{2}}^{H}\Big\} \overline{\mathbf{Z}}_{m_{1},n_{4}} {\bf\Phi}_{n_{4}}\overline{\mathbf{h}}_{n_{4},k}}\\
		=\sqrt{\delta_{m_{1},n_{1}}\delta_{m_{1},n_{4}} \varepsilon_{n_{1},k}\varepsilon_{n_{4},k}} N_{r}
		{f_{m_{1},n_{1},k}^{H}({\bf \Phi}) 
			f_{m_{1},n_{4},k}({\bf \Phi}) 
			{\bf a}^{H}_{M_{b}}(m_{1},n_{1})    
			{\bf a}_{M_{b}}(m_{1},n_{4})}.
	\end{array}
\end{align}

Secondly, consider the terms with $\omega=2$. When $\psi=1$, we have
\begin{align}
	\begin{array}{l}
		\mathbb{E}\Bigg\{  % 2   1   1   2
		\Big(\mathbf{g}_{m_{1},n_{1},k}^{2}\Big)^{H}\mathbf{g}_{m_{1},n_{2},k}^{1}
		\Big(\mathbf{g}_{m_{2},n_{3},k}^{1}\Big)^{H}
		\mathbf{g}_{m_{2},n_{4},k}^{2}
		\Bigg\} \\
		\overset{\substack{(e)}}{=}
		\sqrt{\delta_{m_{1},n_{1}}\delta_{m_{1},n_{2}} \delta_{m_{2},n_{3}}\delta_{m_{2},n_{1}} \varepsilon_{n_{2},k}\varepsilon_{n_{3},k}} 
		\\ %{\xlongequal[m_{1}=m_{2}]{n_{2}=n_{3}}}
		\qquad{\mathbb{E}\Big\{\tilde{\mathbf{h}}_{n_{1},k}^{H}  {\bf\Phi}^{H}_{n_{1}} \overline{\mathbf{Z}}_{m_{1},n_{1}}^{H} \overline{\mathbf{Z}}_{m_{1},n_{2}} {\bf\Phi}_{n_{2}} \overline{\mathbf{h}}_{n_{2},k} \overline{\mathbf{h}}_{n_{3},k}^{H} {\bf\Phi}^{H}_{n_{3}}  \overline{\mathbf{Z}}_{m_{2},n_{3}}^{H} \overline{\mathbf{Z}}_{m_{2},n_{1}} {\bf\Phi}_{n_{1}}\tilde{\mathbf{h}}_{n_{1},k}\Big\}}   \\
		=\sqrt{\delta_{m_{1},n_{1}}\delta_{m_{1},n_{2}} \delta_{m_{2},n_{3}}\delta_{m_{2},n_{1}} \varepsilon_{n_{2},k}\varepsilon_{n_{3},k}} 
		\\
		\qquad{\overline{\mathbf{h}}_{n_{3},k}^{H}  {\bf\Phi}^{H}_{n_{3}} \overline{\mathbf{Z}}_{m_{2},n_{3}}^{H} \overline{\mathbf{Z}}_{m_{2},n_{1}} {\bf\Phi}_{n_{1}} \mathbb{E}\Big\{\tilde{\mathbf{h}}_{n_{1},k} \tilde{\mathbf{h}}_{n_{1},k}^{H}\Big\} {\bf\Phi}^{H}_{n_{1}}  \overline{\mathbf{Z}}_{m_{1},n_{1}}^{H} \overline{\mathbf{Z}}_{m_{1},n_{2}} {\bf\Phi}_{n_{2}}\overline{\mathbf{h}}_{n_{2},k}}\\
		=\sqrt{\delta_{m_{1},n_{1}}\delta_{m_{1},n_{2}} \delta_{m_{2},n_{3}}\delta_{m_{2},n_{1}} \varepsilon_{n_{2},k}\varepsilon_{n_{3},k}} 
		f_{m_{2},n_{3},k}^{H}({\bf \Phi}) 
		f_{m_{1},n_{2},k}({\bf \Phi}) 
		\\
		\qquad{{\bf a}^{H}_{M_{b}}(m_{2},n_{3})    
			{\bf a}_{M_{b}}(m_{2},n_{1})
			{\bf a}^{H}_{N_{r}}(m_{2},n_{1})    
			{\bf a}_{N_{r}}(m_{1},n_{1})	
			{\bf a}^{H}_{M_{b}}(m_{1},n_{1})    
			{\bf a}_{M_{b}}(m_{1},n_{2})}.
	\end{array}
\end{align}

When $\psi=2$, the term is the addition of the following two parts
\\part 1:
\begin{align}
	\begin{array}{l}
		%2 2 2 2    (1)
		\Bigg(\mathbb{E}\Bigg\{  
		\Big(\mathbf{g}_{m_{1},n_{1},k}^{2}\Big)^{H}\mathbf{g}_{m_{1},n_{2},k}^{2}
		\Big(\mathbf{g}_{m_{2},n_{3},k}^{2}\Big)^{H}
		\mathbf{g}_{m_{2},n_{4},k}^{2}
		\Bigg\} \Bigg)_{1}\\
		\overset{\substack{(f)}}{=}
		\sqrt{\delta_{m_{1},n_{1}}\delta_{m_{1},n_{2}} \delta_{m_{2},n_{1}}\delta_{m_{2},n_{2}}} 
		\\ 
		\qquad{\mathbb{E}\Big\{\tilde{\mathbf{h}}_{n_{1},k}^{H}  {\bf\Phi}^{H}_{n_{1}} \overline{\mathbf{Z}}_{m_{1},n_{1}}^{H} \overline{\mathbf{Z}}_{m_{1},n_{2}} {\bf\Phi}_{n_{2}} \mathbb{E}\Big\{\tilde{\mathbf{h}}_{n_{2},k} \tilde{\mathbf{h}}_{n_{2},k}^{H}\Big\} {\bf\Phi}^{H}_{n_{2}}  \overline{\mathbf{Z}}_{m_{2},n_{2}}^{H} \overline{\mathbf{Z}}_{m_{2},n_{1}} {\bf\Phi}_{n_{1}}\tilde{\mathbf{h}}_{n_{1},k}\Big\}}   \\
		=\sqrt{\delta_{m_{1},n_{1}}\delta_{m_{1},n_{2}} \delta_{m_{2},n_{1}}\delta_{m_{2},n_{2}}} 
		{\mathbb{E}\Big\{\tilde{\mathbf{h}}_{n_{1},k}^{H}{\bf\Phi}^{H}_{n_{1}} \overline{\mathbf{Z}}_{m_{1},n_{1}}^{H} \overline{\mathbf{Z}}_{m_{1},n_{2}}  \overline{\mathbf{Z}}_{m_{2},n_{2}}^{H} \overline{\mathbf{Z}}_{m_{2},n_{1}} {\bf\Phi}_{n_{1}}\tilde{\mathbf{h}}_{n_{1},k}\Big\}}   \\
		=\sqrt{\delta_{m_{1},n_{1}}\delta_{m_{1},n_{2}} \delta_{m_{2},n_{1}}\delta_{m_{2},n_{2}}} 
		{\mathrm{Tr}\Big\{ \overline{\mathbf{Z}}_{m_{2},n_{2}}^{H} \overline{\mathbf{Z}}_{m_{2},n_{1}}{\bf\Phi}_{n_{1}} \mathbb{E}\Big\{\tilde{\mathbf{h}}_{n_{1},k} \tilde{\mathbf{h}}_{n_{1},k}^{H} \Big\}{\bf\Phi}^{H}_{n_{1}} \overline{\mathbf{Z}}_{m_{1},n_{1}}^{H} \overline{\mathbf{Z}}_{m_{1},n_{2}} \Big\}}   
		\\
		=\sqrt{\delta_{m_{1},n_{1}}\delta_{m_{1},n_{2}} \delta_{m_{2},n_{1}}\delta_{m_{2},n_{2}}} 
		{\mathrm{Tr}\Big\{ \overline{\mathbf{Z}}_{m_{2},n_{2}}^{H} \overline{\mathbf{Z}}_{m_{2},n_{1}}  \overline{\mathbf{Z}}_{m_{1},n_{1}}^{H} \overline{\mathbf{Z}}_{m_{1},n_{2}} \Big\}}   
		\\
		=\sqrt{\delta_{m_{1},n_{1}}\delta_{m_{1},n_{2}} \delta_{m_{2},n_{1}}\delta_{m_{2},n_{2}}}  
		{\bf a}^{H}_{M_{b}}(m_{1},n_{1})    
		{\bf a}_{M_{b}}(m_{1},n_{2})
		{\bf a}^{H}_{N_{r}}(m_{1},n_{2})    
		{\bf a}_{N_{r}}(m_{2},n_{2})\\
		\qquad{\bf a}^{H}_{M_{b}}(m_{2},n_{2})    
		{\bf a}_{M_{b}}(m_{2},n_{1})
		{\bf a}^{H}_{N_{r}}(m_{2},n_{1})    
		{\bf a}_{N_{r}}(m_{1},n_{1}),
	\end{array}
\end{align}
part 2:
\begin{align}
	\begin{array}{l} 
		%2 2 2 2  (2)
		\Bigg(\mathbb{E}\Bigg\{  
		\Big(\mathbf{g}_{m_{1},n_{1},k}^{2}\Big)^{H}\mathbf{g}_{m_{1},n_{2},k}^{2}
		\Big(\mathbf{g}_{m_{2},n_{3},k}^{2}\Big)^{H}
		\mathbf{g}_{m_{2},n_{4},k}^{2}
		\Bigg\}\Bigg)_{2} \\
		\overset{\substack{(i)}}{=}
		\delta_{m_{1},n_{1}}\delta_{m_{2},n_{3}} 
		{\mathbb{E}\Big\{\tilde{\mathbf{h}}_{n_{1},k}^{H}  {\bf\Phi}^{H}_{n_{1}} \overline{\mathbf{Z}}_{m_{1},n_{1}}^{H} \overline{\mathbf{Z}}_{m_{1},n_{1}} {\bf\Phi}_{n_{1}} \tilde{\mathbf{h}}_{n_{1},k}\Big\} \mathbb{E}\Big\{\tilde{\mathbf{h}}_{n_{3},k}^{H} {\bf\Phi}^{H}_{n_{3}}  \overline{\mathbf{Z}}_{m_{2},n_{3}}^{H} \overline{\mathbf{Z}}_{m_{2},n_{3}} {\bf\Phi}_{n_{3}}\tilde{\mathbf{h}}_{n_{3},k}\Big\}}   
		\\
		=\delta_{m_{1},n_{1}}\delta_{m_{2},n_{3}} 
		{\mathrm{Tr}\Big\{{\bf\Phi}^{H}_{n_{1}} \overline{\mathbf{Z}}_{m_{1},n_{1}}^{H} \overline{\mathbf{Z}}_{m_{1},n_{1}} {\bf\Phi}_{n_{1}} \Big\} 
			\mathrm{Tr}\Big\{ {\bf\Phi}^{H}_{n_{3}}  \overline{\mathbf{Z}}_{m_{2},n_{3}}^{H} \overline{\mathbf{Z}}_{m_{2},n_{3}} {\bf\Phi}_{n_{3}}\Big\}}   
		\\
		=\delta_{m_{1},n_{1}}\delta_{m_{2},n_{3}} 
		M_{b}^{2}  N_{r}^{2}.
	\end{array}
\end{align}

When $\psi=3$, we have
\begin{align}
	\begin{array}{l}     
		%  2 3 3 2
		\mathbb{E}\Bigg\{  
		\Big(\mathbf{g}_{m_{1},n_{1},k}^{2}\Big)^{H}\mathbf{g}_{m_{1},n_{2},k}^{3}
		\Big(\mathbf{g}_{m_{2},n_{3},k}^{3}\Big)^{H}
		\mathbf{g}_{m_{2},n_{4},k}^{2}
		\Bigg\} \\
		\overset{\substack{(g)}}{=} \delta_{m_{1},n_{1}}\varepsilon_{n_{2},k}
		{\mathbb{E}\Big\{\tilde{\mathbf{h}}_{n_{1},k}^{H}{\bf\Phi}^{H}_{n_{1}} \overline{\mathbf{Z}}_{m_{1},n_{1}}^{H} \mathbb{E}\Big\{\tilde{\mathbf{Z}}_{m_{1},n_{2}} {\bf\Phi}_{n_{2}} \overline{\mathbf{h}}_{n_{2},k}\overline{\mathbf{h}}_{n_{2},k}^{H}{\bf\Phi}^{H}_{n_{2}}  \tilde{\mathbf{Z}}_{m_{1},n_{2}}^{H}\Big\}  \overline{\mathbf{Z}}_{m_{1},n_{1}} {\bf\Phi}_{n_{1}}\tilde{\mathbf{h}}_{n_{1},k}\Big\}}   \\
		=\delta_{m_{1},n_{1}}\varepsilon_{n_{2},k} 
		{\mathbb{E}\Big\{\tilde{\mathbf{h}}_{n_{1},k}^{H}{\bf\Phi}^{H}_{n_{1}} \overline{\mathbf{Z}}_{m_{1},n_{1}}^{H}  \overline{\mathbf{Z}}_{m_{1},n_{1}} {\bf\Phi}_{n_{1}}\tilde{\mathbf{h}}_{n_{1},k}\Big\} \mathrm{Tr}\Big\{ {\bf\Phi}_{n_{2}} \overline{\mathbf{h}}_{n_{2},k}\overline{\mathbf{h}}_{n_{2},k}^{H}{\bf\Phi}^{H}_{n_{2}} \Big\}}   \\
		=\delta_{m_{1},n_{1}}\varepsilon_{n_{2},k}N_{r}
		{\mathrm{Tr}\Big\{ \overline{\mathbf{Z}}_{m_{1},n_{1}} \overline{\mathbf{Z}}_{m_{1},n_{1}}^{H}\Big\}}  
		\\
		=\delta_{m_{1},n_{1}}\varepsilon_{n_{2},k} M_{b}N_{r}^{2}.
	\end{array}
\end{align}

When $\psi=4$, we derive the term by calculating two parts as follows 
\\part 1:
\begin{align}
	\begin{array}{l}
		%   2  4  4  2  (1)
		\Bigg(\mathbb{E}\Bigg\{  
		\Big(\mathbf{g}_{m_{1},n_{1},k}^{2}\Big)^{H}\mathbf{g}_{m_{1},n_{2},k}^{4}
		\Big(\mathbf{g}_{m_{2},n_{3},k}^{4}\Big)^{H}
		\mathbf{g}_{m_{2},n_{4},k}^{2}
		\Bigg\} \Bigg)_{1}\\
		\overset{\substack{(g)}}{=} \delta_{m_{1},n_{1}} 
		{\mathbb{E}\Big\{\tilde{\mathbf{h}}_{n_{1},k}^{H}{\bf\Phi}^{H}_{n_{1}} \overline{\mathbf{Z}}_{m_{1},n_{1}}^{H} \mathbb{E}\Big\{\tilde{\mathbf{Z}}_{m_{1},n_{2}} {\bf\Phi}_{n_{2}} \mathbb{E}\Big\{\tilde{\mathbf{h}}_{n_{2},k} \tilde{\mathbf{h}}_{n_{2},k}^{H}\Big\} {\bf\Phi}^{H}_{n_{2}}  \tilde{\mathbf{Z}}_{m_{1},n_{2}}^{H}\Big\}  \overline{\mathbf{Z}}_{m_{1},n_{1}} {\bf\Phi}_{n_{1}}\tilde{\mathbf{h}}_{n_{1},k}\Big\}}   \\
		=\delta_{m_{1},n_{1}}N_{r}
		{\mathbb{E}\Big\{\tilde{\mathbf{h}}_{n_{1},k}^{H}{\bf\Phi}^{H}_{n_{1}} \overline{\mathbf{Z}}_{m_{1},n_{1}}^{H}  \overline{\mathbf{Z}}_{m_{1},n_{1}} {\bf\Phi}_{n_{1}}\tilde{\mathbf{h}}_{n_{1},k}\Big\}}   \\
		=\delta_{m_{1},n_{1}}N_{r} 
		{\mathrm{Tr}\Big\{ \overline{\mathbf{Z}}_{m_{1},n_{1}}  \overline{\mathbf{Z}}_{m_{1},n_{1}}^{H}\Big\}}  \\
		=\delta_{m_{1},n_{1}}M_{b}N_{r}^{2} ,
	\end{array}
\end{align}
part 2:
\begin{align}
	\begin{array}{l}
		%   2  4  4  2  (2)
		\Bigg(\mathbb{E}\Bigg\{  
		\Big(\mathbf{g}_{m_{1},n_{1},k}^{2}\Big)^{H}\mathbf{g}_{m_{1},n_{2},k}^{4}
		\Big(\mathbf{g}_{m_{2},n_{3},k}^{4}\Big)^{H}
		\mathbf{g}_{m_{2},n_{4},k}^{2}
		\Bigg\} \Bigg)_{2}\\
		\overset{\substack{(n)}}{=} \delta_{m_{1},n_{1}} 
		{\mathbb{E}_{\tilde{\mathbf{h}}_{n_{1},k}}\Big\{\tilde{\mathbf{h}}_{n_{1},k}^{H}{\bf\Phi}^{H}_{n_{1}} \overline{\mathbf{Z}}_{m_{1},n_{1}}^{H} \mathbb{E}_{\tilde{\mathbf{Z}}_{m_{1},n_{1}}}\Big\{\tilde{\mathbf{Z}}_{m_{1},n_{1}} {\bf\Phi}_{n_{1}} 
			\tilde{\mathbf{h}}_{n_{1},k} \tilde{\mathbf{h}}_{n_{1},k}^{H} {\bf\Phi}^{H}_{n_{1}}  \tilde{\mathbf{Z}}_{m_{1},n_{1}}^{H}\Big\}  \overline{\mathbf{Z}}_{m_{1},n_{1}} {\bf\Phi}_{n_{1}}\tilde{\mathbf{h}}_{n_{1},k}
			\mid\tilde{\mathbf{h}}_{n_{1},k}\Big\}}   
		\\
		= \delta_{m_{1},n_{1}} 
		{\mathbb{E}\Big\{\tilde{\mathbf{h}}_{n_{1},k}^{H}{\bf\Phi}^{H}_{n_{1}} \overline{\mathbf{Z}}_{m_{1},n_{1}}^{H} \mathrm{Tr}\Big\{ {\bf\Phi}_{n_{1}} 
			\tilde{\mathbf{h}}_{n_{1},k} \tilde{\mathbf{h}}_{n_{1},k}^{H} {\bf\Phi}^{H}_{n_{1}}  \Big\}  \overline{\mathbf{Z}}_{m_{1},n_{1}} {\bf\Phi}_{n_{1}}\tilde{\mathbf{h}}_{n_{1},k}\Big\}}   
		\\
		= \delta_{m_{1},n_{1}} 
		{\mathbb{E}\Big\{\tilde{\mathbf{h}}_{n_{1},k}^{H}{\bf\Phi}^{H}_{n_{1}} \overline{\mathbf{Z}}_{m_{1},n_{1}}^{H}  \overline{\mathbf{Z}}_{m_{1},n_{1}} {\bf\Phi}_{n_{1}}\tilde{\mathbf{h}}_{n_{1},k}	(\tilde{\mathbf{h}}_{n_{1},k}^{H} \tilde{\mathbf{h}}_{n_{1},k})\Big\}}   
		\\
		= \delta_{m_{1},n_{1}} 
		{\mathrm{Tr}\Big\{{\bf\Phi}^{H}_{n_{1}} \overline{\mathbf{Z}}_{m_{1},n_{1}}^{H}  \overline{\mathbf{Z}}_{m_{1},n_{1}} {\bf\Phi}_{n_{1}} \mathbb{E}\Big\{\tilde{\mathbf{h}}_{n_{1},k}\tilde{\mathbf{h}}_{n_{1},k}^{H} \tilde{\mathbf{h}}_{n_{1},k} \tilde{\mathbf{h}}_{n_{1},k}^{H}\Big\}\Big\}}   
		\\
		=\delta_{m_{1},n_{1}} 
		{\mathrm{Tr}\Big\{{\bf\Phi}^{H}_{n_{1}} \overline{\mathbf{Z}}_{m_{1},n_{1}}^{H}  \overline{\mathbf{Z}}_{m_{1},n_{1}} {\bf\Phi}_{n_{1}} 
			(N_{r}+1)\mathbf{I}_{N_{r}}\Big\}}   
		\\
		= \delta_{m_{1},n_{1}} 
		M_{b} N_{r} (N_{r}+1),
	\end{array}
\end{align}
where the part 2 utilizes the property in (\ref{ZZWZZ}) with $\mathbf{W}=\mathbf{I}_{N_{r}}$, and $(n)$ represents $m_{1}=m_{2}$, $n_{1}=n_{2}=n_{3}=n_{4}$ and utilizes the law of total expectation, which calculates the conditional expectation of $\tilde{\mathbf{Z}}_{m_{1},n_{1}}$ given $ \tilde{\mathbf{h}}_{n_{1},k} $, and then calculates the expectation of $ \tilde{\mathbf{h}}_{n_{1},k} $. Since $\tilde{\mathbf{Z}}_{m_{1},n_{1}}$ is independent of $ \tilde{\mathbf{h}}_{n_{1},k} $, the conditional expectation of $\tilde{\mathbf{Z}}_{m_{1},n_{1}}$ given $ \tilde{\mathbf{h}}_{n_{1},k} $ is the same as its unconditional expectation. The second part is the special case of part 1. The term can be derived by combining these two parts, we have 
\begin{align}
\begin{array}{l}
		%   2  4  4  2  
		{\sum\limits_{m_{1}=1}^{M}\sum\limits_{m_{2}=1}^{M}\sum\limits_{n_{1}=1}^{N}\sum\limits_{n_{2}=1}^{N}\sum\limits_{n_{3}=1}^{N}\sum\limits_{n_{4}=1}^{N}}
		\sqrt{c_{m_{1},n_{1},k} c_{m_{1},n_{2},k} c_{m_{2},n_{3},k} c_{m_{2},n_{4},k}}
		\\ \qquad\qquad\qquad
		\mathbb{E}\Bigg\{  
		\Big(\mathbf{g}_{m_{1},n_{1},k}^{2}\Big)^{H}\mathbf{g}_{m_{1},n_{2},k}^{4}
		\Big(\mathbf{g}_{m_{2},n_{3},k}^{4}\Big)^{H}
		\mathbf{g}_{m_{2},n_{4},k}^{2}
		\Bigg\} \\
		={\sum\limits_{m=1}^{M}\sum\limits_{n_{1}=1}^{N}\sum\limits_{n_{2}=1}^{N}}
		c_{m,n_{1},k} c_{m,n_{2},k}  \delta_{m,n_{1}} M_{b} N_{r}^{2} +
		{\sum\limits_{m=1}^{M}\sum\limits_{n=1}^{N}} 	c_{m,n,k}^{2} \delta_{m,n} M_{b} N_{r}.
	\end{array}
\end{align}
Thirdly, consider the terms with $\omega= 3$. When $\psi=1$, we have
\begin{align}
	\begin{array}{l}
		%   3  1  1  3 
		\mathbb{E}\Bigg\{  
		\Big(\mathbf{g}_{m_{1},n_{1},k}^{3}\Big)^{H}\mathbf{g}_{m_{1},n_{2},i}^{1}
		\Big(\mathbf{g}_{m_{2},n_{3},k}^{1}\Big)^{H}
		\mathbf{g}_{m_{2},n_{4},k}^{3}
		\Bigg\} \\
		\overset{\substack{(h)}}{=}\varepsilon_{n_{1},k}
		\sqrt{\delta_{m_{1},n_{2}}\delta_{m_{1},n_{3}} \varepsilon_{n_{2},k}\varepsilon_{n_{3},k}}
		\\
		\qquad{\overline{\mathbf{h}}_{n_{1},k}^{H}{\bf\Phi}^{H}_{n_{1}} \mathbb{E}\Big\{\tilde{\mathbf{Z}}_{m_{1},n_{1}}^{H} \overline{\mathbf{Z}}_{m_{1},n_{2}} {\bf\Phi}_{n_{2}} \overline{\mathbf{h}}_{n_{2},k} \overline{\mathbf{h}}_{n_{3},k}^{H} {\bf\Phi}^{H}_{n_{3}}  \overline{\mathbf{Z}}_{m_{1},n_{3}}^{H} \tilde{\mathbf{Z}}_{m_{1},n_{1}}\Big\}  {\bf\Phi}_{n_{1}}\overline{\mathbf{h}}_{n_{1},k}}\\
		=\varepsilon_{n_{1},k}
		\sqrt{\delta_{m_{1},n_{2}}\delta_{m_{1},n_{3}} \varepsilon_{n_{2},k}\varepsilon_{n_{3},k}}
		{\overline{\mathbf{h}}_{n_{1},k}^{H}{\bf\Phi}^{H}_{n_{1}}{\bf\Phi}_{n_{1}}\overline{\mathbf{h}}_{n_{1},k} \mathrm{Tr}\Big\{\overline{\mathbf{Z}}_{m_{1},n_{2}} {\bf\Phi}_{n_{2}} \overline{\mathbf{h}}_{n_{2},k} \overline{\mathbf{h}}_{n_{3},k}^{H} {\bf\Phi}^{H}_{n_{3}}  \overline{\mathbf{Z}}_{m_{1},n_{3}}^{H}\Big\} }
		\\
		=\varepsilon_{n_{1},k}
		\sqrt{\delta_{m_{1},n_{2}}\delta_{m_{1},n_{3}} \varepsilon_{n_{2},k}\varepsilon_{n_{3},k}}N_{r}
		{\mathrm{Tr}\Big\{\overline{\mathbf{Z}}_{m_{1},n_{2}} {\bf\Phi}_{n_{2}} \overline{\mathbf{h}}_{n_{2},k} \overline{\mathbf{h}}_{n_{3},k}^{H} {\bf\Phi}^{H}_{n_{3}}  \overline{\mathbf{Z}}_{m_{1},n_{3}}^{H}\Big\} }
		\\
		=\varepsilon_{n_{1},k} 
		\sqrt{\delta_{m_{1},n_{2}}\delta_{m_{1},n_{3}} \varepsilon_{n_{2},k}\varepsilon_{n_{3},k}}N_{r}
		\;
		{\overline{\mathbf{h}}_{n_{3},k}^{H} {\bf\Phi}^{H}_{n_{3}}  \overline{\mathbf{Z}}_{m_{1},n_{3}}^{H}
			\overline{\mathbf{Z}}_{m_{1},n_{2}} {\bf\Phi}_{n_{2}} \overline{\mathbf{h}}_{n_{2},k} }
		\\
\end{array}
\end{align}

When $\psi=2$, we have
\begin{align}
	\begin{array}{l}
		%   3  2  2  3 
		\mathbb{E}\Bigg\{  
		\Big(\mathbf{g}_{m_{1},n_{1},k}^{3}\Big)^{H}\mathbf{g}_{m_{1},n_{2},k}^{2}
		\Big(\mathbf{g}_{m_{2},n_{3},k}^{2}\Big)^{H}
		\mathbf{g}_{m_{2},n_{4},k}^{3}
		\Bigg\} \\
		\overset{\substack{(g)}}{=}
		\delta_{m_{1},n_{2}}\varepsilon_{n_{1},k}
		{\overline{\mathbf{h}}_{n_{1},k}^{H}{\bf\Phi}^{H}_{n_{1}} \mathbb{E}\Big\{\tilde{\mathbf{Z}}_{m_{1},n_{1}}^{H} \overline{\mathbf{Z}}_{m_{1},n_{2}} {\bf\Phi}_{n_{2}} \mathbb{E}\Big\{\tilde{\mathbf{h}}_{n_{2},k} \tilde{\mathbf{h}}_{n_{2},k}^{H}\Big\} {\bf\Phi}^{H}_{n_{2}}  \overline{\mathbf{Z}}_{m_{1},n_{2}}^{H} \tilde{\mathbf{Z}}_{m_{1},n_{1}}\Big\}  {\bf\Phi}_{n_{1}}\overline{\mathbf{h}}_{n_{1},k}}\\
		=\delta_{m_{1},n_{2}}\varepsilon_{n_{1},k}
		{\overline{\mathbf{h}}_{n_{1},k}^{H}{\bf\Phi}^{H}_{n_{1}} \mathbb{E}\Big\{\tilde{\mathbf{Z}}_{m_{1},n_{1}}^{H} \overline{\mathbf{Z}}_{m_{1},n_{2}}  \overline{\mathbf{Z}}_{m_{1},n_{2}}^{H} \tilde{\mathbf{Z}}_{m_{1},n_{1}}\Big\}  {\bf\Phi}_{n_{1}}\overline{\mathbf{h}}_{n_{1},k}}\\
		=\delta_{m_{1},n_{2}}\varepsilon_{n_{1},k}
		{\overline{\mathbf{h}}_{n_{1},k}^{H}{\bf\Phi}^{H}_{n_{1}}{\bf\Phi}_{n_{1}}\overline{\mathbf{h}}_{n_{1},k}}\mathrm{Tr}\Big\{ \overline{\mathbf{Z}}_{m_{1},n_{2}}  \overline{\mathbf{Z}}_{m_{1},n_{2}}^{H}\Big\} 
		\\
		=\delta_{m_{1},n_{2}}\varepsilon_{n_{1},k}M_{b} N_{r}^{2}.
	\end{array}
\end{align}

When $\psi=3$, the term is the addition of the following two parts
\\part 1:
\begin{align}
	\begin{array}{l}
		%   3  3  3  3 (1)
		\Bigg(\mathbb{E}\Bigg\{  
		\Big(\mathbf{g}_{m_{1},n_{1},k}^{3}\Big)^{H}\mathbf{g}_{m_{1},n_{2},k}^{3}
		\Big(\mathbf{g}_{m_{2},n_{3},k}^{3}\Big)^{H}
		\mathbf{g}_{m_{2},n_{4},k}^{3}
		\Bigg\} \Bigg)_{1}\\
		\overset{\substack{(g)}}{=}
		\varepsilon_{n_{1},k}\varepsilon_{n_{2},k}
		{\overline{\mathbf{h}}_{n_{1},k}^{H}{\bf\Phi}^{H}_{n_{1}} \mathbb{E}\Big\{\tilde{\mathbf{Z}}_{m_{1},n_{1}}^{H} \mathbb{E}\Big\{\tilde{\mathbf{Z}}_{m_{1},n_{2}} {\bf\Phi}_{n_{2}} \overline{\mathbf{h}}_{n_{2},k} \overline{\mathbf{h}}_{n_{2},k}^{H} {\bf\Phi}^{H}_{n_{2}}  \tilde{\mathbf{Z}}_{m_{1},n_{2}}^{H}\Big\}  \tilde{\mathbf{Z}}_{m_{1},n_{1}}\Big\}  {\bf\Phi}_{n_{1}}\overline{\mathbf{h}}_{n_{1},k}}\\
		=\varepsilon_{n_{1},k}\varepsilon_{n_{2},k}
		{\overline{\mathbf{h}}_{n_{1},k}^{H}{\bf\Phi}^{H}_{n_{1}} \mathbb{E}\Big\{\tilde{\mathbf{Z}}_{m_{1},n_{1}}^{H}  \tilde{\mathbf{Z}}_{m_{1},n_{1}}\Big\}  {\bf\Phi}_{n_{1}}\overline{\mathbf{h}}_{n_{1},k} \mathrm{Tr}\Big\{ {\bf\Phi}_{n_{2}} \overline{\mathbf{h}}_{n_{2},k} \overline{\mathbf{h}}_{n_{2},k}^{H} {\bf\Phi}^{H}_{n_{2}}  \Big\} }
		\\
		=\varepsilon_{n_{1},k}\varepsilon_{n_{2},k}N_{r}
		{\overline{\mathbf{h}}_{n_{1},k}^{H}{\bf\Phi}^{H}_{n_{1}} \mathbb{E}\Big\{\tilde{\mathbf{Z}}_{m_{1},n_{1}}^{H}  \tilde{\mathbf{Z}}_{m_{1},n_{1}}\Big\}  {\bf\Phi}_{n_{1}}\overline{\mathbf{h}}_{n_{1},k}}\\
		=\varepsilon_{n_{1},k}\varepsilon_{n_{2},k}M_{b} N_{r}^{2},
	\end{array}
\end{align}
part 2:
\begin{align}
	\begin{array}{l}
		%   3  3  3  3 (2)
		\Bigg(\mathbb{E}\Bigg\{  
		\Big(\mathbf{g}_{m_{1},n_{1},k}^{3}\Big)^{H}\mathbf{g}_{m_{1},n_{2},k}^{3}
		\Big(\mathbf{g}_{m_{2},n_{3},k}^{3}\Big)^{H}
		\mathbf{g}_{m_{2},n_{4},k}^{3}
		\Bigg\} \Bigg)_{2}\\
		\overset{\substack{(i)}}{=}
		\varepsilon_{n_{1},k}\varepsilon_{n_{3},k}
		{\overline{\mathbf{h}}_{n_{1},k}^{H}{\bf\Phi}^{H}_{n_{1}} \mathbb{E}\Big\{\tilde{\mathbf{Z}}_{m_{1},n_{1}}^{H} \tilde{\mathbf{Z}}_{m_{1},n_{1}}\Big\} {\bf\Phi}_{n_{1}} \overline{\mathbf{h}}_{n_{1},k} \overline{\mathbf{h}}_{n_{3},k}^{H} {\bf\Phi}^{H}_{n_{3}}  \mathbb{E}\Big\{\tilde{\mathbf{Z}}_{m_{2},n_{3}}^{H}  \tilde{\mathbf{Z}}_{m_{2},n_{3}}\Big\}  {\bf\Phi}_{n_{3}}\overline{\mathbf{h}}_{n_{3},k}}
		\\
		=\varepsilon_{n_{1},k}\varepsilon_{n_{3},k}M_{b}^{2}
		{\overline{\mathbf{h}}_{n_{1},k}^{H} \overline{\mathbf{h}}_{n_{1},k} \overline{\mathbf{h}}_{n_{3},k}^{H} \overline{\mathbf{h}}_{n_{3},k}}
		\\
		=\varepsilon_{n_{1},k}\varepsilon_{n_{3},k}M_{b}^{2} N_{r}^{2}.
	\end{array}
\end{align}

When $\psi=4$, the term is the addition of the following two parts
\\part 1:
\begin{align}
	\begin{array}{l}
		\Bigg(%   3  4  4  3 (1)
		\mathbb{E}\Bigg\{  
		\Big(\mathbf{g}_{m_{1},n_{1},k}^{3}\Big)^{H}\mathbf{g}_{m_{1},n_{2},k}^{4}
		\Big(\mathbf{g}_{m_{2},n_{3},k}^{4}\Big)^{H}
		\mathbf{g}_{m_{2},n_{4},k}^{3}
		\Bigg\} \Bigg)_{1}\\
		\overset{\substack{(g)}}{=}
		\varepsilon_{n_{1},k}
		{\overline{\mathbf{h}}_{n_{1},k}^{H}{\bf\Phi}^{H}_{n_{1}} \mathbb{E}\Big\{\tilde{\mathbf{Z}}_{m_{1},n_{1}}^{H} \mathbb{E}\Big\{\tilde{\mathbf{Z}}_{m_{1},n_{2}} {\bf\Phi}_{n_{2}} \mathbb{E}\Big\{\tilde{\mathbf{h}}_{n_{2},k} \tilde{\mathbf{h}}_{n_{2},k}^{H}\Big\} {\bf\Phi}^{H}_{n_{2}}  \tilde{\mathbf{Z}}_{m_{1},n_{2}}^{H}\Big\}  \tilde{\mathbf{Z}}_{m_{1},n_{1}}\Big\}  {\bf\Phi}_{n_{1}}\overline{\mathbf{h}}_{n_{1},k}}\\
		=\varepsilon_{n_{1},k}
		{\overline{\mathbf{h}}_{n_{1},k}^{H}{\bf\Phi}^{H}_{n_{1}} \mathbb{E}\Big\{\tilde{\mathbf{Z}}_{m_{1},n_{1}}^{H} \mathbb{E}\Big\{\tilde{\mathbf{Z}}_{m_{1},n_{2}} \tilde{\mathbf{Z}}_{m_{1},n_{2}}^{H}\Big\}  \tilde{\mathbf{Z}}_{m_{1},n_{1}}\Big\}  {\bf\Phi}_{n_{1}}\overline{\mathbf{h}}_{n_{1},k}}\\
		=\varepsilon_{n_{1},k}N_{r} 
		{\overline{\mathbf{h}}_{n_{1},k}^{H}{\bf\Phi}^{H}_{n_{1}} \mathbb{E}\Big\{\tilde{\mathbf{Z}}_{m_{1},n_{1}}^{H}  \tilde{\mathbf{Z}}_{m_{1},n_{1}}\Big\}  {\bf\Phi}_{n_{1}}\overline{\mathbf{h}}_{n_{1},k}}\\
		=\varepsilon_{n_{1},k}M_{b} N_{r}^{2},
	\end{array}
\end{align}
part 2:
\begin{align}
	\begin{array}{l}
		%   3  4  4  3 (2)
		\Bigg(\mathbb{E}\Bigg\{  
		\Big(\mathbf{g}_{m_{1},n_{1},k}^{3}\Big)^{H}\mathbf{g}_{m_{1},n_{2},k}^{4}
		\Big(\mathbf{g}_{m_{2},n_{3},k}^{4}\Big)^{H}
		\mathbf{g}_{m_{2},n_{4},k}^{3}
		\Bigg\} \Bigg)_{2}\\
		\overset{\substack{(j)}}{=}
		\varepsilon_{n_{1},k}
		{\overline{\mathbf{h}}_{n_{1},k}^{H}{\bf\Phi}^{H}_{n_{1}} \mathbb{E}\Big\{\tilde{\mathbf{Z}}_{m_{1},n_{1}}^{H} \tilde{\mathbf{Z}}_{m_{1},n_{1}}\Big\} {\bf\Phi}_{n_{1}} \mathbb{E}\Big\{\tilde{\mathbf{h}}_{n_{1},k} \tilde{\mathbf{h}}_{n_{1},k}^{H}\Big\} {\bf\Phi}^{H}_{n_{1}}  \mathbb{E}\Big\{\tilde{\mathbf{Z}}_{m_{2},n_{1}}^{H}\tilde{\mathbf{Z}}_{m_{2},n_{1}}\Big\}  {\bf\Phi}_{n_{1}}\overline{\mathbf{h}}_{n_{1},k}}\\
		=\varepsilon_{n_{1},k}M_{b}^{2} 
		{\overline{\mathbf{h}}_{n_{1},k}^{H}{\bf\Phi}^{H}_{n_{1}}  {\bf\Phi}_{n_{1}}  {\bf\Phi}^{H}_{n_{1}}   {\bf\Phi}_{n_{1}}\overline{\mathbf{h}}_{n_{1},k}}\\
		=\varepsilon_{n_{1},k}M_{b}^{2} N_{r}.
	\end{array}
\end{align}

Finally, we consider the terms with $\omega= 4$. When $\psi=1$, we have
\begin{align}
	\begin{array}{l}
		%   4  1  1  4 
		\mathbb{E}\Bigg\{  
		\Big(\mathbf{g}_{m_{1},n_{1},k}^{4}\Big)^{H}\mathbf{g}_{m_{1},n_{2},k}^{1}
		\Big(\mathbf{g}_{m_{2},n_{3},k}^{1}\Big)^{H}
		\mathbf{g}_{m_{2},n_{4},k}^{4}
		\Bigg\} \\
		\overset{\substack{(h)}}{=}
		\sqrt{\delta_{m_{1},n_{2}}\delta_{m_{1},n_{3}} \varepsilon_{n_{2},k}\varepsilon_{n_{3},k}}
		\;{\mathbb{E}\Big\{\tilde{\mathbf{h}}_{n_{1},k}^{H}{\bf\Phi}^{H}_{n_{1}} \mathbb{E}\Big\{\tilde{\mathbf{Z}}_{m_{1},n_{1}}^{H} \overline{\mathbf{Z}}_{m_{1},n_{2}} {\bf\Phi}_{n_{2}} \overline{\mathbf{h}}_{n_{2},k} \overline{\mathbf{h}}_{n_{3},k}^{H}{\bf\Phi}^{H}_{n_{3}}  \overline{\mathbf{Z}}_{m_{1},n_{3}}^{H}\tilde{\mathbf{Z}}_{m_{1},n_{1}}\Big\}  {\bf\Phi}_{n_{1}}\tilde{\mathbf{h}}_{n_{1},k}\Big\} }
		\\
		=\sqrt{\delta_{m_{1},n_{2}}\delta_{m_{1},n_{3}}\varepsilon_{n_{2},k}\varepsilon_{n_{3},k}}
		{\mathbb{E}\Big\{\tilde{\mathbf{h}}_{n_{1},k}^{H}{\bf\Phi}^{H}_{n_{1}}  {\bf\Phi}_{n_{1}}\tilde{\mathbf{h}}_{n_{1},k}\Big\} \mathrm{Tr}\Big\{\overline{\mathbf{Z}}_{m_{1},n_{2}} {\bf\Phi}_{n_{2}} \overline{\mathbf{h}}_{n_{2},k} \overline{\mathbf{h}}_{n_{3},k}^{H}{\bf\Phi}^{H}_{n_{3}}  \overline{\mathbf{Z}}_{m_{1},n_{3}}^{H}\Big\} }
		\\
		=\sqrt{\delta_{m_{1},n_{2}}\delta_{m_{1},n_{3}} \varepsilon_{n_{2},k}\varepsilon_{n_{3},k}}N_{r}
		{f_{m_{1},n_{3},k}^{H}({\bf \Phi}) f_{m_{1},n_{2},k}({\bf \Phi})  
			{\bf a}^{H}_{M_{b}}(m_{1},n_{3}) 
			{\bf a}_{M_{b}}(m_{1},n_{2})  }.
	\end{array}
\end{align}

When $\psi=2$, we derive the term by calculating two parts as follows
\\part 1:
\begin{align}
	\begin{array}{l}
		%   4  2  2  4  (1)
		\Bigg(\mathbb{E}\Bigg\{  
		\Big(\mathbf{g}_{m_{1},n_{1},k}^{4}\Big)^{H}\mathbf{g}_{m_{1},n_{2},k}^{2}
		\Big(\mathbf{g}_{m_{2},n_{3},k}^{2}\Big)^{H}
		\mathbf{g}_{m_{2},n_{4},k}^{4}
		\Bigg\} \Bigg)_{1}\\
		\overset{\substack{(g)}}{=}
		\delta_{m_{1},n_{2}}
		{\mathbb{E}\Big\{\tilde{\mathbf{h}}_{n_{1},k}^{H}{\bf\Phi}^{H}_{n_{1}} \mathbb{E}\Big\{\tilde{\mathbf{Z}}_{m_{1},n_{1}}^{H} \overline{\mathbf{Z}}_{m_{1},n_{2}} {\bf\Phi}_{n_{2}} \mathbb{E}\Big\{\tilde{\mathbf{h}}_{n_{2},k} \tilde{\mathbf{h}}_{n_{2},k}^{H}\Big\}{\bf\Phi}^{H}_{n_{2}}  \overline{\mathbf{Z}}_{m_{1},n_{2}}^{H}\tilde{\mathbf{Z}}_{m_{1},n_{1}}\Big\}  {\bf\Phi}_{n_{1}}\tilde{\mathbf{h}}_{n_{1},k}\Big\} }
		\\
		=\delta_{m_{1},n_{2}}
		{\mathbb{E}\Big\{\tilde{\mathbf{h}}_{n_{1},k}^{H}{\bf\Phi}^{H}_{n_{1}} \mathbb{E}\Big\{\tilde{\mathbf{Z}}_{m_{1},n_{1}}^{H} \overline{\mathbf{Z}}_{m_{1},n_{2}}  \overline{\mathbf{Z}}_{m_{1},n_{2}}^{H}\tilde{\mathbf{Z}}_{m_{1},n_{1}}\Big\}  {\bf\Phi}_{n_{1}}\tilde{\mathbf{h}}_{n_{1},k}\Big\} }
		\\
		=\delta_{m_{1},n_{2}}M_{b}N_{r}^{2},
	\end{array}
\end{align}
part 2:
\begin{align}
	\begin{array}{l}
		%   4  2  2  4  (2)
		\Bigg(\mathbb{E}\Bigg\{  
		\Big(\mathbf{g}_{m_{1},n_{1},k}^{4}\Big)^{H}\mathbf{g}_{m_{1},n_{2},k}^{2}
		\Big(\mathbf{g}_{m_{2},n_{3},k}^{2}\Big)^{H}
		\mathbf{g}_{m_{2},n_{4},k}^{4}
		\Bigg\} \Bigg)_{2}\\
		\overset{\substack{(n)}}{=}
		\delta_{m_{1},n_{1}}
		{\mathbb{E}_{\tilde{\mathbf{h}}_{n_{1},k}}\Big\{\tilde{\mathbf{h}}_{n_{1},k}^{H}{\bf\Phi}^{H}_{n_{1}} \mathbb{E}_{\tilde{\mathbf{Z}}_{m_{1},n_{1}}}\Big\{\tilde{\mathbf{Z}}_{m_{1},n_{1}}^{H} \overline{\mathbf{Z}}_{m_{1},n_{1}} {\bf\Phi}_{n_{1}} \tilde{\mathbf{h}}_{n_{1},k} \tilde{\mathbf{h}}_{n_{1},k}^{H} {\bf\Phi}^{H}_{n_{1}}  \overline{\mathbf{Z}}_{m_{1},n_{1}}^{H}\tilde{\mathbf{Z}}_{m_{1},n_{1}}\Big\}  {\bf\Phi}_{n_{1}}\tilde{\mathbf{h}}_{n_{1},k}
			\mid \tilde{\mathbf{h}}_{n_{1},k} \Big\} }
		\\
		=\delta_{m_{1},n_{1}}
		{\mathbb{E}\Big\{\tilde{\mathbf{h}}_{n_{1},k}^{H}{\bf\Phi}^{H}_{n_{1}} \mathrm{Tr}\Big\{\overline{\mathbf{Z}}_{m_{1},n_{1}} {\bf\Phi}_{n_{1}} \tilde{\mathbf{h}}_{n_{1},k} \tilde{\mathbf{h}}_{n_{1},k}^{H} {\bf\Phi}^{H}_{n_{1}}  \overline{\mathbf{Z}}_{m_{1},n_{1}}^{H}\Big\}  {\bf\Phi}_{n_{1}}\tilde{\mathbf{h}}_{n_{1},k} \Big\} }
		\\
		=\delta_{m_{1},n_{1}}
		{\mathbb{E}\Big\{\tilde{\mathbf{h}}_{n_{1},k}^{H}{\bf\Phi}^{H}_{n_{1}}  {\bf\Phi}_{n_{1}}\tilde{\mathbf{h}}_{n_{1},k}  \tilde{\mathbf{h}}_{n_{1},k}^{H} {\bf\Phi}^{H}_{n_{1}}  \overline{\mathbf{Z}}_{m_{1},n_{1}}^{H} \overline{\mathbf{Z}}_{m_{1},n_{1}} {\bf\Phi}_{n_{1}}\tilde{\mathbf{h}}_{n_{1},k}  \Big\} }
		\\
		=\delta_{m_{1},n_{1}}
		{\mathrm{Tr}\Big\{\mathbb{E}\Big\{\tilde{\mathbf{h}}_{n_{1},k}  \tilde{\mathbf{h}}_{n_{1},k}^{H} {\bf\Phi}^{H}_{n_{1}}  \overline{\mathbf{Z}}_{m_{1},n_{1}}^{H} \overline{\mathbf{Z}}_{m_{1},n_{1}} {\bf\Phi}_{n_{1}}\tilde{\mathbf{h}}_{n_{1},k} \tilde{\mathbf{h}}_{n_{1},k}^{H} \Big\}\Big\} }
		\\
\end{array}
\end{align}
where the part 2 utilizes the property in (\ref{ZZWZZ}). The second part is the special case of part 1. The term can be derived by combining these two parts, we have 
\begin{align}
	\begin{array}{l}
		%   4  2  2  4  
		{\sum\limits_{m_{1}=1}^{M}\sum\limits_{m_{2}=1}^{M}\sum\limits_{n_{1}=1}^{N}\sum\limits_{n_{2}=1}^{N}\sum\limits_{n_{3}=1}^{N}\sum\limits_{n_{4}=1}^{N}}
		\sqrt{c_{m_{1},n_{1},k} c_{m_{1},n_{2},k} c_{m_{2},n_{3},k} c_{m_{2},n_{4},k}}
		\\ \qquad\qquad\qquad
		\mathbb{E}\Bigg\{  
		\Big(\mathbf{g}_{m_{1},n_{1},k}^{4}\Big)^{H}\mathbf{g}_{m_{1},n_{2},k}^{2}
		\Big(\mathbf{g}_{m_{2},n_{3},k}^{2}\Big)^{H}
		\mathbf{g}_{m_{2},n_{4},k}^{4}
		\Bigg\} \\
		={\sum\limits_{m=1}^{M}\sum\limits_{n_{1}=1}^{N}\sum\limits_{n_{2}=1}^{N}}
		c_{m,n_{1},k} c_{m,n_{2},k}  \delta_{m,n_{2}} M_{b} N_{r}^{2} +
		{\sum\limits_{m=1}^{M}\sum\limits_{n=1}^{N}} c_{m,n,k}^{2} \delta_{m,n} M_{b} N_{r}.
	\end{array}
\end{align}

When $\psi=3$, the term is the addition of the following two parts
\\part 1:
\begin{align}
	\begin{array}{l}
		%   4  3  3  4  (1)
		\Bigg(\mathbb{E}\Bigg\{  
		\Big(\mathbf{g}_{m_{1},n_{1},k}^{4}\Big)^{H}\mathbf{g}_{m_{1},n_{2},k}^{3}
		\Big(\mathbf{g}_{m_{2},n_{3},k}^{3}\Big)^{H}
		\mathbf{g}_{m_{2},n_{4},k}^{4}
		\Bigg\} \Bigg)_{1}\\
		\overset{\substack{(g)}}{=}
		\varepsilon_{n_{2},k}
		{\mathbb{E}\Big\{\tilde{\mathbf{h}}_{n_{1},k}^{H}{\bf\Phi}^{H}_{n_{1}} \mathbb{E}\Big\{\tilde{\mathbf{Z}}_{m_{1},n_{1}}^{H} \mathbb{E}\Big\{\tilde{\mathbf{Z}}_{m_{1},n_{2}} {\bf\Phi}_{n_{2}} \overline{\mathbf{h}}_{n_{2},k} \overline{\mathbf{h}}_{n_{2},k}^{H}{\bf\Phi}^{H}_{n_{2}}  \tilde{\mathbf{Z}}_{m_{1},n_{2}}^{H}\Big\}\tilde{\mathbf{Z}}_{m_{1},n_{1}}\Big\}  {\bf\Phi}_{n_{1}}\tilde{\mathbf{h}}_{n_{1},k}\Big\} }
		\\
		=\varepsilon_{n_{2},k}
		{\mathbb{E}\Big\{\tilde{\mathbf{h}}_{n_{1},k}^{H}{\bf\Phi}^{H}_{n_{1}} \mathbb{E}\Big\{\tilde{\mathbf{Z}}_{m_{1},n_{1}}^{H} \tilde{\mathbf{Z}}_{m_{1},n_{1}}\Big\}  {\bf\Phi}_{n_{1}}\tilde{\mathbf{h}}_{n_{1},k}\Big\}\mathrm{Tr}\Big\{{\bf\Phi}_{n_{2}} \overline{\mathbf{h}}_{n_{2},k} \overline{\mathbf{h}}_{n_{2},k}^{H}{\bf\Phi}^{H}_{n_{2}}\Big\} }
		\\
		=\varepsilon_{n_{2},k}M_{b}N_{r}^{2},
	\end{array}
\end{align}
part 2:
\begin{align}
	\begin{array}{l}
		%   4  3  3  4  (2)
		\Bigg(\mathbb{E}\Bigg\{  
		\Big(\mathbf{g}_{m_{1},n_{1},k}^{4}\Big)^{H}\mathbf{g}_{m_{1},n_{2},k}^{3}
		\Big(\mathbf{g}_{m_{2},n_{3},k}^{3}\Big)^{H}
		\mathbf{g}_{m_{2},n_{4},k}^{4}
		\Bigg\} \Bigg)_{2}\\
		\overset{\substack{(j)}}{=}
		\varepsilon_{n_{1},k}
		{\mathbb{E}\Big\{\tilde{\mathbf{h}}_{n_{1},k}^{H}{\bf\Phi}^{H}_{n_{1}} \mathbb{E}\Big\{\tilde{\mathbf{Z}}_{m_{1},n_{1}}^{H} \tilde{\mathbf{Z}}_{m_{1},n_{1}}\Big\} {\bf\Phi}_{n_{1}} \overline{\mathbf{h}}_{n_{1},k} \overline{\mathbf{h}}_{n_{1},k}^{H}{\bf\Phi}^{H}_{n_{1}}  \mathbb{E}\Big\{\tilde{\mathbf{Z}}_{m_{2},n_{1}}^{H}\tilde{\mathbf{Z}}_{m_{2},n_{1}}\Big\}  {\bf\Phi}_{n_{1}}\tilde{\mathbf{h}}_{n_{1},k}\Big\} }
		\\
		=\varepsilon_{n_{1},k}M_{b}^{2} 
		{\mathbb{E}\Big\{\tilde{\mathbf{h}}_{n_{1},k}^{H}{\bf\Phi}^{H}_{n_{1}}  {\bf\Phi}_{n_{1}} \overline{\mathbf{h}}_{n_{1},k} \overline{\mathbf{h}}_{n_{1},k}^{H}{\bf\Phi}^{H}_{n_{1}}   {\bf\Phi}_{n_{1}}\tilde{\mathbf{h}}_{n_{1},k}\Big\} }
		\\
		=\varepsilon_{n_{1},k}M_{b}^{2} 
		\mathbb{E}\Big\{\tilde{\mathbf{h}}_{n_{1},k}^{H}   \overline{\mathbf{h}}_{n_{1},k} \overline{\mathbf{h}}_{n_{1},k}^{H} \tilde{\mathbf{h}}_{n_{1},k} \Big\}
		\\
		=\varepsilon_{n_{1},k}M_{b}^{2} 
		{\overline{\mathbf{h}}_{n_{1},k}^{H}   \mathbb{E}\Big\{\tilde{\mathbf{h}}_{n_{1},k} \tilde{\mathbf{h}}_{n_{1},k}^{H}\Big\} \overline{\mathbf{h}}_{n_{1},k} }
		\\
		=\varepsilon_{n_{1},k} M_{b}^{2} N_{r}.
	\end{array}
\end{align}

When $\psi=4$, based on (\ref{ZZWZZ}), the term is derived by calculating four parts as follows
\\part 1:
\begin{align}
	%
\end{align}
The term can be derived by combining the four parts, we have 
\begin{align}
	\begin{array}{l}
		%   4  4  4  4  
		{\sum\limits_{m_{1}=1}^{M}\sum\limits_{m_{2}=1}^{M}\sum\limits_{n_{1}=1}^{N}\sum\limits_{n_{2}=1}^{N}\sum\limits_{n_{3}=1}^{N}\sum\limits_{n_{4}=1}^{N}}
		\sqrt{c_{m_{1},n_{1},k} c_{m_{1},n_{2},k} c_{m_{2},n_{3},k} c_{m_{2},n_{4},k}}
		\\ \qquad\qquad\qquad
		\mathbb{E}\Bigg\{ 
		\Big(\mathbf{g}_{m_{1},n_{1},k}^{4}\Big)^{H}\mathbf{g}_{m_{1},n_{2},k}^{4}
		\Big(\mathbf{g}_{m_{2},n_{3},k}^{4}\Big)^{H}
		\mathbf{g}_{m_{2},n_{4},k}^{4}
		\Bigg\} \\
		={\sum\limits_{m_{1}=1}^{M}\sum\limits_{m_{2}=1}^{M}\sum\limits_{n_{1}=1}^{N}\sum\limits_{n_{2}=1}^{N}}
		c_{m_{1},n_{1},k} c_{m_{2},n_{2},k}   M_{b}^{2} N_{r}^{2} +
		{\sum\limits_{m_{1}=1}^{M}\sum\limits_{m_{2}=1}^{M}\sum\limits_{n=1}^{N}}c_{m_{1},n,k}c_{m_{2},n,k} M_{b}^{2} N_{r}\\
		\qquad+{\sum\limits_{m=1}^{M}\sum\limits_{n_{1}=1}^{N}\sum\limits_{n_{2}=1}^{N}}c_{m,n_{1},k}c_{m,n_{2},k} M_{b} N_{r}^{2}+
		{\sum\limits_{m=1}^{M}\sum\limits_{n=1}^{N}}c_{m,n,k}^{2} M_{b} N_{r}.
	\end{array}
\end{align}

Similar to the above derivations, we derive the remaining ten parts in (\ref{gk_4}) one by one.

The first term can be derived as follows
\begin{align}
	%
\end{align}

The second term is
\begin{align}
	%
\end{align}

The third term is
\begin{align}
	%
\end{align}

The fourth term is
\begin{align}
	%
\end{align}

The fifth term is
\begin{align}
	%
\end{align}
where $(o)$ represents $m_{1}=m_{2}$, $n_{2}=n_{3}=n_{4}$ and removes the zero terms.

The sixth term is
\begin{align}
	%
\end{align}

The seventh term is
\begin{align}
	%
\end{align}

The eighth term is derived by calculating two parts as follows
\\part 1:
\begin{align}
	%
\end{align}
where part 2 utilizes the property in (\ref{ZZWZZ}).
The second part is the special case of part 1. The term can be derived by combining the two parts, we have 
\begin{align}
	%
\end{align}

The ninth term is
\begin{align}
	%
\end{align}
where $(p)$ represents $m_{1}=m_{2}$, $n_{1}=n_{3}=n_{4}$ and removes the zero terms.

The tenth term is derived by calculating two parts as follows
\\part 1:
\begin{align}
	%
\end{align}
where the part 2 utilizes the property in (\ref{ZZWZZ}).
The second part is the special case of part 1. The term can be derived by combining the two parts, we have 
\begin{align}
	%
\end{align}

Thus, the calculation of the remaining ten parts has been completed. After some simplifications, we can obtain $\mathbb{E}\left\{\left\|\mathbf{g}_{k}\right\|^{4}\right\}$ by combining (\ref{gk4_begin}) $\sim$ (\ref{gk4_end}) with (\ref{gk_4}). With the aid of (\ref{signal_term}) and (\ref{g_k2}), we can complete the calculation of the signal term $\mathbb{E}\left\{\left\|{\bf g}_k+{\bf d}_k\right\|^{4}\right\}$.

\section{}\label{appA}
To begin with, we present some properties which will be utilized in the following derivation.

For the expectation $ \mathbb{E}\left\{{\sum\limits_{m_{1}=1}^M\sum\limits_{m_{2}=1}^M \sum\limits_{n_{1}=1 }^N \sum\limits_{n_{2}=1 }^N}{{f_{m_{1},n_{1},k}^{H}({\bf\Phi})}{f_{m_{2},n_{2},k}({\bf\Phi})}}\right\}$ with respect to random phase shifts $\bf \Phi$, using the same method as \cite[Corollary 4]{9743440}, we have
\begin{align}\label{B_begin}

\end{align} 
which exploits the independence and the zero-mean properties of ${\bf\Phi}_{n_{1}}$ and ${\bf\Phi}_{n_{2}}$, $\forall n_{1} \neq n_{2}$.

For the expectation $ \mathbb{E}\left\{{\sum\limits_{m_{1}=1}^M\sum\limits_{m_{2}=1}^M \sum\limits_{n_{1}=1 }^N \sum\limits_{n_{2}=1 }^N}{{f_{m_{1},n_{1},k}^{H}({\bf\Phi})}{f_{m_{2},n_{2},i}({\bf\Phi})}}\right\}$, we have
\begin{align}
	%
\end{align} 

Based on the independence and the zero-mean properties of ${\bf\Phi}_{n_{1}}$ and ${\bf\Phi}_{n_{2}}$, we can ignore the zero term in the next derivation. For the expectation $ \mathbb{E}\bigg\{{\sum\limits_{m_{1}=1}^M\sum\limits_{m_{2}=1}^M \sum\limits_{n_{1}=1}^N\sum\limits_{n_{2}=1}^N \sum\limits_{n_{3}=1 }^N \sum\limits_{n_{4}=1 }^N} {{f_{m_{1},n_{1},k}^{H}({\bf\Phi})}}$$\\{{f_{m_{1},n_{2},i}({\bf\Phi})}}{{f_{m_{2},n_{3},i}^{H}({\bf\Phi})}{f_{m_{2},n_{4},k}({\bf\Phi})}}\bigg\}$, we have 
\begin{align}
	%
\end{align} 
and all the terms
\begin{align}
	%
\end{align}
are bounded when $N_{r}\to\infty$.

Similarly, the expectation $ \mathbb{E}\bigg\{{\sum\limits_{m_{1}=1}^M\sum\limits_{m_{2}=1}^M \sum\limits_{n_{1}=1}^N\sum\limits_{n_{2}=1}^N \sum\limits_{n_{3}=1 }^N \sum\limits_{n_{4}=1 }^N} {{f_{m_{1},n_{1},k}^{H}({\bf\Phi})}{f_{m_{1},n_{2},k}({\bf\Phi})}}{{f_{m_{2},n_{3},k}^{H}({\bf\Phi})}}\\{f_{m_{2},n_{4},k}({\bf\Phi})}\bigg\}$ is derived as
\begin{align}
	%
\end{align}

Next, based on (\ref{B_begin}) and (\ref{B_end}), we can calculate the expectation of terms involving $\bf\Phi$ in (\ref{rate}). Considering a large value of $N_{r}$, we can ignore the insignificant terms which are not on the order of $\mathcal{O}(N_{r})$. 
%The derivation will be carried out in three parts
We first derive the terms involving $\bf\Phi$ in the noise term $E_{k}^{(noise)}({\bf\Phi})$ as
%we need to provide some necessary preliminary results.
\begin{align}
	%
\end{align} 

For the signal term ${E}_{k}^{(signal)}({\bf\Phi})$, there are seven terms involving $\bf\Phi$ which will be derived one by one. The first term is derived as
\begin{align}
	%
\end{align}

The second term can be derived as follows
\begin{align}
	%
\end{align}

The third term can be derived as follows
\begin{align}
	%
\end{align}

The fourth term can be derived as follows
\begin{align}
	%
%\end{align}

The fifth term is
\begin{align}
	%
\end{align}

The sixth term is
\begin{align}
	%
\end{align}

The seventh term is
\begin{align}
	%
\end{align}

For the interference term $I_{ki}({\bf\Phi})$, there are seven parts involving $\bf\Phi$ which will be derived in sequence.
The first term is 
\begin{align}
	%
\end{align}

The second term is
\begin{align}
	%
\end{align}
which is bounded when $N_{r}\to\infty$.

The third term is
\begin{align}
	%
\end{align}

The fourth term is
\begin{align}
	%
\end{align}

The fifth term is
\begin{align}
	%
\end{align}

The sixth term is
\begin{align}
	%
\end{align}

The seventh term is
\begin{align}
	%
\end{align}

Thus, we have completed the calculation of the expectation of terms involving $\bf\Phi$ in (\ref{rate}). When $N_{r}\to\infty$ or $M_{b}\to\infty$, we can ignore the insignificant terms which are not on the order of $\mathcal{O}(N_{r}^2)$ or $\mathcal{O}(M_{b}^2)$. After some direct simplifications, we can derive the data rate expression in (\ref{SINK_rm_1}).

\end{appendices}

%%%%%%%%%%%%%%%%%%%%%%%%%%%%%%%%%%%%% Reference
\bibliographystyle{IEEEtran}
\vspace{-6pt}
\bibliography{myref.bib}
%\end{thebibliography}

\end{document}